\documentclass[11pt]{article}
\usepackage{amsmath}
\usepackage{amsthm}
\allowdisplaybreaks
\usepackage{amssymb}
\usepackage{algorithm}
\usepackage{subfig}
\usepackage{color}
\usepackage[dvipsnames]{xcolor}
\usepackage{graphicx}
\usepackage{wrapfig,epsfig}
\usepackage{epstopdf}
\usepackage{url}
\usepackage{graphicx}
\usepackage{color}
\usepackage{epstopdf}
\usepackage{algpseudocode}
\usepackage{scrextend}
\usepackage[T1]{fontenc}
\usepackage{bbm}
\usepackage{comment}

\usepackage{multicol}
\usepackage{dsfont}
\usepackage{mathtools}
\usepackage{enumitem}

\let\C\relax
\usepackage{tikz}
\usepackage{hyperref}
\hypersetup{colorlinks=true,citecolor=red,linkcolor=blue}

\usetikzlibrary{arrows}
\usepackage[margin=1in]{geometry}
\graphicspath{{./figs/}}

\definecolor{b2}{RGB}{51,153,255}
\definecolor{mygreen}{RGB}{80,180,0}

\theoremstyle{plain}
\newtheorem{theorem}{Theorem}[section]
\newtheorem{lemma}[theorem]{Lemma}

\newtheorem{fact}[theorem]{Fact}
\newtheorem{claim}[theorem]{Claim}
\newtheorem{corollary}[theorem]{Corollary}

\newtheorem{question}[theorem]{Question}

\newtheorem{remark}[theorem]{Remark}

\theoremstyle{definition}
\newtheorem{definition}[theorem]{Definition}

\newtheorem{task}[theorem]{Task}

\newcommand{\wh}{\widehat}
\newcommand{\wt}{\widetilde}
\newcommand{\ov}{\overline}

\renewcommand{\epsilon}{\varepsilon}
\renewcommand{\phi}{\varphi}

\newcommand{\N}{\mathcal{N}}
\newcommand{\R}{\mathbb{R}}
\newcommand{\C}{\mathbb{C}}

\renewcommand{\tilde}{\wt}
\renewcommand{\hat}{\wh}

\newcommand{\poly}{\mathrm{poly}}
\newcommand{\tr}{\mathrm{tr}}
\newcommand{\Tmat}{{\cal T}_{\mathrm{mat}}}

\DeclareMathOperator{\pr}{Pr}
\newcommand{\E}{\mathbb{E}}

\makeatletter
\newcommand*{\RN}[1]{\expandafter\@slowromancap\romannumeral #1@}
\makeatother

\newcommand{\afn}{\mathsf{AFN}}

\newcommand{\fail}{$\mathsf{fail}$}
 
\newcommand{\vect}{\mathrm{vec}}
\newcommand{\minip}{\mathsf{Min}\text{-}\mathsf{IP}}
\newcommand{\aipe}{\mathsf{AIPE}} 
\newcommand{\ade}{\mathsf{ADE}} 
\newcommand{\dfn}{\mathsf{DFN}}
\newcommand{\bw}{\mathrm{bw}}
\newcommand{\maxip}{\mathsf{Max}\text{-}\mathsf{IP}}
\newcommand{\tensorsketch}{\mathsf{TensorSketch}}
\newcommand{\tensorsrht}{\mathsf{TensorSRHT}}
\newcommand{\tensorsparse}{\mathsf{TensorSparse}}
\newcommand{\nnz}{\mathrm{nnz}}
\title{Speeding Up Sparsification using Inner Product Search \\Data Structures}

\author{
Zhao Song\thanks{\texttt{zsong@adobe.com}. Adobe Research.}
\and
Zhaozhuo Xu\thanks{\texttt{zx22@rice.edu}. Rice University.}
\and
Lichen Zhang\thanks{\texttt{lichenz@andrew.cmu.edu}. Carnegie Mellon University.}
}

\date{}

\begin{document}

\begin{titlepage}
    \maketitle
    \begin{abstract}
        We present a general framework that utilizes different efficient data structures to improve various sparsification problems involving an iterative process. We also provide insights and characterization for different iterative process, and answer that when should we use which data structures in what type of problem. We obtain improved running time for the following problems.

\begin{itemize}
    \item For constructing linear-sized spectral sparsifier, 
    all the existing deterministic algorithms require $\Omega(d^4)$ time \cite{bss12,z12}. In this work, we provide the first deterministic algorithm that breaks that barrier which runs in $O(d^{\omega+1})$ time, where $\omega$ is the exponent of matrix multiplication.
    \item For one-sided Kadison-Singer-typed discrepancy problem~\cite{w13}, we give fast algorithms for both small and large number of iterations.
    \item For experimental design problem~\cite{azlsw20}, we speed up a key swapping process.
\end{itemize}

In the heart of our work is the design of a variety of different inner product search data structures that have efficient initialization, query and update time, compatible to dimensionality reduction and robust against adaptive adversary.

    \end{abstract}
    \thispagestyle{empty}
\end{titlepage}


\section{Introduction}\label{sec:intro}

Speeding up iterative process and obtaining faster algorithms has always been a central topic in theoretical computer science. In recent years of development, various breakthroughs have been achieved in improving the running time of Laplacian solver~\cite{st11,kosz13,ckmp+14},
linear programming~\cite{v89,cls19,jswz21,dly21}, empirical risk minimization~\cite{lsz19}, semi-definite programming~\cite{lsw15,jlsw20,hjstz21} and sum-of-squares method~\cite{jnw22}. A key ingredient in these improvements is the use of efficient data structures to reduce the cost per iteration of the iterative process. Data structures not only speed up the algorithm, but also expose the inherent structure of the problem to solve. While data structures are prevalent in many continuous optimization problems, they are rarely used when solving a large class of problems revolving around \emph{inner product}, such as linear-sized spectral sparsifier~\cite{bss12,z12,azlo15,ls15,ls17}, restricted invertibility and its variant~\cite{s10,w13} and experimental design problem~\cite{azlsw20,lz20}. In all these problems, one typically gives a set of vectors $V:=\{v_1,\ldots,v_m\}$ and at each iteration, one forms a query matrix $A$, the goal is to search a vector $v_i$ such that the inner product $\langle v_iv_i^\top, A\rangle$ satisfies certain constraints. Standard techniques to speed up these iterative processes include using a more refined potential function to reduce the number of iterations~\cite{azlo15,ls15} or a more powerful solver at each iteration for stronger objectives~\cite{ls17}. From a data structure design perspective, one wishes to develop efficient, adaptive and high accuracy data structures for inner product type queries and combines them into the iterative process to reduce the cost per iteration. The inner product queries we need to handle including the following: given a query matrix $A$, find a vector $v_i$ such that $\langle v_iv_i^\top, A\rangle > 0$ or find the $v_i$ that (approximates) minimizes the inner product $\langle v_iv_i^\top, A\rangle$ in $V$. To this end, we show that 1).\ In the spectral sparsifier task, we use simple yet highly efficient and effective deterministic data structures to speed up the algorithm posed in~\cite{bss12}. We obtain the fastest deterministic algorithm for constructing a BSS sparsifier. Our algorithm is conceptually simple and easy to implement compared to the SDP-based solution of~\cite{ls17}. 2).\ In the task of~\cite{w13,azlsw20}, we develop data structures that solve the minimum inner product search problem and gain significant speedup from~\cite{w13,azlsw20} for both small and large numbers of iterations. To support the deployment of our approximated data structure, we provide a robust analysis on the quality of solution of~\cite{w13,azlsw20}. Our work can be viewed as a combination of efficient data structure and robust analysis of the spectrum potential~\cite{s10,bss12}.

\subsection{Related Work}

\paragraph{Speeding Up Iterations via Efficient Data Structures.} Given an optimization problem that involves an iterative process, we can decompose the running time into two-folds: 1).\ Number of iterations and 2).\ Cost per iteration. Reduce the number of iterations has led to significant breakthroughs for various problems, such as maximum flow~\cite{ds08,ckmst11, m13_flow,m16} and linear programming~\cite{ks06,ls14}. In recent years, however, more efforts have been dedicated to reduce the cost per iteration via data structures, which lead to the fastest known algorithms for various problems~\cite{v89,lsw15,cls19,lsz19,jlsw20,jswz21,sy21,y21,dly21,hjstz21,jnw22}. While data structures have been playing an important roles for these results, most of them are complicated, cumbersome and adapted in a black-box manner. It is also imperative to make them robust against adaptive queries, since in an iterative process, subsequent queries can well depend on the result outputted by the data structure from prior iterations. Efforts have been made to design generic adaptive data structures for norm estimations, but extra slowdown seems to be inevitably due to the necessity of handling adaptive queries. One important direction in this area is to simplify these algorithms with simpler data structures and analysis, similar to the simplification of Laplacian solver via simple, combinatorial data structures as in~\cite{kosz13}.

\paragraph{Spectral Sparsification and Algorithms via Spectrum Potential.} Given a matrix $V\in \R^{m\times d}$ in which $m\gg d$, the goal is to select a subset of $s$ rescaled rows where $s\ll m$ to form a new matrix $\wt V\in \R^{s\times d}$, such that $(1-\epsilon) V^\top V\preceq \wt V^\top \wt V^\top \preceq (1+\epsilon) V^\top V$. Leverage score sampling~\cite{ss11} gives a fast algorithm to find $s=\Theta(\epsilon^{-2}d\log d)$ such rows, and similar idea has been investigated for graph in the semi-streaming setting~\cite{kl12}. The optimal result regarding $s$ is obtained by Batson, Spielman and Srivastava~\cite{bss12} in which $s=\Theta(\epsilon^{-2}d)$. In the setting where $V^\top V$ is a graph Laplacian matrix, this produces a spectral sparsifier with only $\Theta(\epsilon^{-2}n)$ edges. Unlike leverage score sampling, the original algorithm in~\cite{bss12} is rather slow. To speed up this process,~\cite{azlo15,ls15} adapt a new potential function that reduces the number of iterations required in the expense of a worse size of sparsifier ($s=\Theta(\epsilon^{-2}qd)$ for $q\geq 10$ being an integer). By using an SDP-based solver,~\cite{ls17} achieves a nearly linear running time for graph and a nearly optimal running time $\epsilon^{-O(1)} ((\sum_{i\in [m]}\nnz(v_i)^2)+d^\omega)$ for general matrices where $\omega$ is the exponent for matrix multiplication~\cite{w12,l14,aw21}. However, all these methods are randomized, the only known deterministic construction faster than~\cite{bss12} is due to~\cite{z12}, in which it obtains an algorithm that runs in $\wt O(\epsilon^{-2}md^2+\epsilon^{-4}d^4)$. Apart from graph spectral sparsifier, it also finds applications in various numerical linear algebra tasks, such as constrained linear regression, multi-response regression~\cite{bdm13}. It also has important usage when one looks for a low rank approximation of $V$ using its own rescaled rows and columns, such as matrix CUR decomposition and tensor CURT decomposition~\cite{bw14,swz17,swz19_soda}. The sampling distribution described by the BSS process is also useful for combating the presence of noise in Fourier signal interpolation task~\cite{cp19}.

The potential function developed in~\cite{bss12} has a wide range of other applications~\cite{s10,w13,azlsw20,lz20}, e.g., in~\cite{w13}, we are given a matrix $V\in \R^{m\times d}$ with $V^\top V=I$ and each row has $\ell_2$ norm $\frac{1}{\sqrt N}$. The goal is to pick a subset $S$ of the rows such that $\|V_S^\top V_S\| \leq \frac{n}{m}+O(\frac{1}{\sqrt N})$, where $n$ is the cardinality of $S$. In the experimental design problem of~\cite{azlsw20}, they obtain a rounding algorithm by swapping vectors based on the potential defined by the vectors, and the randomized version of their method has been used in spectral network design~\cite{lz20}.

\section{Data Structures for Inner Product Query}

To develop efficient data structures for optimization, we first abstract the objective into designing data structures for inner product queries. Specifically, consider the following three tasks:

\begin{task}[Positive Inner Product Search]\label{task:threshold}
Let $X=\{x_1,\ldots,x_m\} \in (\R^d)^m$. Given a query $Q\in \R^{d\times d}$ with the promise $\sum_{i\in [m]}\langle Q,x_ix_i^\top\rangle > 0$, we aim at finding a $x_i\in X$ such that $\langle Q,x_ix_i^\top\rangle > 0$.
\end{task}

\begin{task}[Minimum Inner Product Search]\label{task:minip}
Let $X=\{x_1,\ldots,x_m\} \in (\R^d)^m$. Given a query $Q\in \R^{d\times d}$ with the promise $\forall i\in [m], \langle Q,x_ix_i^\top\rangle \geq 0$, we aim at finding a $\arg \min_{x\in X} \langle Q,xx^\top \rangle$.
\end{task}

Efficient Inner product search is a challenging task~\cite{arw17,c18,w18,cw19,a19}. Moreover, 
apart from efficiency, we also need to care about the robustness of our data structures against adaptive queries. In an iterative process, the query vector we generated usually depends on the output from last iteration, or more concretely, from the output of the data structure of last iteration. This means that if we are using a Monte Carlo data structure, then the success probability needs to be against a sequence of adaptive queries. There are two general strategies for this purpose: use a deterministic data structure, or augment an oblivious data structure to handle adaptive queries. We show that for the positive search task, one can use simple yet highly effective deterministic data structures. For the last task, we present two Monte Carlo data structures that are robust against adaptive queries. Based on the query length, these two data structures have their own strengths.

\subsection{Positive Inner Product Search}
\label{sec:pos_ip}
The data structure we use to solve Task~\ref{task:threshold} is a simple, deterministic data structure that makes use of a search tree. The idea is to build up a tree in which the leaf nodes store the matrix $x_ix_i^\top$, and for each internal node, it stores the sum of all outer products in its subtree. It is not hard to see that the root of the tree stores the overall sum $\sum_{i\in [m]} x_ix_i^\top$. During query, one starts with the root, computes the inner product in $O(d^2)$ time, for positive search, one chooses the leaf with a non-negative inner product to recurse. Hence, to generate a target vector $x_i$, one only needs to pay $O(d \log m)$ time, with an initialization time of $O(\nnz(X^2))$. 

We summarize the result in the following theorem.

\begin{theorem}[Informal version of Theorem~\ref{thm:matrix_sample_tree}]
\label{thm:matrix_ss_intro}
There exists a deterministic data structure for Task~\ref{task:threshold} 
with space $O(\nnz(X^2))$, initialization time $O(\nnz(X^2))$, query and update time $O(d^2 \log m)$. Moreover, the data structure is robust against adaptive adversary.
\end{theorem}

The search tree itself is general enough to handle input given as matrices. However, the tasks themselves only consider the input as a list of vectors, hence we can exploit more structures on inputs. We design a tree that batches $d$ vectors together as a leaf node, so that the tree itself only has $m/d$ leafs. Hence, during initialization, it is enough to compute $m/d$ matrix-matrix multiplications of $d\times d$ matrices, gives a better initialization time $O(md^{\omega-1})$ time for dense matrices. During query, one reaches a leaf node consisting of a sum in the form of $\sum_{i\in S, |S|=d} x_ix_i^\top$, to either compute the threshold query or sampling probability, we compute the matrix $X_S^\top Q X_S$, the diagonal entries of this matrix product is exactly what we want. Therefore, the query time of the data structure is $O(d^2 \log (m/d)+d^\omega)$. For certain applications in which one has to pay $d^\omega$ time alongside with query, the batch tree data structure gives a better performance. It also only uses $O(md)$ space instead of $O(\nnz(X^2))$ space, when the input is dense, the latter becomes $O(md^2)$.

\begin{theorem}[Informal version of Theorem~\ref{thm:vector_sample_tree}]
\label{thm:vector_ss_intro}
There exists a deterministic data structure for Task~\ref{task:threshold}  
with space $O(md)$, initialization time $O(md^{\omega-1})$, query time $O(d^2 \log (m/d)+d^\omega)$ and update time $O(d^2 \log (m/d))$. Moreover, the data structure is robust against adaptive adversary. 
\end{theorem}

\subsection{Minimum Inner Product Search} 
To implement fast and robust minimum inner product search data structure, search trees are no longer sufficient, since we care about an ordering related to the query matrix $Q$. One idea is to use deterministic high-dimensional search trees~\cite{b75}, however, such data structures typically suffer from the curse of dimensionality (initialization time exponential in $d$). To resolve such issues, we consider two different data structures, where one uses the duality between minimum inner product search and furthest neighbor search, and the other estimates all Euclidean distances efficiently and robustly. To simplify our discussion, we consider the minimum inner product search search between vectors $X=\{x_1,\ldots,x_m\}\subset \R^d$ and query vector $q\in \R^d$, note that the task of finding $\arg\min_{x\in X} \langle q,x\rangle$ is equivalent of $\arg\max_{x\in X} \|q-x\|_2$, which in words, is to find the vector that is the furthest neighbor of $q$. A natural idea is to use randomized data structures geared towards approximately finding furthest neighbor~\cite{i03} which has sublinear query time in $m$. An alternative solution is to use adaptive distance estimation data structure~\cite{cn20,cn22} that approximates $\|q-x_i\|_2$ for all $i\in [m]$, then perform a linear scan to find the desired vector. 

While the adaptive distance estimation data structures~\cite{cn20,cn22} are inherently robust against adaptive queries, the same does not hold for the furthest neighbor search data structure. In fact, in a standard high-dimensional search pipeline, one typically applies a Johnson-Lindenstrauss transform~\cite{jl84} to reduce the dimension of the dataset and query vectors, however, in an adaptive setting, even this step needs to be modified. The Johnson-Lindenstrauss transform assumes the query vectors are oblivious with respect to the randomness of the JL matrix, then queries depend on the randomness of the JL, the guarantee no longer holds. Our first order of business is to augment JL to make it robust. To achieve this objective, we note that it is enough to prove a dimension of JL so that it preserves the length of \emph{all} vectors. One possibility is to use sketching matrices with \emph{subspace embedding property}~\cite{s06}, however, the dimension of such matrices are too large for our applications. Inspired by the robustness construction of adaptive distance estimation and its applications~\cite{cn20,cn21}, we use many independent sketches of smaller dimensions, during query time, we only need to sample $\wt O(1)$ of them and output the optimal estimates. Given an $d$-dimensional dataset, we still need to use $\wt O(d)$ independent sketches, but each of dimension $\wt O(1)$, hence, the downstream furthest neighbor search task operates on much smaller dimensions, enabling us to use a much simpler net argument to robustify the data structure. We summarize the two results below and both succeed with high probability ($1/\poly(m, d)$):

\begin{theorem}[Informal version of Theorem~\ref{thm:aipe}]
\label{thm:intro_aipe}
Let $\tau\in (0,1)$ and $c\in (\tau, \frac{1.01\tau}{\tau+0.01})$. Given a set of $m$-points $V\subset \mathbb{S}^{d-1}$ on the sphere, one can build a data structure with preprocessing time ${\cal T}_{{\rm init}}=\wt O(md)$ so that for every query $q\in \mathbb{S}^{d-1}$ in an adaptive sequence $Q=\{q_1,\ldots,q_T\}$, the query time is $\wt O(m+d)$ and update time is $\wt O(d)$, with the following guarantee:
\begin{itemize}
    \item Let $v^*\in V$ be the vector such that $\langle v^*,q\rangle$ is minimized among all $v\in V$ and $\langle v^*,q\rangle\geq \tau$, then we output a vector $\wh v\in V$ such that $\langle \wh v,q\rangle\leq \frac{1}{c}\cdot \tau$.
    \item Otherwise, we output {\fail}.
\end{itemize}
\end{theorem}

\begin{theorem}[Informal version of Theorem~\ref{thm:robust_minip}]
\label{thm:intro_robust_minip}
Let ${\cal T}_S(x)$ to denote the time of applying a JL transform matrix $S\in \R^{s\times d}$ to a vector $x\in \R^d$. Let $\tau \in (0,1)$ and $c\in (\tau, \frac{400\tau}{\tau+399})$.

Given a set of $m$-points $V\subset \mathbb{S}^{d-1}$ on the sphere, one can build a data structure with preprocessing time ${\cal T}_{\mathrm{init}}=\wt O(dm^{1.01}+d\cdot {\cal T}_S(V))$ so that for every query $q\in \mathbb{S}^{d-1}$ in an adaptive sequence $Q=\{q_1,\ldots,q_T\}$, the query time is $\wt O(m^{0.01}+{\cal T}_S(q))$ and update time is $\wt O(dm^{0.01}+d\cdot {\cal T}_S(q))$, with the following guarantee:
\begin{itemize}
    \item Let $v^*\in V$ be the vector such that $\langle v^*,q\rangle$ is minimized among all $v\in V$ and $\langle v^*,q\rangle\geq \tau$, then we output a vector $\wh v\in V$ such that $\langle \wh v,q\rangle\leq \frac{1}{c}\cdot \tau+O(\frac{1}{m^{O(1)}})$.
    \item Otherwise, we output {\fail}.
\end{itemize}
\end{theorem}

We remark that the above two results have their own strengths and weaknesses, while Theorem~\ref{thm:intro_aipe} has a linear dependence on $m$ in its query time, it has a better initialization time, and the slow query typically does not impose a problem when number of iterations is relatively small. On the other hand, Theorem~\ref{thm:intro_robust_minip} has a sublinear query time, but worse initialization and update time. When number of iteration gets larger, it finds more applications. 

Note that the vector we are to apply the JL transform is of the form $x_i\otimes x_i\in \R^{d^2}$\footnote{We use $x\otimes y$ to denote the tensor product of $x$ and $y$. Note that $x\otimes y=\vect(xy^\top)$.}, in which ${\cal T}_S(V)$ takes at least $O(md^2)$ time. To leverage the structure that the input vectors are in the form of tensor products, we develop JL matrices that can be applied to tensor-typed inputs fast. Specifically, we generalize the sparse embedding construction of~\cite{dks10,kn10,kn14,cjn18} to handle tensor product of two vectors $u$ and $v$. The idea is to construct a polynomial approximation for the tensor product, and observe that the polynomial can be computed via FFT algorithm. As a result, we design a sparse JL matrix that can be applied to vectors $u$ and $v$ in nearly $O(\nnz(u)+\nnz(v))$ time. We call the matrix of interest the $\tensorsparse$ matrix.

\begin{theorem}[Informal version of Theorem~\ref{thm:tensor_sparse_JLT}]\label{thm:intro_tensor_sparse_JLT}
Let $V=\{v_1,\ldots,v_m\}\in (\R^d)^m$, then the $\tensorsparse$ matrix $S$ with $k=\Theta(\epsilon^{-2}\log(m/\delta))$ rows and each column has $s=\Theta(\epsilon^{-1}\log(m/\delta))$ sparsity has the property that, for any $v_i,v_j\in V$, 
\begin{align*}
    (1-\epsilon)\|v_i\otimes v_i-v_j\otimes v_j\|_2 \leq \|S(v_i\otimes v_i)-S(v_j\otimes v_j)\|_2 \leq (1+\epsilon)\|v_i\otimes v_i-v_j\otimes v_j\|_2
\end{align*}
with probability at least $1-\delta$. Moreover, $S(v_i\otimes v_j)$ can be computed in time $\wt O(\nnz(v_i)+\nnz(v_j))$.
\end{theorem}

Though we develop and utilize $\tensorsparse$ mainly for the purpose of our data structure task, the family of matrices itself might be of independent interest, e.g., in designing subspace embedding for polynomial kernels and improving various downstream tasks, such as sketching Gaussian kernels, $p$-convergent kernels and neural tangent kernels.

\section{Speeding Up Iterations via Inner Product Query Data Structures}

Now that we have enough tools in our toolkit, we will see how to speed up iterations of different discrete optimization problems using these data structures.

\subsection{Linear-Sized Spectral Sparsifier}

\subsubsection{Our Result}

Given a matrix $V\in \R^{m\times d}$ with $m\gg d$, the goal is to pick $s\ll m$ rescaled rows of $V$ to form a matrix $\wt V\in \R^{s\times d}$ such that $(1-\epsilon) V^\top V \preceq \wt V^\top \wt V\preceq (1+\epsilon) V^\top V$. This is the well-known spectral sparsification problem. We present an algorithm that solves this problem deterministically and efficiently.

\begin{theorem}[Informal version of Theorem~\ref{thm:bss_det}]\label{thm:intro_bss}
Let $V=\{v_1,\ldots,v_m\}\in (\R^d)^m$ such that $\sum_{i=1}^m v_iv_i^\top=I$. There exists a \emph{deterministic} algorithm to find a set of weights $\{s_i\}_{i=1}^m$ such that 
$
    (1- \epsilon)I \preceq \sum_{i=1}^m s_iv_iv_i^\top \preceq (1+\epsilon)I
$
and $|\{s_i:s_i\neq 0\}|=\Theta(\epsilon^{-2}d)$. Moreover, the running time of this algorithm is  
\begin{align*}
    \wt O(\min\{ \sum_{i\in [m]}\nnz(v_i)^2 , md^{\omega-1} \}+\epsilon^{-2}d^{\omega+1}).
\end{align*}
\end{theorem}

\begin{remark}
To the best of our knowledge, our algorithm is the first deterministic spectral sparsification construction that achieves the optimal size $s=\Theta(\epsilon^{-2}d)$ and breaks the $\Omega(d^4)$ barrier (when $m=d^2$) of~\cite{bss12,z12}. 

For the situation where $\omega\approx 2$ and $m \geq d^2$, our algorithm is optimal since it matches the input size of the problem.
\end{remark}

\begin{table}[!ht]
    \centering
    \small
    \begin{tabular}{|l|l|l|l|l|} \hline
     {\bf References} & {\bf Time for Sparse Instance} & {\bf Time for Dense Instance}  & {\bf D.}/{\bf R.} 
       \\ \hline \hline
       \cite{bss12}&  $\epsilon^{-2}md^3+\epsilon^{-2}d^{\omega+1}$ &  $\epsilon^{-2}md^3+\epsilon^{-2}d^{\omega+1}$  & {\bf D.} \\ \hline 
       \cite{z12} & $\epsilon^{-2}md^2+\epsilon^{-4}d^4$ &  $\epsilon^{-2}md^2+\epsilon^{-4}d^4$ & {\bf D.} \\ \hline
       \cite{azlo15} & $\epsilon^{-5}md^2+\epsilon^{-4}d^{3+1/q}$ & $\epsilon^{-5}md^2+\epsilon^{-4}d^{3+1/q}$  & {\bf R.}  \\ \hline 
        \cite{ls15}   & $\epsilon^{-2}qm d^{\omega-1+3/q}$ & $\epsilon^{-2}qm d^{\omega-1+3/q}$  & {\bf R.} \\ \hline 
         \cite{ls17}& $ \epsilon^{-O(1)}(\nnz(V^2)+d^\omega)$ & $ \epsilon^{-O(1)} (md^2+d^\omega)$  & {\bf R.} \\ \hline 
          Theorem~\ref{thm:intro_bss}  &$ \nnz(V^2)+\epsilon^{-2}d^{\omega+1}$ &  $md^{\omega-1}+\epsilon^{-2}d^{\omega+1}$ &  {\bf D.}  \\ \hline 
    \end{tabular}
    \caption{Main Results for BSS Sparsifier. $\nnz(V^2)=\sum_{i=1}^m \nnz(v_i)^2$. For simplicity, we ignore the $\wt O(\cdot)$ notation. Note that \cite{bss12,z12} are  deterministic algorithms, and \cite{azlo15,ls15,ls17} are randomized algorithms. We use {\bf D.} to denote deterministic algorithm, and {\bf R.} to denote randomized algorithm. In fact, any known randomized construction of matrix spectral sparsifier requires either sampling from the leverage score distribution~\cite{ss11}, use a modified potential function for batch sampling~\cite{azlo15,ls15,ls17} or randomized sketching techniques to bootstrap~\cite{bw14,swz17,swz19_soda}.}
    \label{tab:bss_main}
\end{table}

\subsubsection{Previous Techniques}
In the next few paragraphs, we summarize all the previous algorithms. For deterministic algorithms \cite{bss12,z12}, we explain why they are inherently slow. For randomized algorithms \cite{azlo15,ls15,ls17}, we explain which step do they mandate randomness (see Section~\ref{sec:bss_maxip_comp_app} for a more detailed discussion).

\paragraph{\cite{bss12}} From an algorithmic perspective,~\cite{bss12} needs to maintain two barrier matrices $L_t \in \R^{d\times d}$ and $U_t\in \R^{d\times d}$ at each iteration, then search for the vector $v_j$ such that $v_j^\top L_t v_j\geq 1/t\geq v_j^\top U_t v_j$ where $\{v_j\}_{j=1}^m$ is the collection of vectors we are given at the beginning and $t$ is some positive value. By the choice of parameters and the two barrier matrices, such $v_j$ is guaranteed to exist. At the beginning, they initialize $A$ to a zero matrix. In each iteration $t$, they then use a multiple of $v_jv_j^\top$ to update the matrix $A$. After $\epsilon^{-2}d$ iterations, the matrix $A$ has the desired spectral property. The algorithm itself is inherently deterministic, since it just needs to form barrier matrices $L_t$ and $U_t$ then perform a search over all vectors. Without any techniques to speedup, it has a slow running time of $O(\epsilon^{-2}md^3)$.

\paragraph{\cite{z12}} One key observation of the~\cite{bss12} algorithm is that one can first use leverage score sampling to perform a crude sparsification of $V$, and reduce the problem to find $\epsilon^{-2}d$ rows from a matrix of only $\epsilon^{-2}d\log d$ rows. By using hyperbolic cosine as a new potential function,~\cite{z12} develops an algorithm that can construct a sparsifier with similar quality of a leverage score sampling ($O(\epsilon^{-2}d\log d)$ rows) but deterministically. However, their bootstrap step is slow ($\wt O(\epsilon^{-2}md^2)$) due to the computation of hyperbolic potential over all $m$ rows.

When $m=d^2$, it is not hard to see that all prior deterministic algorithms will require $\Omega(d^4)$ time. Next, we discuss more efficient randomized algorithms.

\paragraph{\cite{azlo15}} An alternative view of spectral sparsifier construction is via regret minimization, in which one player wants to minimize the eigenvalue and the other player wants to maximize it. Inspired by this idea and in conjunction with a novel (improved) potential function from multiplicative weights update,~\cite{azlo15} presents a mirror descent-based algorithm. However, the major reason they gain speedup from~\cite{z12} is by using Johnson-Lindenstrauss to accelerate the search process similar to~\cite{ss11}, which benefits from the robustness provided by their new potential function. Without the JL step, their algorithm has the same running time as~\cite{bss12}.

\paragraph{\cite{ls15}} The new potential function of~\cite{azlo15} has some bonus structures to potentially reduce the number of iterations required, and~\cite{ls15} extensively exploits this feature. It makes use the new potential function coupling with a sampling procedure that samples a batch of vectors within a single iteration. By using fast matrix multiplication to compute the distribution for only $\epsilon^{-2}qd^{3/q}$ iterations, they obtain a running time of $\wt O(\epsilon^{-2}qmd^{\omega-1+3/q})$. Their algorithm can be viewed as a faster implementation and more refined analysis of the randomized variant of~\cite{bss12} and the correctness of algorithm is heavily reliant on the sampling step. It is unclear how to derandomize/remove that sampling step while still preserving the structure of their analysis.

\paragraph{\cite{ls17}} To approach the optimal time of constructing linear-sized spectral sparsifier,~\cite{ls17} further reduces the number of iterations to $\epsilon^{-2}$, but the sparsifier itself has size $\epsilon^{-2}d$, hence one has to use a much more powerful solver at each iteration to add many rows.~\cite{ls17} uses a positive SDP solver for each iteration. To derandomize their algorithm, one first needs to derandomize the SDP solver of~\cite{azlo16}, which is unclear how to do it efficiently.

\subsubsection{Our Techniques}
We take a completely different approach from prior works, by viewing the~\cite{bss12} algorithm as a data structure problem. To better describe the iterative process invented in~\cite{bss12}, we define the following two barrier functions: given a symmetric matrix $A\in \R^{d\times d}$ and two reals $u$ and $\ell$, we define 
$
    \Phi^u(A) := \tr[(uI-A)^{-1}] =\sum_{i=1}^d \frac{1}{u-\lambda_i},
    \Phi_\ell(A) := \tr[(A-\ell I)^{-1}]=\sum_{i=1}^d \frac{1}{\lambda_i-\ell}.
$

The BSS sparsifier maintains two initial barriers $u_0=d/\epsilon$ and $\ell_0=-d/\epsilon$ and an initial matrix $A_0={\bf 0}_{d\times d}$, then at each iteration $t\in [T]$, the two barriers are incremented respectively: $u_t=u_{t-1}+\delta_U, \ell_t=\ell_{t-1}+\delta_L$, and define the following quantities:
$
    L_t =  \frac{(A_{t-1}-\ell_t I)^{-2}}{\Phi_{\ell_t}(A_{t-1})-\Phi_{\ell_{t-1}}(A_{t-1})}-(A_{t-1}-\ell_t I)^{-1}, 
U_t = \frac{(u_t I-A_{t-1})^{-2}}{\Phi^{u_{t-1}}(A_{t-1})-\Phi^{u_t}(A_{t-1})}+(u_t I-A_{t-1})^{-1}, 
$
the algorithm proceeds by finding an index $j\in [m]$ that \emph{witnesses the gap between lower and upper barriers}, i.e., $v_j^\top L_t v_j\geq 1/t \geq v_j^\top U_t v_j$ for some positive value $t$. The core result proved in~\cite{bss12} is that if we set $\delta_U,\delta_L$ properly, then such condition is always satisfied.

After finding such $v_j$, one then uses $c\cdot v_jv_j^\top$ to update $A$, where $c=1/t$. After $T=\Theta(\epsilon^{-2}d)$ iterations, the resulting matrix $A_T/d$ satisfies the desired property. 

\paragraph{Turning Vector Threshold Search to Data Structure Problem.}

To turn this into a data structure problem, we first notice that if we only aim to find a vector $v_j^\top L_t v_j\geq 1/t$ for some positive value $t$, then the positive inner product search tree data structure (See Section~\ref{sec:pos_ip} and Section~\ref{sec:weighted_sample_ds}) does not work, since it relies on the fact that given a list of number whose sum is at least 0, then there must exist a number itself is at least 0. In fact, if one wants to design a simple, deterministic yet efficient search tree for a general threshold, it is unclear to us how to generalize our data structure to accomplish this goal. For example, if we are given the promise that the sum of numbers are at least 0 and the goal is to find some value that is at least some positive threshold $\tau$, it is possible that the target value lives in a subtree whose sum is negative, and hence we will never touch that subtree and find the correct value. The threshold 0 is a simpler task than general threshold search and enables the design of fast deterministic data structure.

\paragraph{Strengthening the Analysis to Support Positive Inner Product Search.}
Our key observation is the analytical framework of~\cite{bss12} gives more power than this ``one-sided'' search. Namely, it is enough to search for a $v_j$ such that $v_j^\top (L_t-U_t)v_j > 0$.

While their original argument requires to look for
\begin{align*}
v_j^\top L_t v_j\geq 1/t \text{~~~and~~~} v_j^\top U_t v_j \leq 1/t,
\end{align*}
it can be generalized as follows (see Section~\ref{sec:general_sum_rank1}):
\begin{align*}\sum_{i=1}^m v_i^\top L_t v_i\geq \frac{1}{\delta_L}-\epsilon_L  \text{~~~and~~~} \sum_{i=1}^m v_i^\top U_t v_i \leq \frac{1}{\delta_U}+\epsilon_U
\end{align*}
. 

By choosing parameters $\epsilon_L, \epsilon_U, \delta_L, \delta_U$ to ensure that $\frac{1}{\delta_L}-\epsilon_L \geq \frac{1}{\delta_U}+\epsilon_U$, then we know $\sum_{i=1}^m v_i^\top L_t v_i\geq \sum_{i=1}^m v_i^\top U_t v_i$ and the $v_j$ indeed exists via an averaging argument. Hence, there's no need to perform a one-sided search, rather, one can pack the matrix $L_t$ and $U_t$ together and search for the vector $v_j$ with $v_j^\top (L_t-U_t)v_j \geq 0$. By the above argument, such $v_j$ must exist, and we can choose $c={v_j^\top (L_t+U_t)v_j}/{2}$ to update $A$. In fact, by careful choices of $\epsilon_L, \epsilon_U, \delta_L, \delta_U$, we can make sure that both $\sum_{i=1}^m v_i^\top (L_t-U_t)v_i>0$ and hence there exists $v_j$ such that $v_j^\top (L_t-U_t) v_j>0$.

This reduces one iteration of the BSS algorithm to that of the positive inner product search (Task~\ref{task:threshold}), i.e., we form a query matrix $L_t-U_t$ at each iteration, then use a data structure to decide which vector $v_i$ has the property $\langle v_iv_i^\top, L_t-U_t\rangle \geq 0$. By using the search tree we introduced before, we can construct the linear-sized sparsifier \emph{deterministically} in time 
\begin{align*}
    \wt O(\min\{ \sum_{i\in [m]}\nnz(v_i)^2, md^{\omega-1} \}+\epsilon^{-2}d^{\omega+1}).
\end{align*}

\paragraph{Discussions.}
We make several observations regarding our results. For both~\cite{ls15} and~\cite{ls17}, the algorithm needs to read the input \emph{for each iteration}. This is acceptable for their algorithms, since by using different potential functions,~\cite{ls15} reduces the number of iterations from $\epsilon^{-2}d$ to $\epsilon^{-2}d^{3/q}$,~\cite{ls17} further improves the iteration count to $\epsilon^{-2}$. Due to the extremely low iteration count, their algorithms can afford read the input for each iteration. However, this is clearly far from optimal. As we have shown in Theorem~\ref{thm:intro_bss}, it is possible to only read the input once using a carefully-designed data structure and further improves the complexity per iteration. 

Another important advantage of our algorithm is its simplicity. The data structure itself exploits the fundamental property of the linear-sized sparsifier: it constructs a $d\times d$ matrix that measures how far the matrix we've constructed so far are away from the lower and upper barriers, then it searches for a row vector whose inner product makes sure that the algorithm ``progresses'' since it witnesses the gap between lower and upper barrier. Hence, any algorithms that make use of the potential function and its variants defined in~\cite{s10,bss12} can be viewed as performing either inner product sampling or searching, and be transformed into a data structure problem that admits highly efficient and effective data structure solution. 

Finally, our algorithm is completely deterministic, this is the first improvement of deterministic spectral sparsifier since~\cite{z12}. While randomized algorithms are typically much more efficient than their deterministic counterpart, when themselves are used as a subroutine in another iterative process, the randomness itself poses a challenge. Consider the problem of maintaining a dynamic spectral sparsifier against an adaptive adversary in which the adversary can observe the internal randomness of the data structure based on the output of query, it is highly nontrivial to turn a randomized static algorithm into dynamic. On the other hand, deterministic algorithms are guaranteed to succeed against an adaptive adversary, therefore, obtaining an efficient deterministic algorithm for spectral sparsification has more interesting implications.

\subsection{One-Sided Kadison-Singer Problem}
Given a matrix $V\in \R^{m\times d}$ in which each row $v_i$ has $\|v_i\|_2=\frac{1}{\sqrt{N}}$ for some positive value $N$ and $V^\top V=I$, the task is to find a subset $S\subseteq [m]$ with $|S|=n$ rows such that $\|V_S^\top V_S\| \leq \frac{n}{m}+O(\frac{1}{\sqrt N})$. This problem can be viewed as a dual problem of the restricted invertibility problem~\cite{s10} in which one requires a lower bound on the min eigenvalue. In~\cite{w13}, this problem is called the one-sided Kadison-Singer problem, since the Kadison-Singer problem requires upper bound and lower bound simultaneously. 

To solve this problem, we adapt a similar approach as that of~\cite{s10}, i.e., using only one barrier functions instead of two as in~\cite{bss12}. Specifically, we use the upper barrier function $\Phi^u(A)=\tr[(uI-A)^{-1}]$ to progress. At each iteration, the algorithm looks for an index $j\in [m]$ such that $v_j^\top U_t v_j\leq 1$, where $U_t$ is defined as $\frac{(u_t I-A_{t-1})^{-2}}{\Phi_{u_{t-1}}(A_{t-1})-\Phi_{u_{t}}(A_{t-1})}+(u_t I-A_{t-1})^{-1}$. By using a one-sided argument as in~\cite{s10,w13}, one can guarantee that such an index always exists.

We can formulate this problem as a \emph{minimum inner product search} problem (Task~\ref{task:minip}), where we first preprocess all vectors $v_i\otimes v_i$, then at each iteration we form the matrix $U_t$ and use it as a query to the $\minip$ data structure. Depends on the number of rows $n$ we wish to pick, one can either use Theorem~\ref{thm:intro_aipe} or Theorem~\ref{thm:intro_robust_minip}.

\begin{theorem}[Informal of Theorem~\ref{thm:ks_small_iter} and~\ref{thm:alg4}]\label{thm:alg4:intro}
Let $\tau,c\in (0,1)$ and $N\in \mathbb{N}_+$, if $V:=\{v_1,\ldots,v_m \}$ is a finite sequence of vectors in $\R^d$ satisfying $\|v_i\|_2=\frac{1}{\sqrt N},\forall i\in [m]$ and
$
    \sum_{i=1}^m v_iv_i^\top =  I
$. Then for any $n<m$, there exists a randomized algorithm (success with high probability) that takes time ${\cal T}$ to find a set $S$ $(|S|=n)$ such that
\begin{align*}
    \left\|\sum_{i\in S} v_iv_i^\top\right\|  \leq & ~ \frac{1}{c}\cdot (\frac{n}{m}+O(\frac{1}{\sqrt N})).
\end{align*}
Further, we have that, 
\begin{itemize}
    \item if $c\in (\tau,\frac{400\tau}{399+\tau})$, then 
    $
        {\cal T}=  \wt O((m^{1.01}+\nnz(V))\cdot d^2+n \cdot (m^{0.01}d^2+d^\omega) ).
    $
    \item if $c\in (\tau, \frac{1.01\tau}{0.01+\tau})$, then
    $
        {\cal T} =  \wt O(md^2+n\cdot (m+d^\omega)).
    $
\end{itemize}
\end{theorem}

\begin{table}[h]
    \centering
    \small
    \begin{tabular}{|l|l|} \hline
     {\bf References} & {\bf Running Time}  \\ \hline\hline
      \cite{w13} &  $n\cdot (md^2+d^\omega)$\\ \hline 
       Theorem~\ref{thm:alg4:intro}   &    $(m^{1.01}+\nnz(V))\cdot d^2+n \cdot (m^{0.01}d^2+d^\omega) $\\ \hline
      Theorem~\ref{thm:alg4:intro}   &   $ md^2+n \cdot (m+d^\omega) $\\ \hline
    \end{tabular}
    \caption{Main Results for One-Sided Kadison-Singer Problem.}
\end{table}

We first remark that due to the nature of minimum inner product search and its duality with approximate furthest neighbor search (see Section~\ref{sec:prelim_afn_minip} for a more detailed discussion), one can set $\tau$ to some small constant and hence the quality of the final solution has only been blowup by some constant factor. We also note that depending on the value of $n$, the two different data structures provide very different runtime behavior. For the sake of illustration, consider $n=O(1)$, in this case the initialization time dominates. The second running time is $\wt O(md^2)$ while for the first data structure, one has to pay $\wt O(\nnz(V)d^2)$ for initialization. On the flip side, when number of iterations grows larger (say $n=\frac{m}{2}$), the iteration cost dominates and linear scan becomes too expensive: the first data structure gives a total cost of $\wt O(m^{1.01}d^2)$ while the second one becomes $\wt O(m^2)$.

\subsection{Experimental Design via Regret Minimization}

In the work by Allen-Zhu, Li, Singh and Wang~\cite{azlsw20}, they introduce a unified framework to solve the experimental design problem, which concerns the following problem: given $V:=\{v_1,\ldots,v_m\}\in (\R^d)^m$, the goal is to select $n$ of them from $V$ so that the statistical efficiency is maximized when regressed on the $n$ selected points. To approach this problem, they first use variants of mirror descent algorithm to solve a continuous relaxation, then round the solution via a regret minimization framework. More concretely, let $\pi \in [0,1]^m$ be a fractional solution satisfying $\|\pi\|_1\leq n$, the goal is to round $\pi$ into a vector $s\in \{0,1\}^m$ such that $\sum_{i=1}^m s_i\cdot v_iv_i^\top \succeq (1-\epsilon) \sum_{i=1}^m \pi_i\cdot v_iv_i^\top$ and $\|s\|_1=n$. To implement the rounding, they develop a swapping algorithm, which involves initiating a set $S_0$ randomly of cardinality $n$, then at iteration $t$, we construct the following matrix: $A_t=(c_tI-\alpha \sum_{i\in S_{t-1}}v_iv_i^\top)^{-2}$ where $c_t\in \R$ is the constant such that $A_t\succ 0$ for $\alpha=\sqrt{d}/\epsilon$. Then we are to find two indices:
\begin{align*}
    i_t =  ~ \underset{i\in S_{t-1}, 2\alpha \langle A_t^{1/2}, v_iv_i^\top\rangle<1}{\arg\min} ~ B^{-}(v_i), \text{~~~and~~~}
    j_t =  ~ \underset{j\in [m]\setminus S_{t-1}}{\arg\max} ~ B^+(v_j)   .
\end{align*}
where $B^{-}(v_i):=\frac{\langle A_t,v_iv_i^\top\rangle}{1-2\alpha \langle A_t^{1/2}, v_iv_i^\top\rangle}$ and $B^+(v_j):=\frac{\langle A_t,v_jv_j^\top\rangle}{1+2\alpha \langle A_t^{1/2}, v_jv_j^\top\rangle}$. Then we set $S_t=S_{t-1}\setminus \{i_t\}\cup \{j_t\}$. 

In order to perform inner product search with this framework, it is necessary to exhibit an upper bound on $B^-(v_i)$ and a lower bound on $B^+(v_j)$, which is also critical for the correctness proof. By an averaging argument,~\cite{azlsw20} showed that one will always have $B^-(v_i)\leq \frac{1-\epsilon}{n}$ and $B^+(v_j)\geq \frac{1}{n}$. This enables us to reduce this problem into an $\minip$ search in the set $S_{t-1}$. One might consider to use a variant of $\maxip$ search on the set $\ov S_{t-1}$, however, since the potential inner product is small ($\sim\frac{1}{n}$), to achieve a high accuracy $\maxip$ search, one has to set the accuracy parameter in $\maxip$ to be proportional to $\frac{1}{n}$, renders the algorithm inefficient. Hence, we simply perform linear scan over the set $\ov S_{t-1}$.

\begin{theorem}[Informal version of Theorem~\ref{thm:zlsw_main}]\label{thm:intro_zlsw_main}
Let $\pi\in [0,1]^m$ with $\|\pi\|_1\leq n$ and $\sum_{i=1}^m \pi_i x_ix_i^\top=I_d$. Let $\gamma\geq 3$ and $\epsilon\in (0,\frac{1}{\gamma}]$. Then, there exists a subset $S\subset [m]$ with $|S|\leq n$ such that
\begin{align*}
    \lambda_{\min} (\sum_{i\in S} x_ix_i^\top)\geq & ~ 1-\gamma\cdot \epsilon.
\end{align*}
Let $\tau\in (0,1)$ and $c\in (\frac{1}{\gamma-1},1)$. If $n\geq \frac{6d/\epsilon^2}{\gamma-1-1/c}$
, then there exists a randomized algorithm (success with high probability) that takes time ${\cal T}$ to find such $S$. Furthermore,
\begin{itemize}
    \item If $c\in (\tau,\frac{1.01\tau}{0.01+\tau})$, then ${\cal T}=\tilde O(nd^2+\epsilon^{-1}n\cdot (n+d^\omega+(m-n)d^2));$
   \item 
 If $c\in (\tau,\frac{400\tau}{399+\tau})$, then 
$
      {\cal T}=  \tilde O((n^{1.01}+\nnz(X))d^2 +\epsilon^{-1}n\cdot (n^{0.01}d^2+d^\omega+(m-n)d^2)).
$

\end{itemize}
\end{theorem}

\begin{table}[!th]
    \centering
    \small
    \begin{tabular}{|l|l|} \hline
      {\bf References} &  {\bf Running Time}  \\ \hline\hline
      \cite{azlsw20} vanilla & $\epsilon^{-1}mnd^2$\\ \hline 
      \cite{azlsw20} warm restart &  $mnd^2$\\ \hline 
      Theorem~\ref{thm:intro_zlsw_main}  &    $ nd^2 +\epsilon^{-1}n(n+d^\omega+(m-n)d^2)$\\ \hline
      Theorem~\ref{thm:intro_zlsw_main}  & $(n^{1.01}+\nnz(X))d^2+\epsilon^{-1}n((n^{0.01}+z)d^2+d^\omega+(m-n)d^2)$ \\ \hline
    \end{tabular}
    \caption{Main Results for Experimental Design via Regret Minimization. $z=\max_{i\in [m]} \nnz(x_i)$.}
\end{table}

\newpage

\paragraph{Roadmap.} In Section~\ref{sec:prelim}, we give a preliminary on notations, definitions, some useful facts and probabilistic tools used in this paper. 

In Section~\ref{sec:weighted_sample_ds}, we present our positive inner product search tree data structure. In Section~\ref{sec:AIPE}, we show how to perform efficient adaptive inner product estimation via adaptive distance estimation. In Section~\ref{sec:sketching}, we introduce the efficient sketchings for tensors that is robust to adaptive adversary.  In Section~\ref{sec:minip_ds}, we provide the efficient minimum inner product search data structure. In Section~\ref{sec:general_sum_rank1}, we introduce our fast deterministic algorithm for linear-sized spectral sparsification problem.  In Section~\ref{sec:ks}, we present our algorithmic result for one-sided Kadison-Singer problem with approximate guarantee. In Section~\ref{sec:zlsw}, we utilize our algorithmic framework on the rounding up of experimental design problem.

\section{Preliminaries}
\label{sec:prelim}
This section gives some preliminary background definitions and facts. 
\begin{itemize}
    \item In Section~\ref{sec:prelim_notation}, we introduce notations used across this paper. 
    \item In Section~\ref{sec:prelim_jl}, we recall the definition of Johnson-Lindenstrauss transform.
    \item In Section~\ref{sec:prelim_afn_minip}, we formulate the $\minip$ and $\afn$ problem, and further show they are dual to each other.
    \item In Section~\ref{sec:prelim_facts}, we record some useful facts for our later proof.
    \item In Section~\ref{sec:prelim_probtools}, we introduce the probability tools used in the paper.
\end{itemize}

\subsection{Notations}
\label{sec:prelim_notation}

We introduce some notations and definitions we will use throughout this paper.

For a positive integer $n$, we use $[n]$ to denote the set $\{1,2,\cdots, n\}.$
For a vector $x$, we use $\| x \|_2$ to denote its $\ell_2$ norm. For a matrix $A$, we use $\| A \|$ to denote its spectral norm. For a square matrix $A$, we use $\tr[A]$ to denote its trace. For a square and full rank matrix $A$, we use $A^{-1}$ to denote its inverse. 

We say a symmetric matrix $A \in \R^{n \times n}$ is positive semi-definite (PSD, denoted as $A \succeq 0$) if for any vector $x \in \R^n$, $x^{\top} A x \geq 0$. We say a symmetric matrix $A \in \R^{n \times n}$ is positive definite (PD, denoted as $A \succ 0$) if for any vector $x \in \R^n$, $x^{\top} A x > 0$.

For a real positive semi-definite matrix $A$, we define its square root $A^{1/2}$ to be the unique positive semi-definite matrix such that $(A^{1/2})^\top A^{1/2}=A$.

For two conforming matrices $A$ and $B$, we have $\tr[AB] = \tr[BA]$.

For a real symmetric matrix $A$, we use $\lambda_{\max}(A)$ to denote its largest eigenvalue and $\lambda_{\min}(A)$ to denote its smallest eigenvalue.

We define $\Tmat(a,b,c)$ to be the time of multiplying an $a \times b$ matrix with another $b \times c$ matrix. Note that $\Tmat(a,b,c) = O(\Tmat(a,c,b)) = O(\Tmat(b,a,c))$.

For real symmetric matrices $A$ and $B$ of the same size, we use $A\approx_\epsilon B$ if $(1-\epsilon)B\preceq A\preceq (1+\epsilon)B$.

\subsection{Johnson-Lindenstrauss Transform}\label{sec:prelim_jl}
We consider the well-known Johnson-Lindenstrauss transform~\cite{jl84}, throughout this paper, we will make use of various sketching matrices that satisfy the Johnson-Lindenstrauss lemma. We introduce the following definition.

\begin{definition}[Johnson-Lindenstrauss transform (JLT)]
\label{def:JLT}
Let $\{x_1,\ldots,x_m\}\in (\R^d)^m$, we say a distribution $\Pi$ over $s\times d$ matrices is a $(m,\epsilon,\delta)$-JLT if for any $S\sim \Pi$, we have
\begin{align*}
    \Pr[\|S(x_i-x_j)\|_2^2\leq (1\pm\epsilon) \|x_i-x_j\|_2^2] \geq & ~ 1-\delta, ~~ \forall (i,j)\in [m]\times m.
\end{align*}
\end{definition}

We remark that in order to obtain this property for all $m^2$ pairs of point, it suffices to obtain the following guarantee for any fixed point $x\in \R^d$:
\begin{align*}
    \Pr[\|Sx\|_2^2 \leq (1\pm \epsilon) \|x\|_2^2] \geq & ~ 1-\delta,
\end{align*}
then union bound over all $m^2$ pairs of points, we are done.

It is a common practice to use JLT to reduce the dimension of input points for similarity search data structures, then feed into low dimensional vectors into the data structures. We will later show this idea is very powerful when designing task-specific data structures.

\subsection{Approximate Furthest Neighbor and Minimum Inner Product}
\label{sec:prelim_afn_minip}
One of the key tools in this paper is to use Approximate Furthest Neighbor ($\afn$) data structure to solve the Minimum Inner Product Search ($\minip$) problem. We define the corresponding problems here.

\begin{definition}[$\minip$]\label{def:minip}
Given an $n$-point dataset $P \subset \mathbb{S}^{d-1}$ on the sphere, the goal of the Minimum Inner Product Search ({$\minip$}) is to build a data structure that, given a query $q \in \mathbb{S}^{d-1}$, retrieve the solution of $\arg\min_{p\in P}\langle p , q \rangle$.
\end{definition}

The naive brutal force algorithm solves $\minip$ in $O(nd)$ time. However, there exists algorithms that achieve time complexity sublinear in $n$ with relaxation on the retrieved vector. These algorithms aim at solving the approximate $\minip$ problem.

\begin{definition}[Approximate $\minip$]\label{def:aminip}
Let $c \in (0,1)$ and $\tau \in (0,1)$.
Given an $n$-point dataset $P \subset \mathbb{S}^{d-1}$ on the sphere, the goal of the $(c,\tau)$-Minimum Inner Product Search ({$\minip$}) is to build a data structure that, given a query $q \in \mathbb{S}^{d-1}$ with the promise that $\min_{p\in P} \langle p , q \rangle \leq \tau$, it reports a point $p' \in P$ with similarity $\langle p' , q \rangle \leq \tau/c$.
\end{definition}

The approximate $\minip$ has a dual problem: approximate furthest neighbor ($\afn$). We could solve approximate $\minip$ via solving $\afn$. To illustrate this, we first present the definition of $\afn$.

\begin{definition}[Approximate Furthest Neighbor ($\afn$)]\label{def:afn}
Let $\ov{c} >1$ and $r \in (0,2)$. Given an $n$-point dataset $P \subset \mathbb{S}^{d-1}$ on the sphere, the goal of the $(\ov{c},r)$-Approximate Furthest-Neighbor ($\afn$) problem is to build a data structure that, given a query $q \in \mathbb{S}^{d-1}$ with the promise that $\max_{p\in P}\| p - q \|_2 \geq r$, it reports a point $p' \in P$ with distance $\| p' - q \|_2\geq r/\ov{c} $.
\end{definition}

Next, we show the connection between approximate $\minip$ and $\afn$. In this discussion, we assume all vectors are unit vectors, later we'll see a transformation realizes this guarantee.

\begin{lemma}\label{lem:minip_to_afn}
Given an $n$-point dataset $P\subset \mathbb{S}^{d-1}$ and a query point $q\in \mathbb{S}^{d-1}$, suppose for some $\ov c>1$ and $r\in (0,2)$, we have a $(\ov c, r)$-$\afn$ data structure, then we can solve the $(c,\tau)$-$\minip$ problem for
\begin{align*}
    \tau = 1-0.5r^2, c=\frac{1-0.5r^2}{1-0.5r^2/\ov c^2}.
\end{align*}
\end{lemma}

\begin{proof}

For any two points $x,y$ with $\| x \|_2 = \|y \|_2=1$, we have $\|x-y\|_2^2= 2 - 2\langle x , y\rangle$. This implies that if we have $\|x_i-q\|_2^2\geq r^2$, then we have $\langle x_i,q\rangle\leq 1-0.5r^2$. Moreover, if we find a $x_j$ such that $\|x_j-q\|_2^2\geq r^2/\ov{c}^2$, then we have $\langle x_j,q\rangle\leq 1-0.5r^2/\ov{c}^2$. If we set $\tau=1-0.5 r^2$ and 
$c = \frac{1-0.5  r^2}{1 - 0.5 r^2/\ov{c}^2}$, then the above inner product guarantee becomes
\begin{align*}
    \langle x_j, q \rangle \leq & ~ 1-0.5r^2/\ov c^2 \\
    = & ~ 1-0.5r^2+(1-1/\ov c^2)0.5r^2 \\
    = & ~ \tau+(1-\frac{c-\tau}{c(1-\tau)})(1-\tau) \\
    = & ~ \tau/c
\end{align*}
where the second-to-last line is because 
\begin{align}\label{eq:ovc_delta_c_tau}
\ov{c}^2
=&~\frac{cr^2}{2c-2+  r^2}\notag\\
=&~\frac{c(2-2\tau)}{2c-2+  (2-2\tau)}\notag\\
=&~\frac{c(1-\tau)}{c-\tau}.
\end{align}

This indicates that if we have a data structure for $(\ov c,r)$-$\afn$, it automatically becomes a data structure for $(c,\tau)$-$\minip$ with $\tau$ and $c$ chosen as above.
\end{proof}

Next, we explore some structures on the function $\frac{c(1-\tau)}{c-\tau}$. We show that it increases as $\tau$ increases.

\begin{lemma}\label{lem:c_tau_greater_4}
Let $c\in(0,1)$ and $\tau\in (0,1)$, we show that function $f(c,\tau):=\frac{c(1-\tau)}{c-\tau}$ is decreasing as $c$ increase and increasing as $\tau$ increase.
\end{lemma}
\begin{proof}

We take the derivative of $f(c,\tau)$ over $c$ and get
\begin{align*}
    \frac{\partial}{\partial c} f(c,\tau)
    =&~\frac{(\tau-1)\tau}{(c-\tau)^2}<0
\end{align*}
where the second step follows from $c>\tau$ and $\tau<1$.

Therefore, $f(c,\tau):=\frac{c(1-\tau)}{c-\tau}$ is decreasing as $c$ increase.

We take the derivative of $f(c,\tau)$ over $\tau$ and get
\begin{align*}
    \frac{\partial}{\partial \tau} f(c,\tau)
    =&~\frac{c(\tau^2-2c\tau+c)}{(c-\tau)^2}\\
    =&~\frac{c((\tau-c)\tau+c(1-\tau))}{(c-\tau)^2}>0
\end{align*}
where the second step follows from $c>\tau$ and $\tau<1$.

Therefore, $f(c,\tau):=\frac{c(1-\tau)}{c-\tau}$ is increasing as $\tau$ increases.
\end{proof}

In most applications, query and data vectors are usually not unit vectors. Therefore, we need to transform them into unit vectors without breaking the $\minip$ solution. We consider the following pair of asymmetric transformations:

\begin{definition}[\cite{ns15}]\label{def:transform}
Given the query set $X\subset \R^d$ and a dataset $Y\subset \R^d$, we performs the following transformations for any $x\in X$ and $y\in Y$. 
\begin{align*}
    \phi(x) = \begin{bmatrix}\frac{x^\top}{D_X}  & 0 & \sqrt{1-\frac{\|x\|_2^2}{D_X^2}} \end{bmatrix}^\top,  \psi(y) = \begin{bmatrix} \frac{y^\top}{D_Y} & \sqrt{1-\frac{\|y\|_2^2}{D_Y^2}}  & 0 \end{bmatrix}^\top,
\end{align*}
where $D_X$ is larger than the maximum diameter of $X$ and and $D_Y$ is larger than the maximum diameter of $Y$. 
In this way, we map $x\in X$ and $y\in Y$ to unit vectors. In this way, we have $\|\phi(x)-\psi(y)\|_2^2=2-2\langle \phi(x),\psi(y) \rangle$. Moreover, we have $\arg\min_{y\in Y} \|\phi(x)-\psi(y)\|_2 =\arg\max_{y\in Y} \langle x,y \rangle$ and $\arg\max_{y\in Y}- \|\phi(x)-\psi(y)\|_2 =\arg\min_{y\in Y} \langle x,y \rangle$  
\end{definition}
\begin{remark}
In our later applications, we implicitly assume all points have undergone such transformations in preprocessing phase. We also remark that in query phase, the set $Y$ consists of a single query point, it suffices to pick $D_Y$ as $\|y\|_2$, in this case, the transformation can be viewed as normalizing the query vector. If computing the inner product between $x$ and $y$ is required, we can retrieve the original $x$ and $y$ by its first $d$ dimension, and by storing $D_X$ as a variable in the data structure. Moreover, $D_X$ and $D_Y$ play a role in controlling the value of parameter $\tau$ in approximate $\minip$ (see Definition~\ref{def:aminip}) for better efficiency.
\end{remark}

Given a $(c, \tau)$-$\minip$ on unit sphere after transformation, it also gives guarantee on the inner product prior to transformation. We formalize it in the following lemma.

\begin{lemma}
Let $X\subset \R^d$ be a dataset and $q\in \R^d$ be a query vector. Suppose $X$ and $q$ undergo the transformation in Definition~\ref{def:transform} and we are given a $(\ov c, r)$-$\afn$ data structure on the transformed dataset. Then, we can output a $(c, \tau)$-$\minip$  
\begin{align*}
    \tau\geq D_X^{-1}D_Y^{-1}\cdot \minip (q,X), & ~ c\leq \frac{\ov{c}^2 \tau}{\ov{c}^2-1+\tau},
\end{align*}
here we overload the definition of $\minip (q,X) = \min_{x\in X}\langle q,x\rangle$.
\end{lemma}

Note that if the product $D_XD_Y$ is large, then we have to set $\tau$ to be small, this will cause $c$ to be small and the approximation ratio $1/c$ to be large. Especially, if the product $D_XD_Y\sim m$, then the approximation ratio can be as bad as $\sim m$. In our applications, we show that $D_XD_Y$ is at most some constant.

We augment the $\minip$ definition to tolerate additive errors. 
\begin{definition}[Additive approximate $\minip$]\label{def:ada_minip}
Let $c\in (0,1)$ and $\tau\in (0,1)$. Let $\lambda\geq 0$. Given an $n$-point dataset $Y\subset \mathbb{S}^{d-1}$, the goal of the $(c,\tau,\lambda)$-$\minip$ is to build a data structure, given a query $x\in \mathbb{S}^{d-1}$ with the promise that $\min_{p\in P} \langle p , q \rangle \leq \tau$, it reports a data point $z\in Y$ such that $\langle x,z\rangle\leq c^{-1}\min_{p\in P} \langle p , q \rangle+\lambda$.
\end{definition}

\subsection{Useful Facts}
\label{sec:prelim_facts}
We list and prove some useful facts regarding matrices.

\begin{fact}\label{fac:trace_sqrt}
For any PSD matrix $Z \in \R^{d \times d}$, we have $\tr[Z^{1/2}] \leq \sqrt{d \cdot \tr[Z]}$.
\end{fact}
\begin{proof}
Note that for any $d \times d$ positive semi-definite matrix $Z \succeq 0$, $\tr[Z^{1/2}] \leq \sqrt{d \cdot \tr[Z]}$ due to Cauchy-Schwartz inequality applied to the non-negative spectrum of $Z^{1/2}$.
\end{proof}

\begin{fact}[Matrix Woodbury identity, \cite{w49,w50}]\label{fac:woodbury}
For matrices $M\in \R^{n\times n}$, $U\in \R^{n\times d}$, $C\in \R^{d\times d}$, $V\in \R^{d\times n}$,
    \begin{align*}
        ( M + U C V )^{-1} = M^{-1} - M^{-1} U (C^{-1} + V M^{-1} U )^{-1} V M^{-1} .
    \end{align*}
\end{fact}

\begin{fact}\label{fac:trace_diff_diagonal}
Let $A$ and $B$ denote two diagonal matrices in $\R^{d \times d}$. Suppose $\forall i\neq j\in [n]$, we have $\beta_i-\alpha_i=\beta_j-\alpha_j$, and let $\gamma = \beta_i - \alpha_i$. We have
\begin{align*}
    \tr[ A^{-1} - B^{-1} ] = \gamma \cdot \tr[ A^{-1} B^{-1} ].
\end{align*}
\end{fact}
\begin{proof}
We have
\begin{align*}
    \tr[ A^{-1} - B^{-1} ]
    = & ~ \sum_{i=1}^k \frac{1}{\alpha_i} - \frac{1}{\beta_i} \\
    = & ~ \sum_{i=1}^k \frac{\beta_i - \alpha_i}{ \alpha_i \beta_i } \\
    = & ~ \gamma \sum_{i=1}^k \frac{1}{\alpha_i \beta_i} \\
    = & ~ \gamma \cdot \tr[ A^{-1} B^{-1} ]
\end{align*}
Thus, we complete the proof.
\end{proof}

\begin{fact}[Inequality for two monotone sequences]
\label{fact:sorting}
Suppose $a_1 \geq a_2 \geq \cdots \geq a_n \geq 0$, $b_1 \geq \cdots \geq b_n \geq 0$, then we have
\begin{align*}
    \sum_{i=1}^{n}a_i b_{n-i} \leq \frac{1}{n}\sum_{i=1}^{n} a_i \sum_{j=1}^{n} b_j
\end{align*}
\end{fact}

\subsection{Probability Tools}\label{sec:prelim_probtools}
In this section, we present some probability tools.

We start with the standard 2-stable Gaussian distribution. We refer the readers to~\cite{diim04} for more details.
\begin{fact}[Standard Gaussian is 2-stable]
\label{fact:2stable}
Let $Z,X_1,X_2,\ldots,X_k\sim {\cal N}(0,1)$ and $v\in \R^k$, then $\sum_{i=1}^k v_iX_i$ and $\|v\|_2\cdot Z$ have the same distribution.
\end{fact}

Next, we present a concentration and anti-concentration bound for Gaussian distribution.
\begin{fact}[Gaussian concentration bound]
\label{fact:gauss_concentration_bd}
Let $X\sim {\cal N}(0,1)$ and $t>0$, then we have

\begin{itemize}
    \item {\bf Part 1 Concentration.} 
$
    \pr[|X|\geq t] \leq 2\exp(-t^2/2)/t
$.
    \item {\bf Part 2 Anti-Concentration.}
There exists a constant $B>0$ such that
\begin{align*}
    \pr[|X|\geq t] \geq & ~ 2B\cdot \exp(-t^2/2)/\max\{1,t\}.
\end{align*}
\end{itemize}

\end{fact}

\begin{definition}
Let $X$ be a random variable, we use $\|X\|_{L_q}$ to denote $(\E[|X|^q])^{1/q}$. By Minkowski's inequality, $\|\cdot \|_{L_q}$ is a norm when $q\geq 1$.
\end{definition}

\begin{lemma}[Hanson-Wright inequality \cite{hw87}]\label{lem:hanson_wright}
For $\sigma_1,\sigma_n$ independent Rademachers and $A\in \R^{n\times n}$, for all $q\geq 1$,
\begin{align*}
    \|\sigma^\top A\sigma-\E[\sigma^\top A\sigma]\|_{L_q} \leq & ~ O(1)\cdot (\sqrt{q}\cdot \|A\|_F+q\cdot \|A\|).
\end{align*}
\end{lemma}

\begin{lemma}\label{lem:binomial_p_moment}
For $Y$ distributed as $\mathsf{Binomial}(N,\alpha)$ for integer $N\geq 1$ and $\alpha\in (0,1)$, let $1\leq p\leq N$ and define $B:=p/(\alpha N)$. Then
\begin{align*}
    \|Y\|_{L_p} \leq & ~ \begin{cases}
    \frac{p}{\log B}, & \text{if $B\geq e$} \\
    \frac{p}{B}, & \text{if $B<e$}.
    \end{cases}
\end{align*}
\end{lemma}

\section{Positive Inner Product Search Tree: Crude and Refined Computations}\label{sec:weighted_sample_ds}

In this section, we present two data structures that can solve the positive inner product search task (Task~\ref{task:threshold}).

\begin{itemize}
    \item In Section~\ref{sec:matrixSS}, we show a data structure when input is given as a list of matrices.
    \item In Section~\ref{sec:vectorSS}, we present a data structure when input is given as a list of vectors.
\end{itemize}

\subsection{Matrix Search Tree: Input Sparsity Time Initialization and Fast Query}
\label{sec:matrixSS}

Given a list of matrices $\{M_1,\ldots,M_m\}\subset \R^{d\times d}$, we design a data structure to solve Task~\ref{task:threshold}. The data structure proprocesses the list of matrices in input sparsity time, i.e., $\sum_{i\in [m]}\nnz(M_i)$. When query, it takes inner product between a query matrix $A$ and a partial sum matrix stored at a tree node in $O(d^2)$ time and only traverses one path from root to a leaf. Note that when we are dealing with vector inputs, we need to spend $O(\sum_{i\in [m]} \nnz(v_i)^2)$ time forming the outer products $v_iv_i^\top$.

\begin{algorithm}[!ht]\caption{Matrix Positive Search}\label{alg:matrix_sampling_p1}
\begin{algorithmic}[1]
\State {\bf data structure} \textsc{MatrixPS} \Comment{Theorem~\ref{thm:matrix_sample_tree}}
\State {\bf members}
\State \hspace{4mm} $M_1,M_2,\cdots,M_m \subset \R^{d\times d}$ (matrix of each index)
\State \hspace{4mm} $S_0,S_1,S_2,\cdots,S_n \subset \R^{d\times d}$ (partial sum of each node)
\State \hspace{4mm} Binary tree $T$ (each node is a tuple $(i_1,i_2,S)$ where $i_1<i_2$ are indices and $S=\sum_{i=i_1}^{i_2}S_i$)
\State {\bf end members}
\State 
\Procedure{Init}{$M_1,M_2,\cdots,M_m \subset \R^d$}
    \State $S_0=0$
    \For {$i=1$ to $m$}
        \State $M_i\gets M_i$
        \State $s_i\gets S_{i-1}+M_i$
    \EndFor
    \State Insert $(1,m,S_m)$ as root of $T$
    \While {exists a leaf $l=(i_1,i_2,S)$ of $T$ such that $i_2-i_1\ge 1$}
        \State $k=\lfloor (i_1+i_2)/2 \rfloor$
        \State Insert $(i_1,k,S_k-S_{i_1-1})$ as left child of $l$
        \State Insert $(k+1,i_2,S_{i_2}-S_k)$ as right child of $l$
    \EndWhile 
\EndProcedure 
\State
\Procedure{QueryPositiveSearch}{$A\in \R^{d\times d}$} \Comment{Lemma~\ref{lem:matrix_sample_tree_query_search}}
    \State $r\gets \text{root of $T$}$
    \While {$r$ is not a leaf of $T$}
        \State $r_1\gets \text{left child of $r$}$, $r_2\gets \text{right child of $r$}$
        \State $M_1\gets$ matrix of $r_1$, $M_2\gets$ matrix of $r_2$
        \State $p_1\gets \langle A, M_1\rangle$, $p_2\gets \langle A, M_2\rangle$
        \If {$p_1 > 0$}
            \State $r\gets r_1$
        \Else \Comment{$p_2 > 0$}
            \State $r\gets r_2$
        \EndIf
    \EndWhile 
    \State \Return index of $r$
\EndProcedure
\end{algorithmic}
\end{algorithm}

We summarize the correctness and running time of Algorithm~\ref{alg:matrix_sampling_p1} as follows:

\begin{theorem}[Formal version of Theorem~\ref{thm:matrix_ss_intro}]\label{thm:matrix_sample_tree}
There exists a data structure with the following procedures:
\begin{itemize}
    \item \textsc{Init}$(\{M_1,M_2,\cdots,M_m\}\subseteq\mathbb{R}^{d\times d})$. It takes a sequence of matrices $M_1,M_2,\cdots,M_m$ as input, and preprocesses in time $O(\sum_{i=1}^m \nnz(M_i))$.
    \item \textsc{QueryPositiveSearch}$(A\in \R^{d\times d})$. Given a matrix $A$ with the promise that $\sum_{i=1}^m \langle M_i, A\rangle > 0$, it returns an index $i$ such that $\langle M_i, A\rangle>0$ in time $O(d^2 \log m)$.
\end{itemize}
\end{theorem}

\begin{proof}
We prove the data structure (see Algorithm~\ref{alg:matrix_sampling_p1}) satisfies the requirements. In \textsc{Init}, every node $(i_1,i_2, M)$ stores the partial sum of matrices $\sum_{j=i_1}^{i_2} M_j$, the number of nodes is $O(m)$, then the preprocess time is $O(\sum_{i=1}^m\nnz(M_i))$ accounts for the sparsity of the input.

For \textsc{QueryPositiveSearch}, see Lemma~\ref{lem:matrix_sample_tree_query_search}.

\end{proof}

\begin{lemma}[Positive Search]\label{lem:matrix_sample_tree_query_search}
Given a matrix $A$ with the promise that $\sum_{i=1}^m \langle M_i,A\rangle > 0$, \textsc{QuerySearch} returns an index $i$ such that $\langle M_i,A\rangle>0$ in time $O(d^2 \log m)$.
\end{lemma}
\begin{proof}
For \textsc{QuerySearch}, note that the correctness holds obviously: given a node and its two children, suppose we know the inner product at the node is greater than 0, then it must be the case that at least one of its two children has value greater than 0. For the running time, each inner product takes $O(d^2)$ time, and we traverse a path on the tree of depth $O(\log m)$, so it takes $O(d^2\log m)$ time in total.
\end{proof}

\subsection{Speeding Up Initialization via Fast Matrix Multiplication and Batching}
\label{sec:vectorSS}
We note that the \textsc{MatrixPS} data structure is more general than some of the tasks, in which the input is given as a list of vectors $\{v_1,\ldots,v_m\}\subset \R^d$, we can speed up the initialization via fast matrix multiplication, in the expense of worse query time. In certain tasks we can balance the initialization time and query time to achieve a better overall performance.

The idea is to maintain a tree with only $m/d$ nodes, with each of the leaf is a sum of $d$ outer products $\sum_{i\in S} v_iv_i^\top$ for $S\subset [m]$ and $|S|=d$. During initialization, we can form each leaf in $d^\omega$ time, and since there are $m/d$ leaves in total, it only takes $O(md^{\omega-1})$ time to initialize. We store the $d\times d$ matrix $V$ where each column is $v_i$. During query, when we reach the leaf node, we can perform the matrix multiplication $V^\top A V$ and extract the diagonal entries in time $O(d^\omega)$.

\begin{algorithm}[!ht]\caption{Vector Positive Search}\label{alg:vector_sampling_p1}
\begin{algorithmic}[1]
\State {\bf data structure} \textsc{VectorPS} \Comment{Theorem~\ref{thm:vector_sample_tree}}
\State {\bf members}
\State \hspace{4mm} $v_1,v_2,\cdots,v_m \subset \R^{d}$ (vector of each index)
\State \hspace{4mm} $\{S_{i,j}\}_{i\in \{0,\ldots,\log(m/d)\}, j\in [2^{-i}m/d]}\in \R^{d\times d}$ 
\State {\bf end members}
\State 
\Procedure{Init}{$v_1,v_2,\cdots,v_m \subset \R^d$}
    \For{$i=1 \to m/d$}
        \State $V_i \gets \begin{bmatrix}
        \vert & \vert & \ldots & \vert \\
        v_{(i-1)d+1} & v_{(i-1)d+2} & \ldots & v_{id} \\
        \vert & \vert & \ldots & \vert
        \end{bmatrix}$
        \State $S_{0,i}\gets V_iV_i^\top$
    \EndFor

    \For {$i=1\to \log(m/d)$}
        \For{$j=1\to 2^{-i}m/d$}
            \State $S_{i,j}\gets S_{i-1, 2j-1}+S_{i-1, 2j}$
        \EndFor
    \EndFor
\EndProcedure 
\State
\Procedure{QueryPositiveSearch}{$A\in \R^{d\times d}$} \Comment{Lemma~\ref{lem:vector_sample_tree_search}}
    \State $j\gets 1$
    \For{$i=\log(m/d) \to 0$}
        \State $L\gets S_{i-1, 2j-1}$, $R\gets S_{i-1, 2j}$
        \State $p_1\gets \langle A, L\rangle$, $p_2\gets \langle A, R\rangle$
        \State $j\gets \begin{cases}
            2j-1, & \text{with probability $p_1>0$} \\
            2j, & \text{with probability $p_2>0$}.
        \end{cases}$
    \EndFor
     \State $V\gets V_j$
    \State $B\gets V^\top A V$
    \For{$i = 1 \to d$}
        \If{$B_{i,i} > 0$}
            \State $i^* \gets i$
            \State {\bf break}
        \EndIf
    \EndFor
    \State \Return $i^*$
\EndProcedure
\end{algorithmic}
\end{algorithm}

\begin{theorem}[Formal version of Theorem~\ref{thm:vector_ss_intro}]\label{thm:vector_sample_tree}
There exists a data structure with the following procedures:
\begin{itemize}
    \item \textsc{Init}$(\{v_1,v_2,\cdots,v_m\}\subseteq\mathbb{R}^d)$. It takes a sequence of vectors $v_1,v_2,\cdots,v_m$ as input, and preprocesses in time $O(md^{\omega-1})$ and in space $O(md)$.
    \item \textsc{QueryPositiveSearch}$(A\in \R^{d\times d})$. Given a matrix $A$ with the promise that $\sum_{i=1}^m \langle M_i, A\rangle > 0$, it returns an index $i$ such that $\langle M_i, A\rangle>0$ in time $O(d^2 \log m+d^{\omega})$.
\end{itemize}
\end{theorem}

\begin{proof}
We prove the data structure (see Algorithm~\ref{alg:vector_sampling_p1}) satisfies the requirements. In \textsc{Init}, we will perform $m/d$ matrix multiplications of $d\times d$ matrix, yields a time of $O(md^{\omega-1})$. We then compute $m/d$ sums of $d\times d$ matrices, which takes $O(md)$ time, or $\nnz(V^2)$ time. Note that the space is only $O(md)$, since we have constructed a tree of $O(m/d)$ nodes, with each node stores a $d\times d$ matrix. We note an invariant by our construction: for matrix $S_{i,j}$, it represents the sum of outer products $\sum_{k=i_1}^{i_2} v_kv_k^\top$, where $i_1=2^i(j-1)d+1$ and $i_2=2^ijd$, hence $S_{\log(n/d),1}=\sum_{i=1}^m v_iv_i^\top$.

For \textsc{QuerySearch}, we prove in Lemma~\ref{lem:vector_sample_tree_search}.

\end{proof}

We will show that each matrix $S_{i,j}$ stores the proper sum of $v_iv_i^\top$ over a desired range.

\begin{lemma}\label{lem:partial_sum}
Let $i\in \{0,1,\ldots,\log(m/d)\}$ and $j\in [2^{-i}m/d]$, then we have 
\begin{align*}
    S_{i,j} = & ~ \sum_{k=1}^{2^id} v_{2^ij-2^i+k}v_{2^ij-2^i+k}^\top.
\end{align*}
\end{lemma}

\begin{proof}
We start with the bottom level where $i=0$. Note that $S_{0,j}=V_jV_j^\top$, where $V_j=\begin{bmatrix}
        \vert & \vert & \ldots & \vert \\
        v_{(j-1)d+1} & v_{(j-1)d+2} & \ldots & v_{jd} \\
        \vert & \vert & \ldots & \vert
        \end{bmatrix}$.
Use the outer product formulation of matrix multiplication, we have that
\begin{align*}
    S_{0,j} = & ~ V_jV_j^\top \\
    = & ~ \sum_{k=1}^d (V_j)_{*,k} (V_j)_{*,k}^\top \\
    = & ~ \sum_{k=1}^d v_{(j-1)d+k} v_{(j-1)d+k}^\top \\
    = & ~ v_{(j-1)d+1}v_{(j-1)d+1}^\top+v_{(j-1)d+2}v_{(j-1)d+2}^\top+\ldots+v_{jd}v_{jd}^\top.
\end{align*}
For internal levels, we can show by induction. For $i=1$, note that $S_{1,j}=S_{0,2j-1}+S_{0,2j}$, we know that $S_{0,2j-1}=\sum_{k=1}^d v_{(2j-2)d+k} v_{(2j-2)d+k}^\top$ and $S_{0,2j}=\sum_{k=1}^d v_{(2j-1)d+k} v_{(2j-1)d+k}^\top$, hence 
\begin{align*}
    S_{1,j} = & ~ \sum_{k=1}^{2d} v_{(2j-2)d+k}v_{(2j-2)d+k}^\top.
\end{align*}

Assume this holds up until some level $l$, i.e., $S_{l,j}=\sum_{k=1}^{2^ld} v_{2^{l}j-2^l+k}v_{2^{l}j-2^l+k}^\top$, then
\begin{align*}
    S_{l+1,j} = & ~ S_{l,2j-1}+S_{l,2j} \\
    = & ~ (\sum_{k=1}^{2^ld} v_{2^{l}(2j-1)-2^l+k}v_{2^{l}(2j-1)-2^l+k}^\top)+(\sum_{k=1}^{2^ld} v_{2^{l}(2j)-2^l+k}v_{2^{l}(2j)-2^l+k}^\top) \\
    = & ~ \sum_{k=1}^{2^{l+1}d} v_{2^{l+1}j-2^{l+1}+k}v_{2^{l+1}j-2^{l+1}+k}^\top.
\end{align*}
Hence, we complete the proof. Note that when $i=\log(m/d)$, $j=1$ and 
\begin{align*}
    S_{m/d, 1} = & ~ \sum_{k=1}^{m} v_{(m/d)-(m/d)+k} \\
    = & ~ \sum_{k=1}^m v_{k}.
\end{align*}
\end{proof}

\begin{lemma}[Positive Search]\label{lem:vector_sample_tree_search}
Given a matrix $A$ with the promise that $\sum_{i=1}^m \langle v_iv_i^\top,A\rangle > 0$, \textsc{QueryPositiveSearch} returns an index $i$ such that $\langle v_iv_i^\top,A\rangle>0$ in time $O(d^2 \log (m/d)+d^\omega)$.
\end{lemma}
\begin{proof}
To see the correctness, we note a simple if and only if statement: given numbers $a_1,\ldots,a_m$ such that $\sum_{i=1}^m a_i> 0$, then there must exist an $i$ such that $a_i>0$, otherwise the sum must be negative. For our search procedure, we can prove the correctness inductively: at root, since we know that $\sum_{i=1}^m v_i^\top Av_i>0$, then it must be the case that either $\sum_{i=1}^{m/2} v_i^\top Av_i>0$ or $\sum_{i=m/2+1}^n v_i^\top Av_i>0$, otherwise the root sum must be negative. Suppose this holds to level $k$, and we are deciding where to go for level $k+1$, note by induction hypothesis, for level $k$, the inner product must be positive, then it must be the case that one of its children has a positive inner product, otherwise the sum of inner product will be negative. Also, each node stores the correct partial sum, as shown in Lemma~\ref{lem:partial_sum}.

At the bottom level for leaf node $j$, we compute $B=V_j^\top AV_j$, the claim is the diagonal entry $B_{i,i}=v_{(j-1)d+i}^\top Av_{(j-1)d+i}$, to see this, note that
\begin{align*}
    (V_j^\top A V_j)_{i,i} = & ~ (V_j^\top \begin{bmatrix}
    \vert & \vert & \ldots & \vert \\
    Av_{(j-1)d+1} & Av_{(j-1)d+2} & \ldots & Av_{jd} \\
    \vert & \vert & \ldots & \vert 
    \end{bmatrix})_{i,i} \\
    = & ~ v_{(j-1)d+i}^\top Av_{(j-1)d+i}.
\end{align*}
This completes the correctness proof.

 For the running time, each inner product takes $O(d^2)$ time, and we traverse a path on the tree of depth $O(\log (m/d))$, for the leaf, it takes $O(d^\omega)$ time. This concludes our proof.
\end{proof}

\begin{remark}
The \textsc{VectorPS} data structure can be viewed as using a crude estimation for all levels above the bottom level, and for the bottom level, we use a more refined computation to exactly estimate $v_i^\top Av_i$. This means we have to spend more time at the bottom level, but this is fine since we also gain speedup from the initialization. In the setting of a dense graph or a matrix with $m\geq d^2$ rows, we achieve a initialization time of $m d^{\omega-1}$ and overall iteration cost $\epsilon^{-2}d^{\omega+1}$, these two terms balance out. In contrast, with the \textsc{MatrixPS} data structure, it might incur $md^2 \approx d^4$ time for initialization, which is no faster than the other known deterministic spectral sparsification algorithm~\cite{z12}. Such a high-level idea of the tradeoff between crude and refined computation has also been utilized in balancing sample complexity in completely different field (see sparse Fourier transform in the continuous setting~\cite{ps15}). Our case is a different scenario, since we care about the running time perspective of this tradeoff.
\end{remark}
\section{Adaptive Inner Product Estimation via Adaptive Distance Estimation}
\label{sec:AIPE}
In this section, we consider to use the adaptive distance estimation~($\ade$) data structure~\cite{cn20,cn22} to perform adaptive inner product estimation~($\aipe$), which means it is efficient and robust against adaptive adversary. 

\begin{itemize}
    \item In Section~\ref{sec:alg_AIPE}, we present our $\aipe$ and $\afn$ algorithm based on $\ade$ of~\cite{cn22}.
    \item In Section~\ref{sec:thm_AIPE}, we prove the correctness and runtime of the algorithm in the prior section.
\end{itemize}

We start with the core definition of this section.

\begin{definition}[Adaptive Inner Product Estimation~($\aipe$)]
\label{def:AIPE}
Let $X=\{x_1,\ldots,x_m\}\subset (\R^d)^m$ be a dataset of dimension $d$ and radius $D$ and let $q\in \R^d$ be a query point in unit Euclidean ball. The \emph{Adaptive Inner Product Estimation} ($\aipe$) data structure, $D$, has the following guarantee: with probability at least $1-\delta$ we have for any $i\in [m]$,
\begin{align*}
    (1+\epsilon)\langle x_i-q\rangle-D\epsilon \leq  w_i \leq (1-\epsilon)\langle x_i-q\rangle+D\epsilon,
\end{align*}
where $w_i$ denotes the inner product estimation between $x_i$ and $q$.
\end{definition}

We have the following result from~\cite{cn22}:

\begin{lemma}[Theorem 1.4 of~\cite{cn22}]\label{lem:cn22_ADE}
Let $\epsilon,\delta\in (0,1/2)$. Then, there exists a data structure for Distance Estimation in Euclidean space which is initialized correctly with probability at least $1-\delta$ and supports the following operations:
\begin{itemize}
    \item Output a correct answer to a possibly adaptively chosen distance estimation query with probability at least $1-\delta$, i.e.,
    \begin{align*}
    (1-\epsilon)\|x_i-q\|_2 \leq d_i \leq (1+\epsilon)\|x_i-q\|_2,
    \end{align*}
    where $d_i$ denotes the distance estimation between $x_i$ and $q$.
    \item Add input $x\in \R^d$ to the dataset $X$.
\end{itemize}
Furthermore, the query and update (insert/delete) time of the data structure are $\wt O(\epsilon^{-2}(m+d)\log 1/\delta)$ and $\wt O(\epsilon^{-2}d\log 1/\delta)$ respectively while the data structure is constructed in time $\wt O(\epsilon^{-2}md \log 1/\delta)$.
\end{lemma}

We note that Lemma~\ref{lem:cn22_ADE} provides a more generic data structure that can solve the $\afn$ data structure problem. Specifically, one can prepare an $\ade$ data structure in preprocessing stage. In query stage, one can query the $\ade$ data structure for all-pairs estimations between query $q$ and points in the dataset, then one simply output $x_i$ with the largest $d_i$.

\subsection{Algorithm}
\label{sec:alg_AIPE}
\begin{algorithm}[H]\caption{Adaptive Inner Product Estimation}\label{alg:inner_product_estimation}
\begin{algorithmic}[1]
\State {\bf data structure} \textsc{Adaptive Inner Product Estimation} \Comment{Theorem \ref{thm:aipe}}
\State {\bf members}
\State \hspace{4mm} \textsc{AdaptiveDistanceEstimation ADE}
\State {\bf end members}
\State 
\Procedure{Init}{$x_1,x_2,\cdots,x_m,\epsilon,\delta$} 
    \State \textsc{ADE.Init}$(x_1,x_2,\cdots,x_m,\epsilon,\delta)$
\EndProcedure 
\State 
\Procedure{Insert}{$z \in \R^d$} 
    \State \textsc{ADE.Insert}$(z)$
\EndProcedure 
\State 
\Procedure{Delete}{$i\in [m]$}
    \State \textsc{ADE.Delete}$(i)$
\EndProcedure
\State
\Procedure{Query}{$q \in \R^d$} \Comment{Lemma~\ref{lem:query_AIPE}}
    \State $d_1, d_2, \cdots, d_m$ =\textsc{ADE.Query}$(q)$
    \For {$i=1,2,\cdots,m$}
        \State ${w}_i = 1-\frac{1}{2}{d}_i^2$
    \EndFor 
    \State \Return $\{{w}_i\}_{i=1}^m$
\EndProcedure 
\State 
\Procedure{QueryMin}{$q \in \R^d$} \Comment{Lemma~\ref{lem:query_min_AIPE} }
    \State ${d}_1, {d}_2, \cdots, {d}_m \gets \textsc{ADE.Query}(q)$
    \State $i\gets \arg\max_{i\in [m]}~d_i$
    \State \Return $x_i$
\EndProcedure 
\State {\bf end data structure}
\end{algorithmic}
\end{algorithm}

\subsection{Solve $\aipe$ and $\afn$ with $\ade$}
\label{sec:thm_AIPE}
In this section, we show that the $\aipe$ data structure given in Algorithm~\ref{alg:inner_product_estimation} can solve the $\aipe$ task as in Definition~\ref{def:AIPE} and $\afn$ as in Definition~\ref{def:afn}.

We first show that given an $\ade$ data structure, we can solve the $\afn$ data structure problem. 

\begin{lemma}\label{lem:query_min_AIPE}
Let $X=\{x_1,\ldots,x_m\} \in (\mathbb{S}^{d-1})^m$ be the dataset and $q\in \mathbb{S}^{d-1}$ be a query vector. Suppose for some $r\in (0,2)$, $\max_{x\in X} \|x-q\|_2 \geq r$. Then, procedure \textsc{QueryMin}$(q)$ in Algorithm~\ref{alg:inner_product_estimation} solves the $(1+\epsilon, r)$-$\afn$ data structure problem.
\end{lemma}

\begin{proof}
Let $x\in X$ be the point in $X$ that maximizes the distance with $q$, also, we have $\|x-q\|_2\geq r$.Let $d_x$ denote the distance estimation corresponds to $x$ outputted by the $\ade$ data structure. Suppose for some $y\in X$, $d_y\geq d_x$, then we have
\begin{align*}
    d_y \geq & ~ d_x \\
    \geq & ~ (1-\epsilon)\|x-q\|_2 \\
    \geq & ~ (1-\epsilon) r \\
    \geq & ~ r/(1+2\epsilon),
\end{align*}
this concludes our proof.
\end{proof}

As a corollary, it automatically induces a $\minip$ data structure.

\begin{corollary}
Let $X=\{x_1,\ldots,x_m\} \in (\mathbb{S}^{d-1})^m$ be the dataset and $q\in \mathbb{S}^{d-1}$ be a query vector. Suppose for some $r\in (0,2)$, $\max_{x\in X} \|x-q\|_2 \geq r$. Given a $(1+\epsilon, r)$-$\afn$ data structure, it can solve the $(c, \tau)$-$\minip$ problem with
\begin{align*}
    \tau = 1-0.5r^2, & ~ c = \frac{(1+\epsilon)^2 \tau}{(1+\epsilon)^2-1+\tau}.
\end{align*}
\end{corollary}

Similarly, the $\aipe$ problem can be solved using $\ade$.

\begin{lemma}\label{lem:query_AIPE}
Let $X=\{x_1,\ldots,x_m\}\subset \R^{d-1}$ be the dataset with $m$ points and radius $D$, let $q\in \mathbb{S}^{d-1}$ be the query vector. The procedure \textsc{Query}$(q)$ in Algorithm~\ref{alg:inner_product_estimation} outputs a list of estimates $\{w_i\}_{i=1}^m$ such that
\begin{align*}
    (1+\epsilon)\langle x_i,q\rangle-D\epsilon \leq w_i \leq (1-\epsilon)\langle x_i, q\rangle+D\epsilon.
\end{align*}
\end{lemma}

\begin{proof}
Throughout the proof, we assume transformation $Q$ has been applied to all points $x_i\in X$ and transformation $P$ has been applied to query vector $q$.

By Definition~\ref{def:transform}, we have 
\begin{align}\label{eq:dual_primal}
    \|P(q)-Q(x_i)\|_2^2=2-2\cdot D^{-1}\langle q,x_i\rangle
\end{align}

By Lemma~\ref{lem:cn22_ADE}, we have
\begin{align*}
    (1-\epsilon)^2\|P(q)-Q(x_i)\|_2^2\leq d_i^2 \leq (1+\epsilon)^2\|P(q)-Q(x_i)\|_2^2,\forall i\in[n]
\end{align*}

Then we have
\begin{align*}
    1-\frac{(1+\epsilon)^2\|P(q)-Q(x_i)\|_2^2}{2}\leq 1-\frac{d_i^2}{2} \leq 1-\frac{(1-\epsilon)^2\|P(q)-Q(x_i)\|_2^2}{2}
\end{align*}

Applying Eq.~\eqref{eq:dual_primal} we get
\begin{align*}
    1-\frac{(1+3\epsilon)(2-2\cdot D^{-1}\langle q,x_i\rangle)}{2}\leq 1-\frac{d_i^2}{2} \leq 1-\frac{(1-3\epsilon)(2-2\cdot D^{-1}\langle q,x_i\rangle)}{2}
\end{align*}

Thus, we get
\begin{align*}
    (1+3\epsilon)\langle q,x_i\rangle-3D\epsilon \leq D\cdot(1-\frac{d_i^2}{2}) \leq (1-3\epsilon)\langle q,x_i\rangle+3D\epsilon.
\end{align*}
\end{proof}

We summarize results regarding Algorithm~\ref{alg:inner_product_estimation} in the following main theorem.

\begin{theorem}[Adaptive Inner Product Estimation, formal version of Theorem~\ref{thm:intro_aipe}]\label{thm:aipe}
There is a data structure uses $\tilde{O}(\epsilon^{-2}md\log(1/\delta))$ space for the Adaptive Inner Product Estimation Problem with the following procedures:
\begin{itemize}
    \item \textsc{Init}$(  \{x_1, x_2, \dots, x_m\}\subset \R^d, \epsilon \in (0,1), \delta \in (0,1))$: Given data points $\{x_1, x_2, \dots, x_n\}\subset \R^d$ with radius $D$, an accuracy parameter $\epsilon$ and a failure probability $\delta$ as input, the data structure preprocesses in time $\wt{O}(\epsilon^{-2} m d \log(1/\delta))$.
    \item \textsc{Insert}$(z \in \R^d)$: Given a vector $z$, the data structure insert $z$ in time $\tilde{O}(\epsilon^{-2}d\log(1/\delta))$.
    \item \textsc{Delete}$(i \in [m])$: Given an index $i$, the data structure deletes $x_i$ in time $\tilde{O}(\epsilon^{-2}d\log(1/\delta))$.
    \item \textsc{Query}$(q \in \R^d)$: Given a query point $q \in \R^d$, the \textsc{Query} operation takes $q$ as input and approximately estimates the inner product of $q$ and all the data points $\{x_1, x_2, \dots, x_m\}\subset \R^d$ in time $\tilde{O}(\epsilon^{-2}(m+d)\log(1/\delta))$ i.e. it provides a set of estimates $\{\tilde{w}_i\}_{i=1}^m$ such that:
    \begin{align*}
       \forall i \in[m], (1+\epsilon)\langle q,x_{i}\rangle - D\epsilon \leq \tilde{w}_{i} \leq (1-\epsilon)\langle q,x_{i}\rangle + D\epsilon
    \end{align*}
     with probability at least $1 -\delta$, even for a sequence of adaptively chosen queries.
     \item \textsc{QueryMin}$(q\in \R^d)$: Given a query point $q \in \R^d$, the \textsc{QueryMin} operation takes $q$ as input and solves the $(1+\epsilon, r)$-$\afn$ data structure problem, where $r\in (0,2)$ satisfies $\max_{x\in X}\|x-q\|_2/D\geq r$, in time $\wt O(\epsilon^{-2}d\log(1/\delta))$.
\end{itemize}
\end{theorem}

\begin{proof}

{\bf Proof of \textsc{Init}.} The running time follows from the initialization time of Lemma~\ref{lem:cn22_ADE}.

{\bf Proof of \textsc{Insert} and \textsc{Delete}.} The running time follows from the update time of Lemma~\ref{lem:cn22_ADE}.

{\bf Proof of \textsc{Query}.} The correctness follows from Lemma~\ref{lem:query_AIPE}, for the running time, it follows from Lemma~\ref{lem:cn22_ADE}.

{\bf Proof of \textsc{QueryMin}.} The correctness follows from Lemma~\ref{lem:query_min_AIPE}, for the running time, it follows from Lemma~\ref{lem:cn22_ADE}.

\end{proof}

\begin{remark}
$\ade$ data structure is robust against adaptive queries, which is especially feasible during an iterative process. During query, to find the vector that approximates the minimum inner product, we need to perform a linear scan over all $m$ vectors, this makes it useful when number of iterations is rather small, in which linear scan is affordable. The initialization time of the data structure is also nearly linear in the size of input. 
\end{remark}
\section{Efficient and Adaptive Sketchings for Tensors}\label{sec:sketching}

In this section, we introduce several primitives that perform Johnson-Lindenstrauss transforms efficiently on outer product of vectors, or equivalently, the tensor product on vectors. We will exploit these primitives to design fast Johnson-Lindenstrauss transforms for the matrices in the form of $vv^\top$, then feed in the sketched vectors into our $\minip$ data structures. This yields an improved preprocess and query time of our data structures, which is key to improve the overall running time of several algorithms.

Throughout this section, we will use $b$ to denote the target dimension of sketching, $m$ to denote the number of points we want to preserve their pair-wise distances, $d$ to denote the dimension of original data points, and $s$ to denote the sparsity for each column of a sketching matrix.

This section is organized as below. 
\begin{itemize}
    \item In Section~\ref{sec:tensorsrht}, we introduce the $\tensorsrht$ transform for efficient tensor product. 
    \item In Section~\ref{sec:tensorsparse}, we present the $\tensorsparse$ transform for tensor product in input-sparsity time.
    \item In Section~\ref{sec:robust_sketch}, we provide tools with theoretical guarantees for the sketchings to handle adaptive adversary.
\end{itemize}

\subsection{\texorpdfstring{$\tensorsrht$}{~}: Compute Tensor Product via FFT}\label{sec:tensorsrht}
We first introduce the primitive of $\tensorsrht$ transform~\cite{akk+20,swyz21}, which gives high probability guarantee and nearly-linear time in order to evaluate a tensor product.
\begin{definition}\label{def:tensorsrht}
We define the $\tensorsrht$ $S:\R^d\times \R^d\rightarrow \R^b$ as $S=\frac{1}{\sqrt b} P \cdot (H D_1\times H D_2)$, where each row of $P\in \{0,1\}^{b\times d^2}$ contains only one $1$ at a random coordinate, one can view $P$ as a sampling matrix. $H$ is a $d\times d$ Hadamard matrix, and $D_1,D_2$ are two $d\times d$ independent diagonal matrices with diagonals that are each independently set to be a Rademacher random variable (uniform in $\{-1,1\}$). 
\end{definition}

As the name suggests, one can utilize the structure of Hadamard matrix and use Fast Fourier Transform (FFT) to compute the tensor product of two vectors: $S(x\otimes y)$ can be computed in time $O(d\log d+b)$.

Next, we show that given any fixed pair of vectors $x,y\in \R^d$, $S$ preserves the inner product with high probability:

\begin{lemma}[Theorem 2 of~\cite{akk+20}]
\label{lem:akk_srht}
Let $x,y\in \R^d$ be any fixed pair of vectors. Let $\epsilon\in (0,1)$ be precision parameter and $\delta\in (0,1)$ be success probability. Let $S\in \R^{b\times d^2}$ be a $\tensorsrht$ transform matrix (Def.~\ref{def:tensorsrht}). Suppose $b=\Omega(\frac{\log^3
(1/\epsilon\delta)}{\epsilon^2})$, then we have $S$ is a $(1,\epsilon,\delta)$-JLT (Def.~\ref{def:JLT}).
\end{lemma}

For our purpose, it suffices to preserve such inner products for $\Theta(m^2)$ pairs, hence by a union bound, we shall use a $\tensorsrht$ matrix of size $\Theta(\frac{\log^3 (m/\epsilon\delta)}{\epsilon^2})$. This leads to the following result:

\begin{lemma}\label{lem:JL_srht}
Let $\{x_1,\ldots,x_m\}\in (\R^{d^2})^m$. Let $\epsilon\in (0,1)$ be precision parameter and $\delta\in (0,1)$ be success probability. Let $S\in \R^{b\times d^2}$ be a $\tensorsrht$ transform matrix (Def.~\ref{def:tensorsrht}). Suppose $b=\Omega(\epsilon^{-2}\log^3(m/\epsilon\delta))$, then we have $S$ is an $(m,\epsilon,\delta)$-JLT (Def.~\ref{def:JLT}).

Moreover, if $x=u\otimes v$ for some $u,v\in \R^d$, then $Sx$ can be computed in time $O(d\log d+b)$.
\end{lemma}

\begin{proof}
For the dimension of sketching matrix, it is a consequence of union bounding over all $m^2$ pairs of vectors. For the running time, it follows directly from the structure of $\tensorsrht$.
\end{proof}

One important guarantee given by Lemma~\ref{lem:JL_srht} is it preserves the inner product of any pair of vectors with proper dimensions, but facilitate faster computation for tensor-type computation in the form of $S(u\otimes v)$. For a vector $x\in \R^{d^2}$ that does not have such tensor structure, we can still compute $Sx$ use standard matrix-vector product in time $O(bd^2)=\wt O(d^2)$. This suffices for our application.

Additionally, we prove a result regarding the Frobenius norm of the $\tensorsrht$ matrix. 

\begin{lemma}
Let $S\in \R^{b\times d^2}$ be a $\tensorsrht$ matrix~(Def.~\ref{def:tensorsrht}), then we have 
\begin{align*}
    \|S\|_F \leq & ~ d.
\end{align*}
\end{lemma}

\begin{proof}
We note that a Hadamard matrix has orthonormal columns, and since $D_i$ is a diagonal matrix with $\{\pm 1\}$ on its diagonal, we have that $HD_i$ is also has orthonormal columns for $i\in \{1,2\}$. Hence, we know that $\|HD_i\|_F=\sqrt{d}$. Moreover, we note that $\|HD_1\times HD_2\|_F\leq \|HD_1\|_F\|HD_2\|_F=d$ since $\times$ is the tensor product of two matrices. Finally, note that $P$ is a sampling matrix, it samples $b$ rows from $HD_1\times HD_2$ with replacement, hence $\|P(HD_1\times HD_2)\|_F\leq \|P\|_F \|HD_1\times HD_2\|_F \leq \sqrt{b}d$. The result follows since we need to scale the matrix $P(HD_1\times HD_2)$ by $\frac{1}{\sqrt b}$.
\end{proof}

\subsection{\texorpdfstring{$\tensorsparse$}{~}: Efficient Tensor Product in Input-Sparsity Time}\label{sec:tensorsparse}
We recall the sparse embedding matrix \cite{dks10,kn10,kn14,cjn18}. 
\begin{definition}\label{def:sparse_embedding}
Let $h:[d]\times [s]\rightarrow [b/s]$ be a random $O(\log 1/\delta)$-wise independent hash function and $\sigma:[d]\times [s]\rightarrow \{\pm 1\}$ be $O(\log 1/\delta)$-wise independent. Then $R\in \R^{b\times d}$ is a sparse embedding matrix with sparsity parameter $s$ if we set $R_{(j-1)b/s+h(i,j),i}=\sigma(i,j)/\sqrt{s}$ for all $(i,j)\in [d]\times [s]$ and all other entries to 0.

Alternatively, we can define the following:
\begin{align*}
    R_{r,i} = & ~ \exists k\in [s]: \sigma(i,k)/\sqrt{s}\cdot {\bf 1}[h(i,k)+(k-1)b/s = r]
\end{align*}
\end{definition}

We extend the construction of sparse embedding to handle tensor product of vectors, specifically, our goal is to design a sparse matrix that is similar to Def.~\ref{def:sparse_embedding}, so that we can enjoy certain nice properties, such as it is a $(1,\epsilon,\delta)$-JLT with $b=O(\epsilon^{-2}\log(1/\delta))$, this again enables us to union bound over $m$ points.

\begin{definition}[$\tensorsparse$]\label{def:tensor_sparse}
Let $h_1,h_2:[d]\times [s]\rightarrow [b/s]$ be $O(\log 1/\delta)$-wise independent hash functions and let $\sigma_1,\sigma_2:[d]\times [s]\rightarrow \{\pm 1\}$ be $O(\log 1/\delta)$-wise independent random sign functions. Then, the degree two tensor sparse transform, $R:\R^d\times \R^d\rightarrow \R^b$ is given as: 
\begin{align*}
    R_{r,(i,j)} = & ~ \exists k\in [s]: \sigma_1(i,k)\sigma_2(j,k)/\sqrt{s}\cdot {\bf 1}[ ((h_1(i,k)+h_2(j,k))~\text{mod~}b/s)+(k-1)b/s=r]
\end{align*}
\end{definition}

We will show that for any fixed unit vector $x\in \R^{d^2}$, $Rx$ preserves the length of $x$ with good probability. To do so, we first exhibit some properties of our sketch.

\begin{lemma}\label{lem:delta_property}
The degree two $\tensorsparse$ transform~(Def.~\ref{def:tensor_sparse}) has the following property. We define $\delta_{r,(i,j)}$ as the Bernoulli random variable on whether the entry $R_{r,(i,j)}$ is non-zero or not. Then
\begin{enumerate}
    \item Each column has support size $s$. 
    \item For all $r\in [b]$ and $(i,j)\in [d]\times [d]$, $\E[\delta_{r,(i,j)}]=s/b$.
    \item Negative correlations of $\delta_{r,(i,j)}$'s defined as follows:
    \begin{align*}
    \forall T\subset [b]\times [d]\times [d]~\text{and}~|T|\leq \Theta(\log(1/\delta)), ~~~ \E \Big[ \prod_{r,(i,j)\in T} \delta_{r,(i,j)} \Big] \leq \prod_{r,(i,j)\in T} \E[\delta_{r,(i,j)}]=\left(\frac{s}{b}\right)^{|T|}.
    \end{align*}
\end{enumerate}
\end{lemma}

\begin{proof}
We prove three parts separately.

{\bf Part 1.} To see each column has support size $s$, we partition each column into $s$ blocks, where each block contains $b/s$ entries and then show that each block has exactly 1 non-zero entry. Fix the block to be the $k$-th block and consider the $(i,j)$-th column, then we are looking at the values of hash functions $(h_1(i,k)+h_2(j,k))~\text{mod~}b/s$, since both $h_1$ and $h_2$ have their ranges being $[b/s]$, this means $(h_1(i,k)+h_2(j,k))~\text{mod~}b/s$ must have its value being in the range of $[b/s]$, and its value corresponding to the entry that is non-zero.

{\bf Part 2.} We will again use the block-partition view and consider the $k$-th block of $(i,j)$-th column. For each index $r$, the probability that it is non-zero is equal to the probability that $(h_1(i,k)+h_2(j,k))~\text{mod~}b/s=r-(k-1)b/s$. We first observe that if we are using a single 3-wise independent hashing function, then this probability is naturally $s/m$. Here, the hashing function we are considering is $H(i,j,k):=h_1(i,k)+h_2(j,k)~\text{mod~}b/s$, it is well-known that $H$ is also $\Theta(\log(1/\delta))$-wise independent~\cite{cw79,pt12}. We hence conclude that $\Pr[\delta_{r,(i,j)}=1]=\frac{s}{b}$ and therefore, $\E[\delta_{r,(i,j)}]=\frac{s}{b}$.

{\bf Part 3.} To see the negative correlation, we let $t=|T|$, and we denote the elements in $T$ as $(r_1,l_1),\ldots,(r_t,l_t)$. We define the following indicator random variable: ${\bf 1}[\exists (r_i,l_i),(r_j,l_j)\in T~\text{s.t. $r_i\neq r_j$ belong to the same block and $l_i=l_j$}]$. 

Note that if such event happens, then $\E[\prod_{(r,l)\in T} \delta_{r,l}]=0$ since we can write it as 
\begin{align*}
    \E[\prod_{(r,l)\in T} \delta_{r,l}] = & ~ \Pr[\bigwedge_{r,l\in T} \delta_{r,l}=1] \\
    = & ~ \Pr[\delta_{r_1,l_1}=1\wedge \delta_{r_2,l_2}=1]\cdot \Pr[\bigwedge_{(r,l)\in T, r\neq r_1,r_2, l\neq l_1,l_2} \delta_{r,l}=1\mid \delta_{r_1,l_1}=1\wedge \delta_{r_2,l_2}=1].
\end{align*}
When the above event happens, then we are considering the case that $r_1\neq r_2$ but they belong to the same block, and the column is the same. By construction, for each column, there is exactly one non-zero entry. Hence, $\Pr[\delta_{r_1,l_1}=1\wedge \delta_{r_2,l_2}=1]=0$, and we conclude the expectation is 0.

Suppose the above event does not happen, then we will make use the fact that our hashing function $H$ is $\Theta(\log(1/\delta))$-wise independent, and $\delta_{r,l}=1$ is equivalent to for some $k\in [s]$, we have $H(l,k)=r$. The above event does not happen is equivalent to
\begin{align*}
    \Pr[\bigwedge_{(r,l)\in T} \exists k\in [s], H(l,k)=r] = & ~ \prod_{(r,l)\in T} \Pr[\exists k\in [s], H(l,k)=r] \\
    = & ~ \prod_{(r,l)\in T} \Pr[\delta_{r,l}=1] \\
    = & ~ \prod_{(r,l)\in T} \E[\delta_{r,l}],
\end{align*}
where the first step is due to $H$ is $\Theta(\log(1/\delta))$-wise independence. Therefore, we conclude that the random variables $\delta_{r,l}$'s are negatively correlated.
\end{proof}

\begin{remark}
We note that we only require our hashing function $H$ and sign function $\sigma$ to be $\Theta(\log(1/\delta))$-wise independent, since in our later proofs, we will only consider the $q$-th power of an expression $Z$ which involves the term $\prod_{(r,l)\in T}\delta_{r,l}$ for $|T|\leq q$. Thus, the expectation of $Z$ are term-by-term dominated by the case that all $\delta_{r,l}$ are i.i.d. Bernoulli with expectation $s/b$. This justifies our later use of Lemma~\ref{lem:binomial_p_moment} and Hanson-Wright inequality.
\end{remark}

We will adapt an analysis from~\cite{cjn18} to conclude that $\tensorsparse$ is a JLT:
\begin{lemma}\label{lem:tensor_sparse_JLT}
If $R$ is a $\tensorsparse$ matrix as defined in Def.~\ref{def:tensor_sparse}, with target dimension $m\geq \Omega(\log(1/\delta)/\epsilon^2)$ and sparsity parameter $s=\epsilon m$, then 
\begin{align*}
    \Pr[|\|Rx\|_2^2-1|>\epsilon] \leq & ~ \delta.
\end{align*}
\end{lemma}
\begin{proof}
We first observe that

\begin{align*}
    \|Rx\|_2^2 = & ~ \frac{1}{s}\sum_{r=1}^b \sum_{i,j=1}^{d^2} \delta_{r,i}\delta_{r,j}\sigma_{r,i}\sigma_{r,j}x_ix_j \\
    = & ~ \frac{1}{s}\sum_{r=1}^b \sum_{i=1}^{d^2} \delta_{r,i} x_i^2+\frac{1}{s}\sum_{r=1}^b\sum_{i\neq j}^{d^2} \delta_{r,i}\delta_{r,j}\sigma_{r,i}\sigma_{r,j}x_ix_j,
\end{align*}
for the first term (diagonal term), we have
\begin{align*}
    \frac{1}{s}\sum_{r=1}^b \sum_{i=1}^{d^2} \delta_{r,i} x_i^2 = & ~ \sum_{i=1}^{d^2} x_i^2 (\frac{1}{s}\sum_{r=1}^b \delta_{r,i}) \\
     = & ~ \|x\|_2^2 \\
     = & ~ 1,
\end{align*}
where the second step follows from the fact that each column of $R$ has support size $s$. We define the intermediate variable $Z:=\|Rx\|_2^2-1$, which as shown by proceeding calculations, captures the off-diagonal term. Consider the following terms: we first define $A_{x,\delta}$ which is a block diagonal matrix with $b$ blocks, where the $k$-th block is defined as $\frac{1}{s}x^{(k)}(x^{(k)})^\top$ but with the diagonal zeroed out, with $(x^{(k)})_i=\delta_{k,i}x_i$. Note that by construction, $A_{x,\delta}\in \R^{bd^2\times bd^2}$. We further define the following length $bd^2$ vector $\sigma\in \R^{bd^2}$, where $\sigma_{r,i}$ is the sign generated for the entry $(r,i)$ of $R$. 

It is not hard to see that $Z=\frac{1}{s}\sum_{r=1}^b\sum_{i\neq j}^{d^2} \delta_{r,i}\delta_{r,j}\sigma_{r,i}\sigma_{r,j}x_ix_j=\sigma^\top A_{x,\delta}\sigma$. Let $\|X\|_{L_q}:=(\E[|X|^q])^{1/q}$. Since $\sigma$ is a vector with each entry being independent Rademacher random variable, by Hanson-Wright inequality, we have
\begin{align*}
    \|\sigma^\top A_{x,\delta}\sigma\|_{L_q} \leq & ~ \|\sqrt{q}\cdot \|A_{x,\delta}\|_F+q\cdot \|A_{x,\delta}\|\|_{L_q} \\
    \leq & ~ \sqrt{q}\cdot \| \|A_{x,\delta}\|_F\|_{L_q}+q\cdot \|\|A_{x,\delta}\|\|_{L_q},
\end{align*}
since $A_{x,\delta}$ is block diagonal, its spectral norm is the largest spectral norm of any block. Note that the spectral norm of $k$-th block is
\begin{align*}
    \|\frac{1}{s}\cdot x^{(k)}(x^{(k)})^\top\| \leq & ~ \frac{1}{s}\cdot \|x^{(k)}\|_2^2 \\
    \leq & ~ \frac{1}{s},
\end{align*}
where the first step is the sub-multiplicativity of spectral norm and the spectral norm of a vector is its $\ell_2$ norm, and the second line follows from $\|x^{(k)}\|_2\leq \|x\|_2=1$.

Next, we define $Q_{i,j}=\sum_{r=1}^b \delta_{r,i}\delta_{r,j}$, so
\begin{align*}
    \|A_{x,\delta}\|_F^2 = & ~ \frac{1}{s^2} \sum_{r=1}^b \sum_{i\neq j}^{d^2}\delta_{r,i}\delta_{r,j}x_i^2 x_j^2 \\
    = & ~ \frac{1}{s^2} \sum_{i\neq j}^{d^2} Q_{i,j}x_i^2 x_j^2.
\end{align*}
Recall that for any column $i$ of $R$, there exists exactly $s$ non-zero entries, so we suppose $\delta_{r_t,i}=1$ for all distinct $r_t$.

Consider the event that $\delta_{r_t,j}=1$, and let $Y_t$ be the indicator random variable for this event. By Lemma~\ref{lem:delta_property}, we assume $Y_t$'s are independent, so that the sum $Q_{i,j}=\sum_{t=1}^s Y_t$ has the distribution of $\mathsf{Binomial}(s,s/b)$. Combining with Lemma~\ref{lem:binomial_p_moment}, we have that $\|Q_{i,j}\|_{L_{q/2}}\leq q/2$. Thus,
\begin{align*}
    \|\|A_{x,\delta}\|_F\|_{L_q} = & ~ \|\|A_{x,\delta}\|_F^2\|_{L_{q/2}}^{1/2} \\
    = & ~ \|\frac{1}{s^2}\sum_{i\neq j}x_i^2x_j^2 Q_{i,j}\|_{L_{q/2}}^{1/2} \\
    \leq & ~ \frac{1}{s} (\sum_{i\neq j}x_i^2x_j^2 \|Q_{i,j}\|_{L_{q/2}})^{1/2} \\
    \leq & ~ O\left(\frac{\sqrt q}{s}\right).
\end{align*}
Put things together, we have
\begin{align}
    \|\sigma^\top A_{x,\delta}\sigma\|_{L_q} \leq & ~ O\left(\frac{q}{s}\right).
\end{align}
Set $q=\Theta(\log(1/\delta))=\Theta(s^2/b)$, we have $\|Z\|_{L_q}\leq  O(\frac{s}{b})$, then by Markov inequality, we have
\begin{align*}
    \Pr[|\|Rx\|_2^2-1|>\epsilon] = \Pr[|\sigma^\top A_{x,\delta}\sigma|>\epsilon] <\epsilon^{-q}\cdot C^q(m^{-q/2}+s^{-q})<\delta,
\end{align*}
as desired.
\end{proof}

Note that our construction resembles the $\tensorsketch$ matrix~\cite{p13,anw14}, more specifically, we can view our tensor sparse embedding as $s$ distinct $\tensorsketch$ matrices, each with dimension $b/s\times d^2$. Hence, to compute the tensor product between two vectors, we can run the $\tensorsketch$ algorithm for $s$ blocks, yielding an overall running time of $O(s\cdot (\nnz(x)+\nnz(y))+b\log(b/s))$ for computing $S(x\otimes y)$.

We summarize the JLT result and efficient computation of tensor in the following theorem:
\begin{theorem}[Formal version of Theorem~\ref{thm:intro_tensor_sparse_JLT}]\label{thm:tensor_sparse_JLT}
Let $\{x_1,\ldots,x_m\}\in (\R^{d^2})^m$. Let $\epsilon\in (0,1)$ be precision parameter and $\delta\in (0,1)$ be success probability. Let $R\in \R^{b\times d^2}$ be a $\tensorsparse$ matrix (Def.~\ref{def:tensor_sparse}). Suppose $b=\Omega(\epsilon^{-2}\log(m/\delta))$ and $s=\epsilon m$ be the sparsity parameter, then we have $R$ is an $(m,\epsilon,\delta)$-JLT (Def.~\ref{def:JLT}).

Moreover, if $x=u\otimes v$ for some $u,v\in \R^d$, then $Rx$ can be computed in time $O(s\cdot (\nnz(u)+\nnz(v))+b\log(b/s))$.
\end{theorem}

\begin{proof}
The JLT result is by apply union bound over all $m^2$ pairs of points using Lemma~\ref{lem:tensor_sparse_JLT}. The running time is by using the $\tensorsketch$ algorithm for $s$ blocks.
\end{proof}

For further applications, we prove a simple result regarding the Frobenius norm of $R$.

\begin{lemma}\label{lem:tensor_sparse_F_norm}
Let $R\in \R^{b\times d^2}$ be a $\tensorsparse$ matrix~(Def.~\ref{def:tensor_sparse}), then we have 
\begin{align*}
    \|R\|_F = & ~ d.
\end{align*}
\end{lemma}

\begin{proof}
We observe that each column of $R$ has exactly $s$ non-zero entries, each has magnitude $\frac{1}{\sqrt s}$, hence each column is a unit length vector. There are $d^2$ columns in total, yielding a Frobenius norm of $d$.
\end{proof}

\subsection{Robust Sketches Against Adaptive Adversary}\label{sec:robust_sketch}
We note that the above discussion only applies when we consider an \emph{independent} set of points, i.e., all points we want to preserve using $\tensorsrht$ or $\tensorsparse$ are picked oblivious with respect to the randomness of the sketch. However, this is no longer the case for our application --- specifically, the query we send for iteration $t+1$ is \emph{dependent} on the answer we receive at iteration $t$.

One idea is to require a sketching matrix that preserves the length of \emph{all} vectors in a subspace. Unfortunately, this will result in a sketching dimension of roughly $\Theta(d^2/\epsilon^2)$, which essentially diminishes the necessity of using sketching. To address this problem, we exploit the following idea: we use a number of independent sketches of small dimension, and we show that with high probability, a good fraction of them will do well on a (potentially) adversary query. We will show that the dimension-saving by using lower-dimensional sketching matrices will have to be paid back by the number of sketches required. However, this has one distinctive advantage for our applications: we will then operate our $\afn$ data structures on much lower dimensions, hence the preprocessing time and query time can be significantly improved. 

We prove the following lemma:

\begin{lemma}\label{lem:robust_JL}
Let $V:=\{v_1,\ldots,v_m\}\in (\R^d)^m$, $\epsilon\in (0,1)$ and $\delta\in (0,1)$. Furthermore, let $\{S_i\}_{i=1}^k \subset \R^{b\times d}$ for $k\geq \Omega((d+\log(1/\delta))\log(md))$ such that each $S_i$ is an independent $(m+1,\epsilon,0.99)$-JLT matrix~(Def. \ref{def:JLT}) with $\|S_i\|_F\leq d$. Then we have
\begin{align*}
    \forall q\in \mathbb{S}^{d-1}, \forall v\in V, \sum_{i=1}^k {\bf 1}[\|S_i(q-v)\|_2^2\leq (1\pm O(\epsilon))\|q-v\|_2^2+\alpha]\geq 0.95k
\end{align*}
with probability at least $1-\delta$ and $\alpha\leq O(\frac{1}{(md)^9})$.
\end{lemma}

\begin{proof}
We will prove via a standard $\gamma$-net argument. Let $N$ be a $\gamma$-net of $\mathbb{S}^{d-1}$ with $\gamma=\frac{c}{(md)^{10}}$ for some small enough constant $c$, and it is not hard to see that $|N|\leq (md)^{O(d)}$. Let $u\in N$, define the following event:
\begin{align*}
    W_i(u) = & ~ \|S_iu\|_2^2\leq (1+O(\epsilon))~\text{and}~\forall v_i,v_j\in V, |u^\top S_i^\top S_i(v_i-v_j)-u^\top(v_i-v_j)|\leq O(\epsilon)\|v_i-v_j\|_2,
\end{align*}
i.e., the length of $u$ is preserved by $S_i$ and for any pair of points in $V$, the inner product is also preserved by $S_i$. We note that we only need this property to hold with respect to the set of points $V\cup \{u\}$, since $S_i$ is a $(m+1,\epsilon,0.99)$-JLT, we know this event holds with probability at least $0.99$. 

By an application of Hoeffding's inequality on the random variables $\sum_{i=1}^k W_i(u)$, we have that
\begin{align*}
    \Pr[\sum_{i=1}^k W_i(u)\leq 0.97k] \leq & ~ \exp(-2k),
\end{align*}
we then union bound over all points in $N$:
\begin{align*}
    \Pr[\forall u\in N, \sum_{i=1}^k W_i(u)\leq 0.97k] \leq & ~ \exp(-2k)\cdot (md)^{O(d)} \\
    = & ~ (\frac{1}{md})^{O(d)}\cdot (md)^{O(d)}\cdot \exp(-\log(1/\delta)\log(md)) \\
    \leq & ~ \delta/4.
\end{align*}
We will condition on this event happen throughout the rest of the proof. To extend this bound from all points in $N$ to the entire unit sphere, consider any $q\in \mathbb{S}^{d-1}$ and pick a net point $u\in N$ such that $\|q-u\|_2\leq \gamma$. Let $i\in [k]$ be the index such that $W_i(u)$ happens. We shall bound the term $\|S_i(q-v)\|_2$ for $v\in V$:
\begin{align*}
    \|S_i(q-v)\|_2 \leq & ~ \|S_i(q-u)\|_2+\|S_i(u-v)\|_2 \\
    \leq & ~ d\cdot\gamma+(1\pm O(\epsilon))\|u-v\|_2 \\
    \leq & ~ d\cdot \gamma+(1\pm O(\epsilon))(\|q-v\|_2-\gamma) \\
    = & ~ (1\pm O(\epsilon))\|q-v\|_2+(d-(1\pm O(\epsilon)))\gamma  \\
    \leq & ~ (1\pm O(\epsilon))\|q-v\|_2+\alpha.
\end{align*}

The conclusion of the lemma follows.
\end{proof}

\begin{remark}\label{rem:robust_JL_to_ds}
We note that by using the $\gamma$-net argument, we get a weaker conclusion compared to standard Johnson-Lindenstrauss lemma, namely, we preserve the distance with $(1\pm O(\epsilon))$ relative error and $\alpha$ additive error. Fortunately, the magnitude of $\alpha$ is small enough so that it won't affect the quality of our downstream task too much.

As an example, consider the following adaptive robust $\afn$: we use $k$ different independent data structures where each one has an independent JLT matrix $S_i$. At each query point $q$, we shall sample $\Theta(\log b)$ data structures and output the one with the best quality.

\end{remark}

As a direct consequence, we have the following result with $\tensorsrht$ and $\tensorsparse$:
\begin{corollary}\label{cor:robust_tensorsrht}
Let $V:=\{v_1,\ldots,v_m\}\in (\R^d)^m$, $\epsilon\in (0,1)$ and $\delta\in (0,1)$. Furthermore, let $\{S_i\}_{i=1}^k \in \R^{b\times d}$ for $k\geq \Omega((d+\log(1/\delta))\log(md))$ such that each $S_i$ is an independent $\tensorsrht$ matrix with $b=\Theta( \epsilon^{-2} \log^3(m/\epsilon) )$ rows and $\|S_i\|_F\leq d$. Then we have
\begin{align*}
    \forall q\in \mathbb{S}^{d-1}, \forall v\in V, \sum_{i=1}^k {\bf 1}[\|S_i(q-v)\|_2^2\leq (1\pm O(\epsilon))\|q-v\|_2^2+\alpha]\geq 0.95k
\end{align*}
with probability at least $1-\delta$ and $\alpha\leq O(\frac{1}{(md)^9})$.
\end{corollary}
\begin{proof}
The result follows by combining Lemma~\ref{lem:JL_srht} and Lemma~\ref{lem:robust_JL}.
\end{proof}

\begin{corollary}
\label{cor:robust_tensorsparse}
Let $V:=\{v_1,\ldots,v_m\}\in (\R^d)^m$, $\epsilon\in (0,1)$ and $\delta\in (0,1)$. Furthermore, let $\{R_i\}_{i=1}^k \in \R^{b\times d}$ for $k\geq \Omega((d+\log(1/\delta))\log(md))$ such that each $R_i$ is an independent $\tensorsparse$ matrix with $b=\Theta( \epsilon^{-2} \log m )$ rows and $\|R_i\|_F= d$. Then we have
\begin{align*}
    \forall q\in \mathbb{S}^{d-1}, \forall v\in V, \sum_{i=1}^k {\bf 1}[\|S_i(q-v)\|_2^2\leq (1\pm O(\epsilon))\|q-v\|_2^2+\alpha]\geq 0.95k
\end{align*}
with probability at least $1-\delta$ and $\alpha\leq O(\frac{1}{(md)^9})$.
\end{corollary}
\begin{proof}
The result follows from Theorem~\ref{thm:tensor_sparse_JLT} and Lemma~\ref{lem:robust_JL}.
\end{proof}

\section{Minimum Inner Product Search via Approximate Furthest Neighbor}\label{sec:minip_ds}

In this section, we use an efficient $\afn$ data structure to solve the $\minip$ task. Combining it with adaptive Johnson-Lindenstrauss transform, our data structure is robust and operates on low dimensions.

\begin{itemize}
     \item In Section~\ref{sec:minip_alg_thm}, we reduce the $\minip$ data structure problem by solving $\afn$ problem.
    \item In Section~\ref{sec:minip_ds_afn}, we show how to achieve efficient and robust $\minip$ via robust $\afn$ data structures.
\end{itemize}

Throughout this section, we use $n$ to denote the number of data points and $d$ to denote the dimension of data.

\subsection{Algorithm}
\label{sec:minip_alg_thm}

\begin{algorithm}[H]\caption{Minimum Inner Product Search}\label{alg:minip_via_afn}
\begin{algorithmic}[1]
\State {\bf data structure} \textsc{Minimum Inner Product Search} \Comment{Theorem~\ref{thm:robust_minip}}
\State {\bf members}
\State \hspace{4mm} \textsc{ApproximateFurthestNeighbor AFN}
\State {\bf end members}
\State 
\Procedure{Init}{$x_1,x_2,\cdots,x_n,\ov c,r$} 
    \State \textsc{AFN.Init}$(x_1,x_2,\cdots,x_n,\ov c,r)$
\EndProcedure 
\State 
\Procedure{Insert}{$z \in \R^d$} 
    \State \textsc{AFN.Insert}$(z)$
\EndProcedure 
\State 
\Procedure{Delete}{$i\in [n]$}
    \State \textsc{AFN.Delete}$(i)$
\EndProcedure
\State
\Procedure{QueryMin}{$q \in \R^d$} 
    \State $x_i\gets \textsc{AFN.Query}(q)$
    \State \Return $x_i$
\EndProcedure 
\State {\bf end data structure}
\end{algorithmic}
\end{algorithm}

The $\afn$ data structure we use has similar construction as~\cite{i03}, we include its complete algorithm and correctness proof in Appendix~\ref{app:afn}. We restate Theorem~\ref{thm:afn} here.

\begin{lemma}[Informal version of Theorem~\ref{thm:afn}]
Let $P\subset \R^d$ be an $n$-point dataset, $\ov c>1$, $r>0$ and $\delta>0$. Let $\epsilon=\ov c-1$. There exists a randomized dynamic data structure (against an oblivious adversary) that solves $(\ov c+\delta,r)$-$\afn$ task using space $O((n^{1+1/\ov c^2}\log n+dn^{1/\ov c^2}\log n)\log \log(d/\epsilon\delta)+dn)$ with the following operations:
\begin{itemize}
    \item \textsc{Init}: Preprocess $P$ in $O((n^{1+1/\ov c^2}\log^2 n+dn^{1/\ov c^2}\log n)\log\log (d/\epsilon\delta))$ time;
    \item \textsc{Query}: Given a point $q\in \R^d$, returns a $(\ov c+\delta)$-approximate furthest neighbor $p\in P$ with constant probability in $O(n^{1/\ov c^2}(d+\log n)\log n\log (d/\epsilon\delta)\log\log (d/\epsilon\delta))$ time;
    \item \textsc{Insert}: Insert a point $p\in \R^d$ into the data structure in $O(n^{1/\ov c^2}\log^2 n\log\log (d/\epsilon\delta)+d\log n)$ time;
    \item \textsc{Delete}: Delete a point $p\in \R^d$ from the data structure in $O(n^{1/\ov c^2}\log^2 n\log\log (d/\epsilon\delta)+d\log n)$ time.
\end{itemize}
\end{lemma}

Next, we introduce several corollaries that simplify the time complexity in solving $\afn$. 

\begin{corollary}\label{coro:ov_c2_greater_2}
Let $P\subset \R^d$ be an $n$-point dataset, $\ov c>\sqrt{2}$, and $r>0$. There exists a randomized dynamic data structure (Alg.~\ref{alg:ind03_main}, \ref{alg:ind03_main_2}) that solves $(2\ov c,r)$-$\afn$ with query time $O(n^{0.5}(d+\log n)\log n\log d \log \log d)$, preprocessing time $O((n^{1.5}\log^2 n+dn^{0.5}\log n)\log \log d)$ and space\\ $O((n^{1.5}\log^2 n+dn^{0.5}\log n)\log \log d+nd)$. Moreover, the dynamic data structure supports insert or delete in $O(n^{0.5}\log^2 n \log\log d+d\log n)$ time.
\end{corollary}
\begin{proof}
If $\ov c>\sqrt{2}$, we have $1/\ov c^2 <0.5$. We take this fact into the preprocessing, query, insert and delete time and get the following:

Space 
\begin{align*}
    O((n^{1+1/\ov c^2}\log^2 n+dn^{1/\ov c^2}\log n)\log\log (d/\epsilon\delta)+nd)=O((n^{1.5}\log^2 n+dn^{0.5}\log n)\log \log d+nd)
\end{align*}

Preprocessing time 
\begin{align*}
    O((n^{1+1/\ov c^2}\log^2 n+dn^{1/\ov c^2}\log n)\log\log (d/\epsilon\delta))=O((n^{1.5}\log^2 n+dn^{0.5}\log n)\log \log d)
\end{align*}

Query time
\begin{align*}
    O(n^{1/\ov c^2}(d+\log n)\log n\log (d/\epsilon\delta)\log\log (d/\epsilon\delta))= O(n^{0.5}(d+\log n)\log n\log d \log \log d)
\end{align*}

Insert/delete time
\begin{align*}
    O(n^{1/\ov c^2}\log^2 n\log\log (d/\epsilon\delta)+d\log n)=O(n^{0.5}\log^2 n \log\log d+d\log n)
\end{align*}

\end{proof}

\begin{corollary}\label{coro:ov_c2_greater_100}
Let $P\subset \R^d$ be an $n$-point dataset, $\ov c>10$, and $r>0$. There exists a randomized dynamic data structure (Alg.~\ref{alg:ind03_main}, \ref{alg:ind03_main_2}) that solves $(2\ov c,r)$-$\afn$ with query time $O(n^{0.01}(d+\log n)\log n\log d \log \log d)$, preprocessing time $O((n^{1.01}\log^2 n+dn^{0.01}\log n)\log \log d)$  and space\\ $O((n^{1.01}\log^2 n+dn^{0.01}\log n)\log \log d+nd)$. Moreover, the dynamic data structure supports insert or delete in $O(n^{0.01}\log^2 n \log\log d+d\log n)$ time.
\end{corollary}
\begin{proof}
If $\ov c>10$, we have $1/\ov c^2 <0.01$. We take this fact into the preprocessing, query, insert and delete time and get the following:

Space 
\begin{align*}
    O((n^{1+1/\ov c^2}\log^2 n+dn^{1/\ov c^2}\log n)\log\log (d/\epsilon\delta)+nd)=O((n^{1.01}\log^2 n+dn^{0.01}\log n)\log \log d+nd)
\end{align*}

Preprocessing time 
\begin{align*}
    O((n^{1+1/\ov c^2}\log^2 n+dn^{1/\ov c^2}\log n)\log\log (d/\epsilon\delta))=O((n^{1.01}\log^2 n+dn^{0.01}\log n)\log \log d)
\end{align*}

Query time
\begin{align*}
    O(n^{1/\ov c^2}(d+\log n)\log n\log (d/\epsilon\delta)\log\log (d/\epsilon\delta))= O(n^{0.01}(d+\log n)\log n\log d \log \log d)
\end{align*}

Insert/delete time
\begin{align*}
    O(n^{1/\ov c^2}\log^2 n\log\log (d/\epsilon\delta)+d\log n)=O(n^{0.01}\log^2 n \log\log d+d\log n)
\end{align*}

\end{proof}

\subsection{Guarantees of Approximate \texorpdfstring{$\minip$}{~} via \texorpdfstring{$\afn$}{}}
\label{sec:minip_ds_afn}
In this section, we show how to use $\afn$ data structure to solve approximate $\minip$. We take the adaptive query in iterative optimization algorithm into consideration and design a robust algorithm against adversary. Before proceeding to the main theorem of this section, we first consider a technical lemma regarding quantization.

\begin{lemma}\label{lem:robust_afn}
Let $c\in (0,1)$, $\tau\in (0,1)$ and $\lambda\in (0,1)$. Given a set of $n$-points $Y\subset \mathbb{S}^{d-1}$, one can construct a data structure with ${\cal T}_{\mathrm{init}}\cdot \kappa$ preprocessing time and ${\cal S}_{\mathrm{space}}\cdot \kappa$ space so that for every $x\in \mathbb{S}^{d-1}$ in an adaptive sequence $X=\{x_1,\ldots,x_T\}$, the query time is $\wt O(dn^{0.01}\cdot \kappa)$:
\begin{itemize}
    \item If $\minip(x,Y)\leq \tau$, then we output a vector in $Y$ which is a $(c,\tau,\lambda)$-$\minip$ with respect to $(x,Y)$ with probability at least $1-\delta$.
    \item Otherwise, we output {\fail}.
\end{itemize}
where $\kappa:=d\log(ndD_X/(\lambda\delta))$, and $D_X$ is the diameter of all queries in $X$.
\end{lemma}

\begin{proof}
The failure probability for an adaptive sequence $X$ is equivalent to the probability that at least one query $\hat{q}\in \hat{Q}$ fail in solving all $\kappa$ number of $(c,\tau)$-{$\minip$}. We bound this failure probability as 

\begin{align*}
    \Pr[\exists \hat{q}\in \hat{Q}~~~\textrm{s.t all } ~ (c,\tau)\textsc{-}{\minip}~ \mathsf{fail} ]=n \cdot (\frac{dD_X}{\lambda})^d \cdot (1/10)^{\kappa}\leq \delta
\end{align*}
where the last step follows from $\kappa:= d\log (nd D_X / (\lambda \delta ) ) $.

For the success queries, it introduces a $\lambda$ error in the inner product. Thus, the results is $(c,\tau,\lambda)$-{$\maxip$}.

Then, following Corollary~\ref{coro:ov_c2_greater_100}, we finish the proof.
\end{proof}

\begin{theorem}[Formal version of Theorem~\ref{thm:intro_robust_minip}]\label{thm:robust_minip}
Let $c \in (0,1)$, $\tau \in(0,1)$, $\lambda\in (0,1)$, $\epsilon\in (0,1)$ and $\delta\in (0,1)$. We define the following additional parameters: 
\begin{itemize}
    \item $\alpha=O(\frac{1}{(nd)^9})$, the additive error by Lemma~\ref{lem:robust_JL};
    \item $s\leq d$, the dimension of JLT;
    \item $k=O((d+\log(1/\delta))\log(nd))$, number of independent JLT sketches;
    \item $\kappa=s\log(ns/(\lambda\delta))$;
    \item $\wt \lambda=O(\sqrt{\frac{c-\tau}{c(1-\tau)}})\cdot (\lambda+\alpha)$, the additive error of $\minip$.
\end{itemize}

Let ${\cal T}_S(x)$ denote the time of applying $S$ to a vector $x\in \R^d$. Given a set of $n$-points $Y \subset \mathbb{S}^{d-1}$ on the sphere, one can build a dynamic data structure with preprocessing time ${\cal T}_{\mathsf{init}}\cdot \kappa\cdot k+{\cal T}_S(Y)\cdot k$, space ${\cal S}_{\mathsf{space}}\cdot \kappa\cdot k$ insert time $({\cal T}_{\mathsf{insert}}\cdot \kappa+{\cal T}_S(x))\cdot k$ and delete time  $({\cal T}_{\mathsf{delete}}\cdot \kappa+{\cal T}_S(x))\cdot k$ so that for every query $x \in \mathbb{S}^{d-1}$ in an adaptive sequence $X=\{x_1,\ldots,x_T\}$, the query time is $\wt O({\cal T}_{\mathsf{query}}\cdot\kappa+{\cal T}_S(x))$:
\begin{itemize}
    \item if $ \minip(x,Y)\leq\tau $, then we output a vector in $Y$ that is a $(c,\tau,\wt \lambda)$-$\minip$ with respect to $(x,Y)$.
    \item otherwise, we output {\fail}.
\end{itemize}
Further, 
\begin{itemize}
    \item If $c\in(\tau,\frac{8\tau}{(1-\epsilon)^2\tau+2\epsilon+7})$, we have ${\cal T}_{\mathsf{init}}=O((n^{1.5}\log^2 n+sn^{0.5}\log n)\log \log s)$, \\${\cal S}_{\mathsf{space}} = O((n^{1.5}\log^2 n+sn^{0.5}\log n)\log \log s+ns)$, ${\cal T}_{\mathsf{query}}=O(n^{0.5}(s+\log n)\log n\log s \log \log s)$ and ${\cal T}_{\mathsf{insert}}={\cal T}_{\mathsf{delete}}=O(n^{0.5}\log^2 n \log\log s+s\log n)$
    \item If $c\in(\tau,\frac{400\tau}{(1-\epsilon)^2\tau+2\epsilon+399})$, we have ${\cal T}_{\mathsf{init}}=O((n^{1.01}\log^2 n+sn^{0.01}\log n)\log \log s)$, ${\cal S}_{\mathsf{space}} = O((n^{1.01}\log^2 n+sn^{0.01}\log n)\log \log s+ns)$, ${\cal T}_{\mathsf{query}}=O(n^{0.01}(s+\log n)\log n\log s \log \log s)$ and\\ ${\cal T}_{\mathsf{insert}}={\cal T}_{\mathsf{delete}}=O(n^{0.01}\log^2 n \log\log s+s\log n)$
\end{itemize}
Finally, the probability that all queries succeed is at least $1-\delta$.

\end{theorem}

\begin{proof}
We first use Lemma~\ref{lem:robust_JL} to initiate $k\geq \Omega((d+\log(1/\delta))\log(nd))$ different JLT matrices with parameters $(m+1,\epsilon,\delta/2)$. Then, for each JLT matrix $S_i\in \R^{s\times d}$, we run the quantization process on it. Specifically, this requires us to use $\kappa=s\log (ns/(\lambda\delta))$ independent $\afn$ data structures due to Lemma~\ref{lem:robust_afn}. 

Throughout the proof, we will condition on the event that there exists some $i\in [k]$ such that $S_i$ preserves the pair-wise distances between any query point and points in $X$. To simplify the notation, we use $S$ to denote the corresponding JLT matrix $S_i$.

We consider the following: given a query point $Sx\in \R^s$, we quantize it into a point $\wh x\in \R^s$, then we use $\wh x$ as our query. Let $Sy$ be the furthest neighbor of $\wh x$, the $\afn$ data structure will output a point $Sy'$ with the guarantee that $\|Sy'-\wh x\|_2\geq \|Sy-\wh x\|_2/\ov c$. Towards the end, we wish to have a bound on the term $\|x-y'\|_2$ in terms of $\|x-y\|_2$.
\begin{align*}
    \|Sy'-Sx\|_2 = & ~ \|Sy'-\wh x+\wh x-Sx\|_2 \\
    \geq & ~ \|Sy'-\wh x\|_2-\|Sx -\wh x\|_2 \\
    \geq & ~ \|Sy-\wh x\|_2/\ov c-\lambda \\
    \geq & ~  \|Sy-Sx+Sx-\wh x\|_2/\ov c-\lambda \\
    \geq & ~  (\|Sy-Sx\|_2-\lambda)/\ov c-\lambda \\
    \geq & ~  \ov c^{-1}\cdot ((1-\epsilon)\|y-x\|_2-\alpha-\lambda)-\lambda,
\end{align*}
on the other hand, we know that $\|x-y'\|_2 \geq \frac{\|Sy'-Sx\|_2-\alpha}{1+\epsilon}$, we hence conclude that
\begin{align*}
    \|x-y'\|_2 \geq & ~ \frac{\ov c^{-1}\cdot (1-\epsilon)\|x-y\|_2-(1+\ov c^{-1})\lambda-(1+\ov c^{-1})\alpha}{1+\epsilon} \\
    = & ~ \underbrace{\ov c^{-1}\cdot (1-O(\epsilon))}_{\wt c^{-1}}\|x-y\|_2-\underbrace{(1-O(\epsilon))\cdot ((1+\ov c^{-1})\cdot \lambda+(1+\ov c^{-1})\cdot \alpha)}_{\wt \lambda}.
\end{align*}
By further setting $\wt r=\frac{r}{1-\epsilon}$, we conclude we get a $(\wt c,\wt r)$-$\afn$ data structure with additive error $\wt \lambda$. Moreover, this $(\wt c,\wt r)$-$\afn$ data structure would also be a data structure for $(c, \tau)$-$\minip$  with  $\tau = 1-0.5 \wt r^2$ and $c = \frac{1-0.5  r^2}{1 - 0.5 \wt r^2/\wt c^2}$. Using Eq.~\eqref{eq:ovc_delta_c_tau}, we have $\wt c^2=\frac{c(1-c\tau)}{c-\tau}$.  

Next, we present how to obtain the desired query, preprocessing insert, and delete time complexity in the statement.

\textbf{Part 1.}
Let $\wt c=2\ov c/(1-\epsilon)$, we conclude that $\ov c^2=\frac{c(1-c\tau)(1-\epsilon)^2}{4(c-\tau)}$. If $\tau\in(0,1)$ and $c\in(\tau,\frac{8\tau}{(1-\epsilon)^2\tau+2\epsilon+7})$, we have
\begin{align*}
    \ov{c}^2
    = & ~\frac{c(1-\tau)(1-\epsilon)^2}{4(c-\tau)}\\
    > & ~(1-\epsilon)^2\cdot\frac{8\tau}{(1-\epsilon)^2\tau+2\epsilon+7}) \cdot \frac{1-\tau}{4(\frac{8\tau}{(1-\epsilon)^2\tau+2\epsilon+7})-\tau)}\\
    > & ~(1-\epsilon)^2\cdot\frac{8\tau}{(1-\epsilon)^2\tau-(1-\epsilon)^2+8}) \cdot \frac{1-\tau}{4(\frac{8\tau}{(1-\epsilon)^2\tau-(1-\epsilon)^2+8})-\tau)}\\  
    = & ~ \frac{2(1-\epsilon)^2\tau(1-\tau)}{8\tau-(1-\epsilon)^2\tau^2+(1-\epsilon)^2\tau-8\tau}\\
    = & ~ 2
\end{align*}
where the second and third steps follow from Lemma~\ref{lem:c_tau_greater_4}.

Then, we use Corollary~\ref{coro:ov_c2_greater_2} with $\ov{c}^2>2$ and obtain the  query time $O(n^{0.5}(s+\log n)\log n\log s \log \log s)$, preprocessing time $O((n^{1.5}\log^2 n+sn^{0.5}\log n)\log \log s)$ and space $O((n^{1.5}\log^2 n+sn^{0.5}\log n)\log \log s+ns)$. Moreover, the dynamic data structure supports insert or delete in $O(n^{0.5}\log^2 n \log\log s+s\log n)$ time.

\textbf{Part 2.}
Let $\wt c=2\ov c/(1-\epsilon)$, we conclude that $\ov c^2=\frac{c(1-c\tau)(1-\epsilon)^2}{4(c-\tau)}$. If $\tau\in(0,1)$ and $c\in(\tau,\frac{400\tau}{(1-\epsilon)^2\tau+2\epsilon+399})$, we have
\begin{align*}
    \ov{c}^2
    = & ~\frac{c(1-\tau)(1-\epsilon)^2}{4(c-\tau)}\\
    > & ~(1-\epsilon)^2\cdot\frac{400\tau}{(1-\epsilon)^2\tau+2\epsilon+399}) \cdot \frac{1-\tau}{4(\frac{400\tau}{(1-\epsilon)^2\tau+2\epsilon+399})-\tau)}\\
    > & ~(1-\epsilon)^2\cdot\frac{400\tau}{(1-\epsilon)^2\tau-(1-\epsilon)^2+400}) \cdot \frac{1-\tau}{4(\frac{400\tau}{(1-\epsilon)^2\tau-(1-\epsilon)^2+400})-\tau)}\\  
    = & ~ \frac{100(1-\epsilon)^2\tau(1-\tau)}{400\tau-(1-\epsilon)^2\tau^2+(1-\epsilon)^2\tau-400\tau}\\
    = & ~ 10
\end{align*}
where the second and third steps follow from Lemma~\ref{lem:c_tau_greater_4}.

Then, we use Corollary~\ref{coro:ov_c2_greater_100} with $\ov{c}^2>100$ and obtain the  query time $O(n^{0.01}(s+\log n)\log n\log s \log \log s)$, preprocessing time $O((n^{1.01}\log^2 n+sn^{0.01}\log n)\log \log s)$ and space $O((n^{1.01}\log^2 n+sn^{0.01}\log n)\log \log s+ns)$. Moreover, the dynamic data structure supports insert or delete in $O(n^{0.01}\log^2 n \log\log s+s\log n)$ time.

Next, we analyze the additive error.
Use the relationship $\ov c^2=\frac{c(1-\tau)(1-\epsilon)^2}{4(c-\tau)}$ we derived above, we can further simplify $\wt \lambda$:
\begin{align*}
    (1-O(\epsilon))\cdot ((1+\ov c^{-1})\cdot \lambda+(1+\ov c^{-1})\cdot \alpha) \leq & ~ O(1)
    \cdot \sqrt{\frac{c-\tau}{c(1-\tau)}}\cdot (\lambda+\alpha).
\end{align*}
Therefore, we simplify $\wt \lambda\leq O(\sqrt{\frac{c-\tau}{c(1-\tau)}}\cdot (\lambda+\alpha))$, we conclude that we get a $(c,\tau,\wt \lambda)$-$\minip$.
\end{proof}

\section{Linear-Sized Spectral Sparsification via Positive Inner Product Search}\label{sec:general_sum_rank1}

In this section, we study the linear-sized spectral sparsification problem for a $V\in \R^{m\times d}$ matrix.
\begin{itemize}
    \item In Section~\ref{sec:bss_problem_setup}, we setup the sparsification task for general matrix sparsification.
    \item In Section~\ref{sec:bss_vanilla}, we describe the vanilla BSS algorithm and state several lemmas.
    \item In Section~\ref{sec:bss_maxip}, we present our deterministic algorithm that solves the linear-sized sparsifier problem with positive inner product search data structure.
    \item In Section~\ref{sec:bss_maxip_comp_app}, we compare our method with other fast algorithms to construct spectral sparsifiers and its extension to sparsify higher rank PSD matrices.
\end{itemize}

\subsection{Problem Setup}
\label{sec:bss_problem_setup}
In this section, we setup the problem. Given a full rank matrix $V\in \R^{m\times d}$ with $m\geq d$, the goal is to pick only $s=\Theta(\epsilon^{-2}d)$ rescaled rows of $V$ to form $\wt V\in \R^{s\times d}$ such that $(1-\epsilon)V^\top V\preceq \wt V^\top \wt V (1+\epsilon)V^\top V$. Since $V$ is full rank, we can normalize $V^\top V$ and assume it's identity. The task can be defined as follows:

\begin{definition}
Suppose we are given $m$ vectors $v_i,\ldots,v_m\in \R^d$ satisfying $\sum_{i=1}^m v_iv_i^\top=I$, we want to find scalars $\{s_i\}_{i=1}^m$ satisfying
\begin{align*}
    |\{s_i: s_i\neq 0\}| = & ~ O(d/\epsilon^2),
\end{align*}
such that
\begin{align*}
    (1-\epsilon)\cdot I \preceq \sum_{i=1}^m s_i v_iv_i^\top \preceq (1+\epsilon)\cdot I.
\end{align*}
\end{definition}

We define several notions that will be used extensively in the proof of BSS sparsifier.
\begin{definition}
\label{def:bss_barrier}
Let $A\in \R^{d\times d}$ be a symmetric matrix with eigenvalues $\lambda_1,\ldots,\lambda_d$ and $u,\ell \in \R$, define:
\begin{align*}
    \Phi^u(A) := & ~ \tr[(uI-A)^{-1}] = \sum_{i=1}^d \frac{1}{u-\lambda_i} \\
    \Phi_\ell(A) := & ~ \tr[(A-\ell I)^{-1}] = \sum_{i=1}^d \frac{1}{\lambda_i-\ell}.
\end{align*}
\end{definition}

\subsection{The BSS Algorithm}
\label{sec:bss_vanilla}
The BSS algorithm is as follows: the algorithm starts by maintaining two ``barriers'' of eigenvalues $u_0=\frac{d}{\epsilon}$ and $\ell_0=-\frac{d}{\epsilon}$. Iteratively, the algorithm searches for an index $i\in [m]$ such that the inner product between $v_iv_i^\top$ and a quantity related to lower barrier is large while the inner product related to barrier is small. Then we add this outer product $v_iv_i^\top$ with a scaling into the matrix $A$ we are forming. After $\Theta(d/\epsilon^2)$ iterations, we've constructed a matrix $A$ with the property that $A\approx_\epsilon I$.

We formalize the algorithm as follows:

\begin{algorithm}[H]
\caption{BSS algorithm}
\label{alg:BSS_vanilla}
\begin{algorithmic}[1]
\Procedure{BSS}{$\{ v_1,\ldots,v_m\}\in (\R^d)^m$}
\State $u_0\gets \frac{d}{\epsilon},\ell_0\gets -\frac{d}{\epsilon}$
\State $A_0\gets {\bf 0}_{d\times d}$
\State $T\gets \frac{d}{\epsilon^2}$
\State $\delta_U\gets 1,\delta_L\gets \frac{1}{1+2\epsilon}$
\For{$t=1\to T$}
\State $u_{t}\gets u_{t-1}+\delta_U, \ell_t\gets \ell_{t-1}+\delta_L$
\State $L_t\gets \frac{(A_{t-1}-\ell_t I)^{-2}}{\Phi_{\ell_t}(A_{t-1})-\Phi_{\ell_{t-1}}(A_{t-1})}-(A_{t-1}-\ell_t I)^{-1}$
\State $U_t\gets \frac{(u_t I-A_{t-1})^{-2}}{\Phi_{u_{t-1}}(A_{t-1})-\Phi_{u_t}(A_{t-1})}+(u_t I-A_{t-1})^{-1}$
\State Find an index $j$ such that 
\begin{align*}
    v_j^\top (L_t-U_t) v_j \geq & ~ 0
\end{align*}
\State $c\gets \frac{v_j^\top (L_t+U_t)v_j}{2}$ 
\State $A_t\gets A_{t-1}+\frac{1}{c}\cdot v_jv_j^\top$
\EndFor
\State \Return $A_T/d$
\EndProcedure
\end{algorithmic}
\end{algorithm}
The central lemma that guarantees the BSS algorithm to find a good sparsifier that satisfies both upper and lower bound is the following:

\begin{lemma}[Lemma 3.5 of~\cite{bss12}]
\label{lem:bss_two_barriers}
Suppose $A\in \R^{d\times d}$ satisfying $\ell I\prec A\prec uI$, let $\epsilon\in (0,1)$ and suppose $\Phi_u(A)\leq \epsilon, \Phi_\ell(A)\leq \epsilon$, and $\epsilon,\delta_U,\delta_L$ satisfying
\begin{align*}
    0 \leq \frac{1}{\delta_U}+\epsilon \leq \frac{1}{\delta_L}-\epsilon,
\end{align*}
then we have
\begin{enumerate}
    \item {\bf Lower bounding lower barrier.} 
\begin{align*}
    \sum_{i=1}^m v_i^\top (\frac{(A-(\ell+\delta_L) I)^{-2}}{\Phi_{\ell+\delta_L}(A)-\Phi_\ell(A)}-(A-(\ell+\delta_L)I)^{-1}) v_i \geq & ~ \frac{1}{\delta_L}-\epsilon.
\end{align*}
    \item {\bf Upper bounding upper barrier.}
    \begin{align*}
       \sum_{i=1}^m v_i^\top (\frac{((u+\delta_U)I-A)^{-2}}{\Phi^u(A)-\Phi^{u+\delta_U}(A)}+((u+\delta_U)I-A)^{-1})v_i \leq & ~ \frac{1}{\delta_U}+\epsilon.
    \end{align*}
\end{enumerate}
\end{lemma}

We also record two lemmas that control the growth of lower and upper barriers.

\begin{lemma}[Lemma 3.3 of~\cite{bss12}]
\label{lem:bss_upper_barrier}
Suppose $A\prec uI$ and $v\in \R^d$ is any vector. If
\begin{align*}
    c\geq & ~ v^\top (\frac{((u+\delta_U)I-A)^{-2}}{\Phi^u(A)-\Phi^{u+\delta_U}(A)}+((u+\delta_U)I-A)^{-1})v,
\end{align*}
then
\begin{align*}
    \Phi^{u+\delta_U}(A+\frac{1}{c}\cdot vv^\top)\leq \Phi^u(A)~\text{and}~A+\frac{1}{c}\cdot vv^\top \prec (u+\delta_U)I.
\end{align*}
\end{lemma}

\begin{lemma}[Lemma 3.4 of~\cite{bss12}]
\label{lem:bss_lower_barrier}
Suppose $A\succ \ell I$, $\Phi_\ell(A)\leq 1/\delta_L$ and $v\in \R^d$ is any vector. If
\begin{align*}
    0<c\leq & ~ v^\top (\frac{(A-(\ell+\delta_L) I)^{-2}}{\Phi_{\ell+\delta_L}(A)-\Phi_\ell(A)}-(A-(\ell+\delta_L)I)^{-1})v,
\end{align*}
then
\begin{align*}
    \Phi_{\ell+\delta_L}(A+\frac{1}{c}\cdot vv^\top)\leq \Phi_\ell(A)~\text{and}~A+\frac{1}{c}\cdot vv^\top \succ (\ell+\delta_L)I.
\end{align*}
\end{lemma}

Combining the above 3 lemmas, we derive a lemma that justifies that in each iteration of Alg.~\ref{alg:BSS_vanilla}, we can always find an index $j$ satisfying the inequality on line 10. To simplify notation, we define $L_t:=(\frac{(A-(\ell+\delta_L) I)^{-2}}{\Phi_{\ell+\delta_L}(A)-\Phi_\ell(A)}-(A-(\ell+\delta_L)I)^{-1})$ and $U_t:=(\frac{((u+\delta_U)I-A)^{-2}}{\Phi^u(A)-\Phi^{u+\delta_U}(A)}+((u+\delta_U)I-A)^{-1})$.
\begin{lemma}\label{cor:bss_strict_gap}
Suppose $A\in \R^{d\times d}$ satisfying $\ell I\prec A\prec uI$, let $\epsilon\in (0,1)$ and suppose $\Phi^u(A)\leq \epsilon, \Phi_\ell(A)\leq \epsilon$, and $\epsilon,\delta_U,\delta_L$ satisfying
\begin{align*}
    0 \leq \frac{1}{\delta_U}+\epsilon \leq \frac{1}{\delta_L}-\epsilon,
\end{align*}
then there exists an index $j\in [m]$ and a positive value $c$ such that
\begin{enumerate}
    \item {\bf Witness of gap between lower and upper barriers.} 
    \begin{align*}
        v_j^\top L_t v_j \geq  c \geq v_j^\top U_t v_j.
    \end{align*}
    \item {\bf Spectral property.} 
    \begin{align*}
        (\ell+\delta_L)I \prec A+\frac{1}{c} \cdot v_jv_j^\top \prec (u+\delta_U)I.
    \end{align*}
\end{enumerate}
Moreover, if we further have 
\begin{align*}
    0 \leq \frac{1}{\delta_U}+\epsilon < \frac{1}{\delta_L}-\epsilon,
\end{align*}
then the witness of gap between lower and upper barriers has a strict inequality between the two quantities:
\begin{align*}
    v_j^\top L_t v_j >  c > v_j^\top U_t v_j.
\end{align*}
\end{lemma}
\begin{proof}
By Lemma~\ref{lem:bss_two_barriers}, we have the following:
\begin{align*}
    \sum_{i=1}^m v_i^\top (\frac{(A-(\ell+\delta_L) I)^{-2}}{\Phi_{\ell+\delta_L}(A)-\Phi_\ell(A)}-(A-(\ell+\delta_L)I)^{-1})v_i \geq & ~ \frac{1}{\delta_L}-\epsilon, \\
    \sum_{i=1}^m v_i^\top (\frac{((u+\delta_U)I-A)^{-2}}{\Phi^u(A)-\Phi^{u+\delta_U}(A)}+((u+\delta_U)I-A)^{-1})v_i \leq & ~ \frac{1}{\delta_U}+\epsilon.
\end{align*}
By an averaging argument, there must exist an index $j\in [m]$ that witnesses this inequality, i.e., 
\begin{align*}
    v_j^\top (\frac{(A-(\ell+\delta_L) I)^{-2}}{\Phi_{\ell+\delta_L}(A)-\Phi_\ell(A)}-(A-(\ell+\delta_L)I)^{-1})v_j \geq & ~ v_j^\top (\frac{((u+\delta_U)I-A)^{-2}}{\Phi^u(A)-\Phi^{u+\delta_U}(A)}+((u+\delta_U)I-A)^{-1})v_j.
\end{align*}
The spectral property is guaranteed by Lemma~\ref{lem:bss_lower_barrier} and Lemma~\ref{lem:bss_upper_barrier}.

For the strict inequality, note that if we have $\frac{1}{\delta_L}-\epsilon>\frac{1}{\delta_U}+\epsilon$, then by Lemma~\ref{lem:bss_two_barriers} and again by an averaging argument, we conclude that witness also exhibits a strict inequality.
\end{proof}

We also include a proof for the main Theorem of~\cite{bss12} here, since we will need to later adapt our data structure for this problem.

\begin{lemma}[Theorem 3.1 of~\cite{bss12}]
\label{lem:bss}
Suppose we are given $m$ vectors $v_1,\ldots,v_m\in \R^d$ satisfying $\sum_{i=1}^m v_iv_i^\top=I$, then there exists a deterministic algorithm (Alg.~\ref{alg:BSS_vanilla}) can find scalars $\{s_i\}_{i=1}^m$ satisfying
\begin{align*}
    |\{s_i: s_i\neq 0\}| = & ~ O(d/\epsilon^2),
\end{align*}
such that
\begin{align*}
    (1-\epsilon)\cdot I \preceq \sum_{i=1}^m s_i v_iv_i^\top \preceq (1+\epsilon)I.
\end{align*}
The algorithm (Alg.~\ref{alg:BSS_vanilla}) runs in time $O(md^3/\epsilon^2)$.
\end{lemma}

\begin{proof}
We first prove the correctness. By the update rule of Alg.~\ref{alg:BSS_vanilla}, we know that we maintain the following invariants across all iterations due to Lemma~\ref{lem:bss_lower_barrier} and Lemma~\ref{lem:bss_upper_barrier}
\begin{align*}
    \Phi_{u_t}(A_t) \leq \Phi_{u_{t-1}}(A_{t-1})~\text{and}~\Phi_{\ell_t}(A_t)\leq \Phi_{\ell_{t-1}}(A_{t-1}),
\end{align*}
which means it suffices to examine $\Phi_{u_0}(A_0)$ and $\Phi_{\ell_0}(A_0)$ respectively, recall that we choose $u_0=\frac{d}{\epsilon}$, $\ell_0=-\frac{d}{\epsilon}$, hence we have
\begin{align*}
    \Phi_{u_0}(A_0) = & ~ \sum_{i=1}^d \frac{\epsilon}{d} \\
    = & ~ \epsilon, \\
    \Phi_{\ell_0}(A_0) = & ~ \sum_{i=1}^d \frac{\epsilon}{d} \\
    = & ~ \epsilon.
\end{align*}
To conclude the proof, we shall apply Lemma~\ref{lem:bss_two_barriers} for $T=\Theta(d/\epsilon^2)$ times, so we verify the relations between $\epsilon,\delta_U,\delta_L$:
\begin{align*}
    \frac{1}{\delta_U}+\epsilon = & ~ 1+\epsilon \geq 0, \\
    \frac{1}{\delta_L}-\epsilon = & ~ 1+\epsilon \geq  \frac{1}{\delta_U}+\epsilon.
\end{align*}
This means we can apply Lemma~\ref{lem:bss_two_barriers} and have
\begin{align*}
    (\ell_0+T\delta_L)I\prec A_T \prec (u_0+T\delta_U) I,
\end{align*}
plug in the values of $\ell_0,u_0,\delta_L,\delta_U$ and $T$, we conclude that
\begin{align*}
    (1-\epsilon-2\epsilon^2)I \prec A_T/d \prec (1+\epsilon)I.
\end{align*}
Now we analyze the running time, the algorithm iterates for $T=\Theta(d/\epsilon^2)$ iterations, and for each iteration $t$, we compute $L_t$ and $U_t$ in $O(d^\omega)$ time, and the search for index $j$ takes $O(md^2)$ time. Hence, the total running time is $O(md^3/\epsilon^2)$.
\end{proof}

\subsection{Faster Deterministic Sparsification via Positive Inner Product Search Tree}
\label{sec:bss_maxip}

We observe that the core of the deterministic BSS algorithm is an inner product search step: given a query matrix $L_t-U_t$, we need to find a vector $v_i$ with $\langle v_iv_i^\top, L_t-U_t\rangle$. To speed up this process, we make use of the positive search tree we developed in prior section. In short, we first preprocess all vectors in $\min \{O(\nnz(V^2)), O(md^{\omega-1})\}$ time, then at query time, we simply perform the positive search to find the desired vector, in time $\wt O(d^2)$ or $O(d^\omega)$.

We present our algorithm as follows:
\begin{algorithm}[H]
\caption{Faster Sparsification with Positive Inner Product Search Tree}
\label{alg:BSS_ours}
\begin{algorithmic}[1]
\Procedure{FasterSparsification}{$V=\{ v_1,\ldots,v_m\}\in (\R^d)^m$} \Comment{Theorem~\ref{thm:bss_det}}
\State $u_0\gets \frac{d}{\epsilon},\ell_0\gets -\frac{d}{\epsilon}$
\State $A_0\gets {\bf 0}_{d\times d}$
\State $T\gets \frac{d}{\epsilon^2}$
\State $\delta_U\gets 1,\delta_L\gets \frac{1}{1+3\epsilon}$
\If{$md^{\omega-1}\leq \nnz(V^2)$}
\State $\textsc{DS}\gets \textsc{VectorPS VecPS}$
\Else
\State $\textsc{DS}\gets \textsc{MatrixPS MatPS}$
\EndIf
\State $\textsc{DS.Init}(V)$ 
\For{$t=1\to T$}
\State $u_{t}\gets u_{t-1}+\delta_U, \ell_t\gets \ell_{t-1}+\delta_L$
\State $L_t\gets \frac{(A_{t-1}-\ell_t I)^{-2}}{\Phi_{\ell_t}(A_{t-1})-\Phi_{\ell_{t-1}}(A_{t-1})}-(A_{t-1}-\ell_t I)^{-1}$
\State $U_t\gets \frac{(u_t I-A_{t-1})^{-2}}{\Phi^{u_{t-1}}(A_{t-1})-\Phi^{u_t}(A_{t-1})}+(u_t I-A_{t-1})^{-1}$
\State $Q\gets L_t-U_t$
\State $v_j\gets \textsc{DS.QueryPositiveSearch}(Q)$
\State $c_t\gets \frac{v_j^\top (L_t+U_t)v_j}{2}$
\State $A_t\gets A_{t-1}+\frac{1}{c_t}\cdot v_jv_j^\top$
\EndFor
\State \Return $A_T/d$
\EndProcedure
\end{algorithmic}
\end{algorithm}

The correctness of the above algorithm follows obviously: by using the Positive Inner Product Search Tree, we are guaranteed to find a vector with positive inner product, which suffices to proceed the BSS algorithm. We summarize the running time in the following theorem.

\begin{theorem}[Formal version of Theorem~\ref{thm:intro_bss}]\label{thm:bss_det}
Suppose we have $\sum_{i=1}^m v_iv_i^\top = I$, let $A$ be the output of Algorithm~\ref{alg:BSS_ours}, then
\begin{align*}
    (1-\epsilon)\cdot I \preceq A \preceq (1+\epsilon)\cdot I
\end{align*}
and $A=\sum_{i=1}^m s_iv_iv_i^\top$ for $|\{s_i:s_i\neq 0\}|=\epsilon^{-2}d$.

Moreover, Algorithm~\ref{alg:BSS_ours} has runtime
\begin{align*}
    \min\{\nnz(V^2),md^{\omega-1}\}+\epsilon^{-2}d^{\omega+1}.
\end{align*}
\end{theorem}

\begin{proof}
The correctness follows naturally. To see the running time, note that by Theorem~\ref{thm:matrix_sample_tree} and Theorem~\ref{thm:vector_sample_tree}, the initialization takes $\min\{\nnz(V^2),md^{\omega-1}\}$ time. For each iteration, we need to invert $d\times d$ matrices, which takes $O(d^\omega)$ time, and we need to query the Positive IP Search Tree, which takes $\wt O(d^2)$ (Theorem~\ref{thm:matrix_sample_tree}) or $O(d^\omega)$ (Theorem~\ref{thm:vector_sample_tree}) time. Thus, each iteration takes $ O(d^\omega)$ time, and there are $O(\epsilon^{-2}d)$ iterations in total. Hence, the total running time is
\begin{align*}
    \min\{\nnz(V^2),md^{\omega-1}\}+\epsilon^{-2}d^{\omega+1}.
\end{align*}
\end{proof}

\subsection{Deterministic Sparsification via \textsc{VectorPS}: Comparisons and Extensions}
\label{sec:bss_maxip_comp_app}
We compare our algorithm with known algorithms in the literature, both deterministic and randomized. 

\paragraph{\cite{z12}: Deterministic.} To improve the running time of~\cite{bss12},~\cite{z12} adapts the following strategy: it uses a deterministic lossy construction to first compute a sparsifier of size $\epsilon^{-2}d\log d$, then it runs the BSS algorithm on the matrix with $\epsilon^{-2}d\log d$. The deterministic sparsifier it computes can be viewed as an analogy of the leverage score sampling~\cite{ss11} in deterministic setting. In order to construct the lossy sparsifier, it makes use of the hyperbolic cosine function as a potential to progress. At each iteration, it has to compute the hyperbolic cosine potential over all rows of $V$, incurring a $O(md)$ cost per iteration. Similar to the BSS algorithm, it only selects one vector at each iteration, hence its running time is $\wt O(\epsilon^{-2}md^2)$ for constructing the lossy sparsifier. For the case of $m=d^2$, which is the standard case for a dense graph or a tall skinny matrix, their algorithm has $\Omega(d^4)$ runtime.

\paragraph{\cite{azlo15}: Randomized.} The work by Allen-Zhu, Liao and Orecchia provides an alternative view of constructing the spectral sparsifier, it shows that spectral sparsification can be solved as a regret minimization problem over PSD matrices. While leverage score sampling~\cite{ss11} has inherent connection with matrix multiplicative weights update~\cite{ak07,o11}, the linear-sized sparsifier requires a new view of the problem. Using regret minimization and the popular follow-the-regularized-leader (FTRL) approach, they show that by using a $q$ norm regularizer, one can obtain a linear-sized sparsifier using mirror descent. This also introduces a novel potential function that leads later breakthroughs. From an algorithmic perspective, by using Johnson-Lindenstrauss to reduce the dimension then compute all necessary information at each iteration, they obtain an improved running time of $\wt O(\epsilon^{-O(1)}(md^2+d^{3+1/q}))$. Unfortunately, without the use of Johnson-Lindenstrauss, their algorithm is no faster than~\cite{bss12}.

\paragraph{\cite{ls15}: Randomized.} Motivated by the $q$ norm potential function in~\cite{azlo15}, Lee and Sun show how to further speed up their algorithm via randomized sampling. A key (bonus) result of the $q$ norm potential function is that the iterative process might only run for $O(\epsilon^{-2}qd^{1/q})$ iterations.~\cite{ls15} exploits this feature and ensures that their algorithm only runs for $O(\epsilon^{-2}qd^{3/q})$ iterations, at each iteration, they batch sample many vectors and add them into the target matrix. By using fast matrix multiplication, they achieve a per iteration cost of $md^{\omega-1}$, coupled with $\epsilon^{-2}qd^{3/q}$ iterations, their algorithm has a running time of $\wt O(\epsilon^{-2}qmd^{\omega-1+3/q})$. Since their algorithm heavily relies on the sampling process, it is certain that their algorithm is randomized.

\paragraph{~\cite{ls17}: Randomized.} Note that since~\cite{ls15}, reducing the iteration count has been a main theme of speeding up the construction of linear-sized sparsifier.~\cite{ls17} achieves the optimality by only requires $O(\epsilon^{-2})$ iterations, with a novel potential function that provides more leeway in the analysis. This also means that roughly for each iteration, one needs to select $O(d)$ vectors into the sparsifier. Intuitively,~\cite{ls17} reduces the iteration count by setting up a much stronger objective per iteration. To solve such an objective, they invoke a positive SDP solver~\cite{azlo16}. The correctness of their SDP solver builds upon its internal randomness, hence it is unclear how to derandomize their method and achieve a similar running time. 

\paragraph{Bootstrapping via Sketching: Randomized.} A randomized alternative of~\cite{z12} is to approximate leverage score quickly then run any fast randomized linear-sized algorithm (say,~\cite{ls17}) on the bootstrapped sparsifier. To quickly approximate leverage scores, a popular approach is to use randomized sketching and adaptive sampling~\cite{bw14,swz17,swz19_soda}. Similar to any randomized method we have discussed above, the speed comes from the use of randomness and efficient sketching matrix, which is inherent random. Also, such method typically does not care about the dependence on $d$, since it typically applies a sketching matrix then perform QR decomposition on the sketched matrix, incurring a $\poly(d)$ dependence on the running time.

\paragraph{Comparison with Our Method.} We note that all the faster algorithms that break the $\Omega(d^4)$ barrier of~\cite{z12} are randomized methods, they are either slow when derandomized, or inherently rely on the randomness to progress the algorithm. This poses a challenge when one wants to design dynamic spectral sparsifiers against adaptive adversary based on these primitives. In contrast to their deterministic counterpart, where the robustness against adaptive queries is guaranteed, it is nontrivial to modify a static randomized algorithm for adaptivity without slowdowns. Obtaining efficient deterministic spectral sparsifier has sophisticated implications for various dynamic graph and matrix problems.

In many senses, while previous results~\cite{ls15,ls17} give almost and nearly linear time algorithms for spectral sparsification, they all need to read the input data entirely for each iteration. This is fine when the iteration count is small, and in the case of graph, such $\epsilon^{-2}m$ dependence seems inevitable since one replaces the primitive matrix operations such as inversion with a Laplacian solve. When the target matrix $V$ is a general matrix, it is clear that reading input for each iteration is sub-optimal. From this perspective, our data structure formulation gives the right direction to achieve the \emph{truly} optimal running time for this problem, and various sparsification problem using the potential function of~\cite{bss12}. It also opens up the door to further study of efficient and deterministic spectral sparsifier.

\paragraph{Extensions to Sparsify PSD Matrices.} 
We remark that our framework can be further extended to sparsify sum of PSD matrices, using the regret minimization approach introduced by Allen-Zhu, Liao and Orecchia~\cite{azlo15}. Observe that their mirror descent algorithm can also be viewed as a variant of positive inner product search. 
\section{One-Sided Kadison-Singer via Minimum Inner Product Search}
\label{sec:ks}

In this section, we provide efficient data structure for one-sided Kadison-Singer problem. 
\begin{itemize}
    \item In Section~\ref{sec:ks_setup}, we formally define the one-sided Kadison-Singer problem.
    \item In Section~\ref{sec:ks_correct}, we prove the correctness of a greedy process with an approximate guarantee. 
    \item In Section~\ref{sec:ks_alg1}, we provide an analysis for a straightforward implementation of the greedy process.
    \item In Section~\ref{sec:ks_small_iter}, we use $\aipe$ data structure to handle the case where number of iterations is small.
    \item In Section~\ref{sec:ks_alg4}, we use $\afn$ data structure to handle the case where number of iterations is large.
\end{itemize}

\subsection{Problem Setup}
\label{sec:ks_setup}

We consider a simpler and one-sided version of the well-known Kadison-Singer problem studied by Weaver~\cite{w13}, which is similar to the restricted invertibility problem~\cite{s10}.
\begin{question}
Does there exist a constant $N\in \mathbb{N}$, such that if $\{v_1,\ldots,v_m \}\in (\R^d)^m$ satisfying $\|v_i\|_2=\frac{1}{\sqrt N},\forall i\in [m]$, and
\begin{align*}
    \sum_{i=1}^m v_iv_i^\top = & ~ I,
\end{align*}
then there exists a subset $S\subseteq \{1,\ldots,m\}$ such that for any $q\in (0,1)$, we have
\begin{align*}
    \|\sum_{i\in S} v_iv_i^\top\| \leq & ~ q-\frac{1}{\sqrt N}.
\end{align*}
\end{question}
In Weaver's discrepancy theory II, 2013~\cite{w13}, he presented a polynomial algorithm that has the following guarantee:
\begin{align*}
    \|\sum_{i \in S} v_iv_i^\top \| \leq \frac{n}{m} + O(\frac{1}{\sqrt{N}}).
\end{align*}
Here $n:=|S|$. We will dedicate our efforts to design a faster algorithmic framework to achieve an approximate guarantee as Weaver's result.

\subsection{Approximate Greedy Lemma}
\label{sec:ks_correct}

In this section, we describe and analyze a high level greedy process to construct the set $S$ with the guarantee given in~\cite{w13}. We generalize his analysis by introducing an approximation factor $\beta$, which is particularly valuable when later, we want to use certain approximate data structure to implement the high-level idea. We start with the main lemma of this section.
\begin{lemma}[Approximate greedy lemma]\label{thm:ks_correct}
Let $N\in \R_+$, if $\{v_1, \ldots, v_m\}$ is a finite sequence of vectors in $\R^{d}$ satisfying $\|v_i\|_2 = \frac{1}{\sqrt{N}}, \forall i \in [m]$ and
\begin{align*}
    \sum_{i=1}^m v_iv_i^\top = I.
\end{align*} Then for any $n < m$ and any unit vector $u$, we can find a set $S$ $(|S| = n)$ such that
\begin{align*}
    \|\sum_{i \in S} v_iv_i^\top \|\leq \beta \cdot ( \frac{n}{m} + O(\frac{1}{\sqrt{N}}) ).
\end{align*}
where $\beta \geq 1$.
\end{lemma}
\begin{proof}
Before proceeding to main body of the proof, we observe that for the choice of $N$, we know $\tr [ v_i v_i^{\top} ] = \|v_i\|_2^2 = \frac{1}{N}$ and $\sum_{i=1}^{m} v_i v_i^{\top} = I$, thus we have $m = dN$.

We define the following sequence of numbers
\begin{align*}
    a_i = \frac{1}{\sqrt{N}} + (1 + \frac{1}{\sqrt{N} - 1})\frac{i}{m}, \forall i\in \{0,1,\cdots,n\}.
\end{align*}
Let $S_j$ to be the set we have at round $t$, we also define the matrix $T_j$ as 
\begin{align*}
    T_j:= & ~ \frac{1}{\beta}\cdot \sum_{i\in S_j} v_iv_i^\top
\end{align*}
We are going to find a set of indices $i_1, \ldots, i_n$ such that the following two things hold

\begin{itemize}
    \item $\| T_j \| <  a_j$,
    \item $\Phi^{a_0}(T_0) \geq   \ldots  \geq  \Phi^{a_n}(T_n)$,
\end{itemize}
where $\Phi^a$ is the upper barrier potential as in Def.~\ref{def:bss_barrier}.

Assume the above two conditions hold, then we will have
\begin{align*}
    \left\|\sum_{i\in S_n} v_iv_i^\top \right\| = & ~ \beta\cdot \|T_n\| \\
    < & ~ \beta \cdot a_n \\
    = & ~ \beta\cdot(\frac{1}{\sqrt N}+(1+\frac{1}{\sqrt N-1})\frac{n}{m}) \\
    \leq & ~ \beta\cdot(\frac{n}{m}+O(\frac{1}{\sqrt N})).
\end{align*}

Therefore, it suffices to show how to construct $S_j$ that satisfies above two conditions. We will prove via induction. 
\paragraph{Base case.} For base case, consider $j=0$, note $a_0=\frac{1}{\sqrt N}>0$ and $T_0=0$, so $\|T_0\|<a_0$. For potential, we compute $\Phi^{a_0}(T_0)$:
\begin{align*}
    \Phi^{a_0}(T_0)=\tr[(\frac{1}{\sqrt N}I)^{-1}]=d\sqrt N.
\end{align*}
\paragraph{Inductive hypothesis.} For inductive hypothesis, we suppose for some $j<n$, we have $\|T_j\|<a_j$ and $\Phi^{a_0}(T_0)\geq \ldots \geq \Phi^{a_j}(T_j)$.
\paragraph{Inductive step.} We prove for $j+1$. Suppose $v_1, \ldots v_j$ have been chose and we use $\lambda_1 \leq \cdots \leq \lambda_d$ be the eigenvalue of $T_j$. Then the eigenvalues of $I - T_j$ are $1 - \lambda_1 \geq \cdots \geq 1- \lambda_d$ and the eigenvalues of $(a_{j+1}I - T_j)^{-1}$ are $\frac{1}{a_{j + 1} - \lambda_1} \leq \cdots \leq \frac{1}{a_{j+1}- \lambda_d}$. Note that $T_j$ is a complex symmetric matrix, we can express it using its eigen-decomposition: $T_j = Q_j^{-1} D_j Q_j$, where $D_j\in \C^{d\times d}$ is a diagonal matrix, whose $i$-th entry is $\lambda_i$.

Then we have
\begin{align}\label{eq:try4_2}
    \tr[ (a_{j + 1}I - T_j)^{-1}(I - \beta T_j) ] 
    = & ~ \tr [ Q_{j}^{-1}(a_{j+1}I - D_j)^{-1}( I - \beta D_j)Q_j ] \notag\\
    = & ~ \tr[ (a_{j+1}I - D_j)^{-1}( I - \beta D_j) ] \notag\\
    = & ~ \sum_{l=1}^{d}\frac{1}{a_{j+1} - \lambda_l}(1 - \beta \lambda_l) \notag\\
    \leq & ~ \frac{1}{d}\sum_{l=1}^{d}\frac{1}{a_{j+1} - \lambda_l}\sum_{l=1}^{d}(1 - \beta \lambda_l)\notag\\
    = & ~ \frac{1}{d}\cdot \tr[ (a_{j+1}I - T_j)^{-1} ] \cdot \tr[  I - \beta T_j ] \notag\\
    \leq & ~ \frac{1}{d}\cdot \tr[ (a_{j}I - T_j)^{-1} ] \cdot \tr[  I - \beta T_j ] \notag\\
    = & ~ \frac{1}{d}\Phi^{a_{j}}(T_j) \cdot \tr[ I - \beta T_j ] \notag\\
    \leq & ~ \frac{1}{d} \Phi^{a_0}(T_0) \cdot \tr[  I - \beta T_j ] \notag\\
    = & ~  \sqrt{N} \cdot \tr[  I - \beta T_j ],
\end{align}
where the fourth step follows from sorting inequality~\ref{fact:sorting}, the sixth step follows from $a_{j+1}>a_j$, the eighth step follows from the inductive hypothesis.

Consequently, we have
\begin{align}\label{eq:try1_2}
    \Phi^{a_j}(T_j) - \Phi^{a_{j+1}}(T_j) 
    = & ~ \tr[ (a_j I - T_j)^{-1} - (a_{j+1}I - T_j)^{-1} ]\notag \\
    = & ~ \tr[ Q_j^{-1}(a_j I - D_j)^{-1}Q_j - Q_{j}^{-1} (a_{j+1}I - D_j)^{-1}Q_j ]\notag \\
    = & ~ \tr[ (a_j I - D_j)^{-1} - (a_{j+1}I - D_j)^{-1}]\notag \\
    = & ~ (a_{j+1} - a_j)\tr[ (a_j I - D_j)^{-1}(a_{j+1}I - D_j)^{-1}]\notag \\
    = & ~ (1 + \frac{1}{\sqrt{N}- 1})\frac{1}{m} \cdot \tr[ (a_j I - T_j)^{-1}(a_{j+1}I - T_j)^{-1}] \notag\\
    \geq & ~ (1 + \frac{1}{\sqrt{N}- 1})\frac{1}{m} \cdot \tr[ (a_{j+1} I - T_j)^{-2}]
\end{align}
where the forth step follows from Fact~\ref{fac:trace_diff_diagonal}, and the fifth step follows from $a_{j+1} - a_j = \frac{1}{m}(1 + \frac{1}{\sqrt{N} - 1})$.
The last step follows from 
\[
\frac{1}{(a_{j+1} - \lambda_l)^2} \leq \frac{1}{(a_{j} - \lambda_l)(a_{j+1} - \lambda_l)}.
\]

Furthermore, we have
\begin{align}
    \tr[ (a_{j+1}I - T_j)^{-2}( I - \beta T_j) ] 
    = & ~ \sum_{l=1}^{d}\frac{1}{(a_{j+1} - \lambda_l)^2}(1 - \beta \lambda_l)\notag\\
    \leq &~ \frac{1}{d}\sum_{l=1}^{d}\frac{1}{(a_{j+1} - \lambda_l)^2}\sum_{l=1}^{d}(1 - \beta \lambda_l)\notag\\
    = & ~ \frac{1}{d} \tr[ (a_{j+1}I - T_j)^{-2} ] \cdot \tr[ I- \beta T_j ].\label{eq:try2_2}
\end{align}

Combining Eq.~\eqref{eq:try1_2} and \eqref{eq:try2_2}, we get
\begin{align}\label{eq:try3_2}
    \frac{\tr[ (a_{j+1}I - T_j)^{-2}(I - \beta T_j) ]}{ \Phi^{a_j}(T_j) - \Phi^{a_{j+1}}(T_j) }  
    \leq & ~ \frac{m}{d}\frac{\sqrt{N}-1}{\sqrt{N}} \cdot \tr[I - \beta T_j] \notag \\
    = & ~ N(1 - \frac{1}{\sqrt{N}}) \cdot \tr[ I - \beta T_j ],
\end{align}
where the last step follows from $m/d = N$.

Denote $\ov S = [m]\setminus S$, then we have
\begin{align*}
    & ~ \sum_{i\in \ov S}\left( \frac{v_i^{\top}(a_{j+1}I - T_j)^{-2}v_i}{\Phi^{a}(T_j) - \Phi^{a_{j+1}}(T_j)} + v_i^{\top}(a_{j+1}I - T_j)^{-1}v_i  \right) \\
    = & ~ \frac{\tr[ (a_{j+1}I - T_j)^{-2}(I - \beta T_j) ]}{   \Phi^{a_j}(T_j) - \Phi^{a_{j+1}}(T_j) } + \tr[ (a_{j+1}I - T_j)^{-1}(I - \beta T_j) ] \\
    \leq & ~  N(1 - \frac{1}{\sqrt{N}}) \cdot \tr[ I - \beta T_j ] + \sqrt{N} \cdot \tr [ I - \beta T_j ] \\
    = & ~ N \cdot \tr[ I - \beta T_j ] \\
    = & ~ m-j.
\end{align*}

The first step follows from $\sum_{i\in S'}v_iv_i^{\top} = I - \beta T_j \in \C^{d \times d}$, the second step follows from Eq.~\eqref{eq:try4_2} and \eqref{eq:try3_2}. 
The last step follows from $\tr[\beta T_j ] = j/N$ and $m = Nd$.
Thus we conclude there exists an element of $\ov S$ satisfying

\begin{align*} 
 \frac{v_{i^{\star}}^{\top}(a_{j+1}I - T_j)^{-2}v_{i^{\star}}}{\Phi^{a_j}(T_j) - \Phi^{a_{j+1}}(T_j)} + v_{i^{\star}}^{\top}(a_{j+1}I - T_j)v_{i^{\star}}\leq 1\leq \beta.
\end{align*}
Thus choosing $S_{j+1} = S_j \cup \{i^{\star}\} \subseteq [m]$, and using Lemma~\ref{lem:bss_upper_barrier}, we conclude $\|T_{j + 1}\| < a_{j+1}$ 
and $\Phi^{a_{j+1}}(T_{j+1}) \leq \Phi^{a_j}(T_j)$.
\end{proof}
\begin{remark}
If we choose $\beta=1$, then the above theorem reduces to the original version proved by Weaver~\cite{w13}, which corresponds to the exact algorithms. In our generalized version, we show that if we scale down each copy of $v_iv_i^\top$ by a factor of $\beta$, then the final bound is just worse by a factor of $\beta$, compared to the bound obtained by Weaver. This means that at each step of algorithm, we can tolerate for a vector with only approximately small inner products, as long as we know the approximation ratio, we can scale matrix $T$ down and pay back the factor at the final bound. This inspires the use of data structure that outputs approximate solution.

As another side note, the proof provides an algorithm that runs in $n$ iterations and picks one vector at each iteration. This means the algorithm can either have a few iterations, or a large amount of iterations. Depending on $n$, we provide different algorithms.
\end{remark}

\subsection{An \texorpdfstring{$O(n(md^2+d^\omega))$}{~} Implementation}
\label{sec:ks_alg1}

\begin{algorithm}[ht]\caption{Vanilla greedy algorithm derived from~\cite{w13}, it takes $nm\cdot \Tmat(d,d,d)$ time.}\label{alg:naive}
\begin{algorithmic}[1]

\Procedure{VanillaGreedy}{$\{v_1,\ldots,v_m\},N,n$} \Comment{Theorem~\ref{thm:alg1}}
    \State $T_0 \leftarrow \mathbf{0}_{d\times d}$
    \State $S \leftarrow \emptyset$
    \For{$j =0 \to n$}
        \State $a_j = \frac{1}{\sqrt{N}} + (1 + \frac{1}{ \sqrt{N} - 1 }) \frac{j}{m}$
    \EndFor
    \For{$j = 0 \to n$}
        \For{$i \in [m]\setminus S$}
            \State $c_i \leftarrow (\Phi^{a_{j-1}}(T_j) -\Phi^{a_j}(T_j) )^{-1} \cdot v_i^\top (a_{j} I - T_j)^{-2} v_i +   v_i^\top (a_{j} I - T_j)^{-1} v_i$ \Comment{$\Tmat(d,d,d)$ time}
        \EndFor
        \State $i^* = \arg \min_{i \in [m]\setminus S} c_i $
        \State $T_{j+1} \leftarrow T_j + v_{i^*}  v_{i^*}^\top$
        \State $S \leftarrow S \cup \{ i^* \}$
    \EndFor
    \State \Return $S$
\EndProcedure
\end{algorithmic}
\end{algorithm}
Note that Algorithm~\ref{alg:naive} is a straightforward implementation of the process derived from the proof of Lemma~\ref{thm:ks_correct}.

\begin{theorem}\label{thm:alg1}
Let $N\in \mathbb{N}_+$, if $\{v_1,\ldots,v_m \}$ is a finite sequence of vectors in $\R^d$ satisfying $\|v_i\|_2=\frac{1}{\sqrt N},\forall i\in [m]$ and 
$
    \sum_{i=1}^m v_iv_i^\top = I.
$
Then for any $n<m$, there exists a deterministic algorithm that takes time $O(n(md^2+d^\omega))$ to find a set $S$ with cardinality $n$ such that
\begin{align*}
    \left\|\sum_{i\in S} v_iv_i^\top \right\|\leq & ~ \frac{n}{m}+O(\frac{1}{\sqrt N}),
\end{align*}
\end{theorem}
\begin{proof}
The correctness proof is straightforward, since Algorithm~\ref{alg:naive} implements the greedy process exactly. To analyze the runtime, note the expensive step is to compute quantity $c_i$ at each iteration, where it involves inverting a $d\times d$ matrix, which takes $O(d^\omega)$ time, and compute the quantity in the form of $v^\top A^{-1}v$, which takes $O(d^2)$ time. Note that at each round, we need to compute $c_i$ for at most $m$ vectors, and there are $n$ rounds. Thus, the total running time is
\begin{align*}
    O(n(md^2+d^\omega)).
\end{align*}
\end{proof}

\subsection{Small Iterations via $\aipe$ Data Structure}
\label{sec:ks_small_iter}
We note that the number of iterations in Algorithm~\ref{alg:naive} is determined by the number of vectors in the set $S$, hence, we provide different algorithms for different choices of $n$. In this section, we specifically consider the setting where $n\ll m$. In this case, we use the $\aipe$ data structure with fast preprocessing time but need to linear scan over all vectors at each iteration. This is fine in our setting, since $n$ is small.

\begin{algorithm}[h]\caption{$\aipe$-based Implementation  
}\label{alg:AIPE_ks}
\begin{algorithmic}[1]

\Procedure{AIPE-based}{$d\in \mathbb{N}$, $m\in \mathbb{N}$, $n\in \mathbb{N}$, $V\subset \R^{d} $, $\epsilon\in (0,1)$,$\tau\in (0,1)$} \Comment{Theorem~\ref{thm:alg4}}
    \State $T_0 \leftarrow \mathbf{0}_{d\times d}$
    \State $S \leftarrow \emptyset$
    \State Construct $V$ \Comment{$V=[v_1, v_2, \cdots, v_m]$}
    \For{$j =0 \to n$}
        \State $a_j = \frac{1}{\sqrt{N}} + (1 + \frac{1}{ \sqrt{N} - 1 }) \frac{j}{m} $
    \EndFor
    \State \textsc{AdaptiveInnerProductEstimation} $\textsc{AIPE}$
    \State $\textsc{AIPE}$.\textsc{Init}($V$,$1+\epsilon$,$\delta$)  
    \For{$j = 0 \to n$}
        \State $M_j \leftarrow (a_j I - T_j)^{-1}$ \Comment{$M_j \in \R^{d \times d}$, it takes $d^\omega$ time}
        \State $N_j \leftarrow (a_{j-1} I - T_j)^{-1}$ \Comment{$N_j \in \R^{d \times d}$, it takes $d^\omega$ time}
        \State $q\leftarrow \vect((\tr[N_j] - \tr[M_j] )^{-1}M_j M_j+ M_j))$ 
        \State $i^*\leftarrow \textsc{AIPE}.\textsc{QueryMin}(q)$ 
        \State $T_{j+1} \leftarrow T_j + v_{i^*}  v_{i^*}^\top$
        \State $S \leftarrow S \cup \{ i^* \}$
        \State $\textsc{AIPE}.\textsc{Delete}(i^*)$
    \EndFor
    \State \Return $S$
\EndProcedure
\end{algorithmic}
\end{algorithm}

\begin{theorem}[Formal version of Theorem~\ref{thm:alg4:intro}]\label{thm:ks_small_iter}
Let $\tau,\epsilon,\delta\in (0,1)$ and $N\in \mathbb{N}_+$, if $V=\{v_1,\ldots,v_m \}$ is a finite sequence of vectors in $\R^d$ satisfying $\|v_i\|_2=\frac{1}{\sqrt N},\forall i\in [m]$ and
$
    \sum_{i=1}^m v_iv_i^\top =  I.
$
Then for any $n<m$, there exists a randomized algorithm (Algorithm~\ref{alg:AIPE_ks}) that takes time ${\cal T}$ to find a set $S$ $(|S|=n)$ such that with probability at least $1-\delta$,
\begin{align*}
    \left\|\sum_{i\in S} v_iv_i^\top \right\|\leq & ~ \frac{1}{c}\cdot (\frac{n}{m}+O(\frac{1}{\sqrt N})).
\end{align*}
Further, if $c \in (\tau,\frac{1.01\tau}{0.01+\tau})$, then the running time is ${\cal T}=\wt O(md^2+n\cdot (m+d^\omega))$.
\end{theorem}
\begin{proof}

We first recall that the $\aipe$ data structure provides a $(1+\epsilon, r)$-$\afn$ data structure by Lemma~\ref{lem:query_min_AIPE}, this means that as long as we have $(1+\epsilon)^2=\frac{c-c\tau}{c-\tau}$, then it gives the guarantee for $(c, \tau)$-$\minip$. Note that by the range of $c$, as long as $\tau=O(1)$, we have $\epsilon=O(1)$. From now on, we assume the $\aipe$ data structure produces a $(c, \tau)$-$\minip$ data structure. This implies that
\begin{align*}
    \langle v_{i^*}v_{i^*}^\top,\tau\cdot \frac{(a_{j+1}I-T)^{-2}}{\Phi^{a_j}(T)-\Phi^{a_{j+1}}(T)}+(a_{j+1}I-T)^{-1}\rangle \leq & ~ \frac{\tau}{c} \\
\Rightarrow \langle v_{i^*}v_{i^*}^\top, \frac{(a_{j+1}I-T)^{-2}}{\Phi^{a_j}(T)-\Phi^{a_{j+1}}(T)}+(a_{j+1}I-T)^{-1}\rangle \leq & ~ \frac{1}{c}.
\end{align*}
i.e., we obtain an index with $\frac{1}{c}$ approximation guarantee. As we showed in Theorem~\ref{thm:ks_correct}, if we proceed with adding $c$ copies of $v_{i^*}v_{i^*}^\top$, we will end up with the following guarantee:
\begin{align*}
    \left \| \sum_{i\in S}v_iv_i^\top \right\| \leq & ~ \frac{1}{c}\cdot (\frac{n}{m}+O(\frac{1}{\sqrt N})).
\end{align*}

Regarding the running time the algorithm, we note that it is enough to pick $\epsilon=O(1)$, therefore by Theorem~\ref{thm:aipe}, the initialization takes $\wt O(md^2)$ time. At each iteration, we pay $O(d^\omega)$ to invert matrices, the \textsc{QueryMin} procedure takes $\wt O(m+d^2)$ time and the \textsc{Delete} procedure takes $\wt O(d^2)$ time per Theorem~\ref{thm:aipe}.
\end{proof}

\subsection{Large Iterations via \texorpdfstring{$\afn$}{~} Data Structure}
\label{sec:ks_alg4}
\begin{algorithm}[h]\caption{$\afn$-based implementation. 
}\label{alg:ours}
\begin{algorithmic}[1]

\Procedure{AFN-Based}{$d\in \mathbb{N}$, $m\in \mathbb{N}$, $n\in \mathbb{N}$, $V\subset \R^{d} $, $c\in (0,1)$,$\tau\in (0,1)$} \Comment{Theorem~\ref{thm:alg4}}
    \State $T_0 \leftarrow \mathbf{0}_{d\times d}$
    \State $S \leftarrow \emptyset$
    \State Construct $V$ \Comment{$V=[v_1, v_2, \cdots, v_m]$}
    \For{$j =0 \to n$}
        \State $a_j = \frac{1}{\sqrt{N}} + (1 + \frac{1}{ \sqrt{N} - 1 }) \frac{j}{m} $
    \EndFor
    \State \textsc{MinIP} $\textsc{MI}$
    \State $\textsc{MI}$.\textsc{Init}($d^2$,$m$,$V$,$c$,$\tau$) \Comment{The dimension input to the $\minip$ has been reduced by JLT}
    \For{$j = 0 \to n$}
        \State $M_j \leftarrow (a_j I - T_j)^{-1}$ \Comment{$M_j \in \R^{d \times d}$, it takes $d^\omega$ time}
        \State $N_j \leftarrow (a_{j-1} I - T_j)^{-1}$ \Comment{$N_j \in \R^{d \times d}$, it takes $d^\omega$ time}
        \State $q\leftarrow \vect((\tr[N_j] - \tr[M_j] )^{-1}M_j M_j+ M_j))$ 
        \State $i^*\leftarrow \textsc{MI}.\textsc{QueryMin}(q)$ 
        \State $T_{j+1} \leftarrow T_j + v_{i^*}  v_{i^*}^\top$
        \State $S \leftarrow S \cup \{ i^* \}$
        \State $\textsc{MI}.\textsc{Delete}(\vect(v_{i^*}v_{i^*}^\top))$
    \EndFor
    \State \Return $S$
\EndProcedure
\end{algorithmic}
\end{algorithm}

When number of iterations $n$ becomes large, the linear scan at each round becomes expensive, e.g., if $n=O(m^{0.5})$, then the overall iteration cost becomes $\wt O(m^{1.5})$. To resolve this issue, we utilize the $\afn$-based data structure developed in Section~\ref{sec:minip_ds}, which has a slightly worse initialization time but much improved per iteration cost.

\begin{theorem}[Formal version of Theorem~\ref{thm:alg4:intro}]\label{thm:alg4}
Let $\tau,c,\delta\in (0,1)$ and $N\in \mathbb{N}_+$, if $V=\{v_1,\ldots,v_m \}$ is a finite sequence of vectors in $\R^d$ satisfying $\|v_i\|_2=\frac{1}{\sqrt N},\forall i\in [m]$ and
$
    \sum_{i=1}^m v_iv_i^\top =  I.
$
Then for any $n<m$, there exists a randomized algorithm that takes time ${\cal T}$ to find a set $S$ $(|S|=n)$ such that with probability at least $1-\delta$,
\begin{align*}
    \left\|\sum_{i\in S} v_iv_i^\top \right\|\leq & ~ \frac{2}{c}\cdot (\frac{n}{m}+O(\frac{1}{\sqrt N})).
\end{align*}
Further, we have
\begin{itemize}
    \item If $c\in (\tau,\frac{8\tau}{7+\tau})$, then ${\cal T}=\wt O((m^{1.5}+\nnz(V))d^2+n\cdot (\sqrt{m}d^2+d^\omega))$;
    \item If $c\in (\tau,\frac{400\tau}{399+\tau})$, then ${\cal T}=\wt O((m^{1.01}+\nnz(V))d^2+n\cdot (m^{0.01}d^2+d^\omega))$.
\end{itemize}
\end{theorem}
\begin{proof}
Note that since we are using the approximate $\minip$ data structure with parameter $c$ and $\tau$, we are promised to get an index $i^*$ such that
\begin{align*}
    \langle v_{i^*}v_{i^*}^\top,\tau\cdot \frac{(a_{j+1}I-T)^{-2}}{\Phi^{a_j}(T)-\Phi^{a_{j+1}}(T)}+(a_{j+1}I-T)^{-1}\rangle \leq & ~ \frac{\tau}{c} \\
\Rightarrow \langle v_{i^*}v_{i^*}^\top, \frac{(a_{j+1}I-T)^{-2}}{\Phi^{a_j}(T)-\Phi^{a_{j+1}}(T)}+(a_{j+1}I-T)^{-1}\rangle \leq & ~ \frac{1}{c}.
\end{align*}
i.e., we obtain an index with $\frac{1}{c}$ approximation guarantee. As we showed in Theorem~\ref{thm:ks_correct}, if we proceed with adding $c$ copies of $v_{i^*}v_{i^*}^\top$, we will end up with the following guarantee:
\begin{align*}
    \left \| \sum_{i\in S}v_iv_i^\top \right\| \leq & ~ \frac{1}{c}\cdot (\frac{n}{m}+O(\frac{1}{\sqrt N})).
\end{align*}
It remains to show we can have a data structure with such guarantee, we shall make use of Theorem~\ref{thm:robust_minip} combined with the transformation illustrated in~\ref{def:transform}, we complete the proof of correctness of the data structure. 

Now, we prove the correctness of the running time, which follows directly from Theorem~\ref{thm:robust_minip}. Note that it would incur an additive $\frac{\tau}{c}$ to the guarantee of inner product, which means the quality of approximation becomes $\frac{2}{c}$, with a success probability at least $1-\delta$.

This completes the proof.
\end{proof}

\section{Experimental Design via Minimum Inner Product Search}
\label{sec:zlsw}
In this section, we consider the rounding up task for experimental design problem posed in \cite{azlsw20}.
\begin{itemize}
\item In Section~\ref{sec:zlsw_def}, we introduce definitions and formally state the problem.
\item In Section~\ref{sec:zlsw_facts}, we state some useful facts and tools for later proofs.
\item In Section~\ref{sec:zlsw_alg}, we present our algorithm with $\minip$ data structure.
\item In Section~\ref{sec:zlsw_regret}, we prove an approximate regret lemma, which will provide a lower bound on the eigenvalue. 
\item In Section~\ref{sec:zlsw_swap}, we state the lemma that justifies the correctness of our algorithm. 
\item In Section~\ref{sec:zlsw_swap_1}, we prove the minimum inner product part of swapping algorithm.
\item In Section~\ref{sec:zlsw_swap_2}, we prove the maximum inner product part of swapping algorithm.
\item In Section~\ref{sec:zlsw_main}, we prove the main result of this section.
\end{itemize}
\subsection{Definitions and Problem Setup}
\label{sec:zlsw_def}
\begin{definition}\label{def:Delta_d_d}
Let $\Delta_{d\times d}$ be the class of matrices defined as 
\begin{align*}
    \Delta_{d\times d}:= & ~ \{ A\in \R^{d\times d}: A\succeq 0, \tr[A]=1\}.
\end{align*}
\end{definition}

\begin{definition}
Let $\psi:\R^{d\times d}\rightarrow \R$ be defined as
\begin{align*}
    \psi(A) =  -2\tr[A^{1/2}],
\end{align*}
where $A\in \R^{d\times d}$ is a positive semi-definite matrix.
\end{definition}

\begin{definition}
We define the Bregman divergence function associated with $\psi$, $\Delta_\psi:\R^{d\times d}\times \R^{d\times d}\rightarrow \R$ as 
\begin{align*}
\Delta_\psi(A,B)=\psi(B)-\psi(A)-\langle \nabla \psi(A),B-A\rangle.
\end{align*}
\end{definition}

\begin{definition}\label{def:mirror_descent_matrices}
We define the mirror descent matrices $\tilde A_t\in \R^{d\times d}$ and $A_t\in \R^{d\times d}$ as follows:
\begin{align*}
    \tilde A_t:= & ~ \arg\min_{A\succeq 0}\{\Delta_\psi (A_{t-1},A)+\alpha \langle F_{t-1},A\rangle\}, \\
     A_t:= & ~ \arg\min_{A\in \Delta_{d\times d}} \Delta_\psi(\tilde A_t,A).
\end{align*}
\end{definition}

\begin{definition}\label{def:A_t}
We define a sequence of matrices $A_0,A_1,\ldots\in \R^{d\times d}$ as follows:
\begin{align*}
    A_0 := & ~ (c_0I+\alpha Z_0)^{-2},
\end{align*}
where $c_0\in \R, Z_0\in \R^{d\times d}$ is symmetric and $A_0\succ 0$. We also define $A_t$ as
\begin{align*}
    A_t := & ~ (c_tI+\alpha Z_0+\alpha \sum_{l=0}^{t-1} F_l)^{-2},
\end{align*}
where $c_t\in \R$ is the unique constant such that $A_t\succ 0$ and $\tr[A_t]=1.$
\end{definition}
Note we give two alternative definitions of matrix $A_t$, as shown in Claim~\ref{clm:mirror_descent}, these two definitions are equivalent.

Finally, we formally define the rounding up problem for experimental design.
\begin{question}
Let $\pi\in [0,1]^m$ with $\|\pi\|_1\leq n$ and $\sum_{i=1}^m \pi_i x_ix_i^\top=I_d$. Let $\gamma\geq 3$ and $\epsilon\in (0,\frac{1}{\gamma}]$. Does there exist a subset $S\subset [m]$ with $|S|\leq n$ such that
\begin{align*}
    \lambda_{\min} \big( \sum_{i\in S} x_ix_i^\top \big) \geq & ~ 1-{\color{black}\gamma}\cdot \epsilon?
\end{align*}
\end{question}

\subsection{Useful Facts from Previous Work}
\label{sec:zlsw_facts}
In this section, we list the facts and tools that will be useful for our proof. For the complete proofs of these facts, we refer readers to~\cite{azlsw20}.

\begin{claim}[Lemma 2.7 in \cite{azlsw20}]\label{clm:init_bound}
Let $\Delta_{d \times d}$ be defined as Definition~\ref{def:Delta_d_d}. 
Suppose $A_0=(c_0I+\alpha Z_0)^{-2} \in \R^{d \times d}$, where $c_oI+\alpha Z_0\in \R^{d\times d}$ is positive definite, then for any $U\in \Delta_{d\times d}$,
\begin{align*}
    \Delta_{\psi}(A_0,U)\leq & ~ 2\sqrt d+\alpha \langle Z_0,U\rangle.
\end{align*}
\end{claim}

\begin{claim}[Claim 2.9 in \cite{azlsw20}]\label{clm:mirror_descent}
Let $\tilde A_t, A_t\in \R^{d\times d}$ be the matrices defined in Def.~\ref{def:mirror_descent_matrices}, if 
\begin{align*}
\alpha v_t^\top A_t^{1/2}v_t<1,
\end{align*}
then we have
\begin{align*}
\tilde A_t=(A_{t-1}^{-1/2}+\alpha F_{t-1})^{-2}.
\end{align*}
\end{claim}

\begin{claim}[Claim 2.10 in~\cite{azlsw20}]\label{clm:2_10}
Let $\Delta_{d \times d}$ be defined as in Definition~\ref{def:Delta_d_d}. 
Suppose $P_t^\top A_t^{1/2}P_t=[b~~d;d~~c]\in \R^{2\times 2}$, $J=\mathrm{diag}(1,-1)$, and $2\alpha v_t^\top A_t^{1/2}v_t<1$ for $v_t\in \R^{d}$ and $A_t\in \Delta_{d\times d}$. Then 
\begin{align*}
\Big( J + P_t^\top A_t^{1/2}P_t \Big)^{-1}=  \Big(J+\begin{bmatrix}
b & d \\
d & c
\end{bmatrix} \Big)^{-1}\succeq \Big(J+\begin{bmatrix}
2b & 0 \\
0 & 2c
\end{bmatrix} \Big)^{-1}.
\end{align*}
\end{claim}

\begin{claim}[Claim 2.11 of~\cite{azlsw20}]\label{cla:claim_2.11}
Suppose $Z \succeq 0$ is a $d\times d$ PSD matrix with $\lambda_{\min}(Z) \leq 1$. Let $\alpha>0$ be a parameter and $A= (\alpha Z+ c I)^{-2}\in \R^{d\times d}$, where $c \in \R$ is the unique real number such that $A \succeq 0$ and $\tr[A] = 1$. Then 
\begin{itemize}
	\item $\alpha \langle A^{1/2} , Z \rangle \leq d + \alpha \sqrt{d}$,
	\item $\langle A , Z \rangle \leq \sqrt{d}/\alpha + \lambda_{\min}(Z)$.
\end{itemize}
\end{claim}
\clearpage

\subsection{Algorithm}
\label{sec:zlsw_alg}
\begin{algorithm}[ht]
\caption{Swapping algorithm with $\minip$ data structure}
\label{alg:zlsw17}
\begin{algorithmic}[1]
\Procedure{Swap}{$X\in \R^{m\times d},n \in \mathbb{N}_+, \pi\in [0,1]^m,\epsilon\in (0,1/\gamma]$,$c\in (0,1)$,$\tau\in (0,1)$} \Comment{Theorem~\ref{thm:zlsw_main}}
\State $\alpha\gets \sqrt d\beta /\epsilon$ and $T\gets n/(c\epsilon)$
\State $X\gets X(X^\top \mathrm{diag}(\pi)X)^{-1/2}$ \Comment{Whitening}
\State $S_0\subseteq [m]$ be an arbitrary subset of support $n$
\State $t\gets 1$
\If{$mT<\nnz(V)d^2$} \Comment{Small iterations}
\State $\textsc{DS}\gets \textsc{AdaptiveInnerProductEstimation DS}$
\Else \Comment{Large iterations}
\State $\textsc{DS}\gets \textsc{MinIP DS}$
\EndIf
\State $\textsc{DS}$.\textsc{Init}($d^2$,$m$,$X$,$c$,$\tau$)
\While{$t\leq T$ and $\lambda_{\min}(\sum_{i\in S_{t-1}}x_ix_i^\top)\leq 1-\gamma\epsilon$}
\State Let $c_t$ be the constant s.t. $(c_t I+\alpha \sum_{i\in S_{t-1}}x_ix_i^\top)^{-2}\in \Delta_{d\times d}$ \Comment{Binary search}
\State $A_t\gets (c_t I_d+\alpha \sum_{i\in S_{t-1}}x_ix_i^\top)^{-2}$
\State $q\gets \vect(\frac{A_t}{(1-\epsilon)/n}+2\alpha A^{1/2}_t)$ 
\State {\color{blue} /* Query $q$ */}
\State $i_t\leftarrow \textsc{DS}.\textsc{QueryMin}(q)$ \Comment{If \textsc{DS} is AIPE, then here it is \textsc{QueryMin}}
\State $j_t\gets \arg\max_{j\in \ov S_{t-1}} B^+(x_j)$ \Comment{Def.~\ref{def:B_func} with $\frac{1}{c}$ as $\beta$}
\State $S_t\gets S_{t-1}\cup \{j_t\}\setminus \{i_t\}$ 
\State $t\gets t+1$ \Comment{Increase the counter}
\State {\color{blue} /* Updating data structure by swapping $j_t$ and $i_t$ */}
\State $\textsc{DS}.\textsc{Delete}(x_{i_t}x_{i_t}^\top)$ 
\State $\textsc{DS}.\textsc{Insert}(x_{j_t}x_{j_t}^\top)$
\EndWhile
\State \Return $S_{t-1}$ 
\EndProcedure
\end{algorithmic}
\end{algorithm}
\clearpage

\subsection{Approximate Regret Lemma}
\label{sec:zlsw_regret}
In this section, we prove the approximate regret lemma. The key consequence of this lemma is to provide a lower bound of the eigenvalue $\lambda_{\min}(\sum_{i\in S}x_ix_i^\top)$.
\begin{lemma}[Approximate regret lemma]\label{lem:regret}
Let $\beta\geq 1$. Suppose $F_t=u_tu_t^\top-v_tv_t^\top$ for vectors $u_t,v_t\in \R^d$ and $A_0,\ldots,A_{T-1}\in \Delta_{d\times d}$ are defined in Def.~\ref{def:A_t} some constant $\alpha>0$. Then, if $\alpha v_t^\top A_t^{1/2}v_t<{\color{black}\beta}/2 $ for all $t$, we have for any $U\in \Delta_{d\times d}$,
\begin{align*}
    -\sum_{t=0}^{T-1} \langle F_t,U\rangle\leq \sum_{t=0}^{T-1}(-\frac{\beta u_t^\top A_tu_t}{{\color{black}\beta}+2\alpha u_t^\top A_t^{1/2}u_t}+\frac{\beta v_t^\top A_tv_t}{{\color{black}\beta}-2\alpha v_t^\top A_t^{1/2}v_t})+\frac{\beta \Delta_\psi(A_0,U)}{\alpha}.
\end{align*}
\end{lemma}
\begin{proof}
Throughout the proof, we let $\ov \alpha:=\frac{\alpha}{\beta}$, note that $\ov \alpha$ has the property that $\ov \alpha v_t^\top A_t^{1/2}v_t<1/2$, this enables us to use both Claim~\ref{clm:mirror_descent} and~\ref{clm:2_10}.
The proof relies on the mirror descent matrices $\tilde A_t$ and $A_t$ we defined Def.~\ref{def:mirror_descent_matrices}, we need to modify the definition of $\tilde A_t$ with $\ov \alpha$ instead of $\alpha$. Per Claim~\ref{clm:mirror_descent}, we know that $\tilde A_t=(A_{t-1}^{-1/2}+\ov \alpha F_{t-1})^{-2}$, and because of their definitions, we know that $\nabla \psi(\tilde A_t)-\nabla \psi(A_{t-1})+\ov \alpha F_{t-1}=0$ where the gradient is evaluated at $\tilde A_t$. This means that
\begin{align}\label{eq:clm_29_eq1}
    \langle \alpha F_{t-1},A_{t-1}-U\rangle = & ~ \langle \nabla \psi(A_{t-1})-\nabla \psi(\tilde A_t),A_{t-1}-U\rangle  \notag \\
    = & ~ \Delta_{\psi}(A_{t-1},U)-\Delta_\psi(\tilde A_t,U)+\Delta_\psi(\tilde A_t,A_{t-1}) \notag \\
    \leq & ~ \Delta_\psi(\tilde A_{t-1},U)-\Delta_\psi(\tilde A_t,U)+\Delta_\psi(\tilde A_t,A_{t-1}).
\end{align}

Above, the second inequality and the last inequality follow from standard inequalities and generalized Pythagorean Theorem of Bregman divergence. Now, consider the quantity $\Delta_\psi(\tilde A_t,A_{t-1})$:
\begin{align}\label{eq:clm_29_eq2}
\Delta_\psi(\tilde A_t,A_{t-1}) = & ~ \psi(A_{t-1})-\psi(\tilde A_t)-\langle \nabla \psi(\tilde A_t),A_{t-1}-\tilde A_t\rangle \notag \\
= & ~ -2\tr[A_{t-1}^{-1/2}]+2\tr[\tilde A_t^{1/2}]+\langle \tilde A_t^{-1/2},A_{t-1}-\tilde A_t\rangle\notag  \\
= & ~ \langle \tilde A_{t}^{-1/2},A_{t-1}\rangle+\tr[\tilde A_t^{1/2}]-2\tr[A_{t-1}^{1/2}]\notag \\
= & ~ \langle A^{-1/2}_{t-1}+\ov \alpha F_{t-1},A_{t-1}\rangle+\tr[\tilde A_t^{1/2}]-2\tr[A_{t-1}^{1/2}]\notag \\
= & ~ \ov \alpha \langle F_{t-1},A_{t-1}\rangle+\tr[\tilde A_{t}^{1/2}]-\tr[A_{t-1}^{1/2}].
\end{align}
Combining Eqs.~\eqref{eq:clm_29_eq1} and~\eqref{eq:clm_29_eq2} and telescoping $t$ from 1 to $T$ yields
\begin{align}\label{eq:clm_29_eq3}
-\ov \alpha \sum_{t=0}^{T-1} \langle F_t,U\rangle\leq & ~ \Delta_\psi(A_0,U)-\Delta_\psi(\tilde A_T,U)+\sum_{t=0}^{T-1}\tr[\tilde A_{t+1}^{1/2}]-\tr[A_t^{1/2}]\notag \\
\leq & ~ \Delta_\psi(A_0,U)+\sum_{t=0}^{T-1}\tr[\tilde A_{t+1}^{1/2}]-\tr[A_t^{1/2}],
\end{align}
where the second inequality follows from the non-negativity of Bregman divergence.

It remains to upper bound $\tr[\tilde A_{t+1}^{1/2}]-\tr[A_t^{1/2}]$.

Set $P_t$ as $\sqrt{\ov \alpha} [u_t~~v_t]\in \R^{d\times 2}$ and $J=\mathrm{diag}(1,-1)\in \R^{2\times 2}$, we have $\ov \alpha F_t=P_tJP_t^\top$. By the definition of $\tilde A_{t+1}^{1/2}$ and the matrix Woodbury formula (Fact.~\ref{fac:woodbury}), we have
\begin{align}\label{eq:clm_29_eq4}
\tr[\tilde A_{t+1}^{1/2}]=\tr[(A_t^{-1/2}+P_tJP_t^\top)^{-1}]=\tr[A_t^{1/2}-A_t^{1/2}P_t(J+P_t^\top A_t^{1/2}P_t)^{-1}P_t^\top A_t^{1/2}].
\end{align}
By linearity of trace operator, it suffices to give a spectral lower bound on the $2\times 2$ matrix $(J+P_t^\top A_t^{1/2}P_t)^{-1/2}$. We will use Claim~\ref{clm:2_10} as a lower bound:
\begin{align}\label{eq:clm_29_eq5}
\tr[\tilde A_{t+1}^{1/2}]-\tr[A_t^{1/2}] = & ~ -\tr[-A_t^{1/2}P_t(J+P_t^\top A_t^{1/2}P_t)^{-1}P_t^\top A_t^{1/2}] \notag\\
\leq & ~ -\tr[-A_t^{1/2}P_t(J+
\mathrm{diag}( 2\ov \alpha u_t^\top A_t^{1/2}u_t ,2\ov \alpha v_t^\top A_t^{1/2}v_t )
)^{-1}P_t^\top A_t^{1/2}] \notag\\
= & ~ -\frac{\ov \alpha u_t^\top A_tu_t}{1+2\ov \alpha u_t^\top A_t^{1/2}u_t}+\frac{\ov\alpha v_t^\top A_tv_t}{1-2\ov\alpha v_t^\top A_t^{1/2}v_t}.
\end{align}
Plugging Eq.~\eqref{eq:clm_29_eq5} into Eq.~\eqref{eq:clm_29_eq3}, we arrive at the desired result:
\begin{align*}
    -\sum_{t=0}^{T-1}\langle F_t,U\rangle \leq & ~ \sum_{t=0}^{T-1} (-\frac{\beta u_t^\top A_tu^t}{\beta+2\alpha u_t^\top A_t^{1/2}u_t}+\frac{\beta v_t^\top A_tv_t}{\beta-2\alpha v_t^\top A_t^{1/2}v_t})+\frac{\beta}{\alpha}\Delta_{\psi}(A_0,U).
\end{align*}
\end{proof}

\subsection{Approximate Swapping Lemma}
\label{sec:zlsw_swap}
The goal of this section is to present and prove Lemma~\ref{lem:lemma_2.8}. We start with a helpful definition.
\begin{definition}[$B$ functions]\label{def:B_func}
Let $\alpha, \beta$ denote two fixed parameters. Let $A$ denote a fixed matrix. 
We define function $B^+: \R^d \rightarrow \R$ and $B^- : \R^d \rightarrow \R$ as follows:
\begin{align*}
    B^+(x) = & ~ \frac{ \langle A , x x^\top \rangle }{ \beta + 2 \alpha \langle A^{1/2} , x x^\top \rangle }, \\
    B^-(x) = & ~ \frac{ \langle A , x x^\top \rangle }{ \beta - 2 \alpha \langle A^{1/2} , x x^\top \rangle }. \\
\end{align*}
\end{definition}

\begin{lemma}\label{lem:lemma_2.8}
Let $\beta\in [1,\gamma-1)$ and $\epsilon\in (0,1/{\color{black}\gamma}]$. For every subset $S \subset [m]$ of cardinality $n$ (let $\ov{S}$ denote $[m]\setminus S$), suppose $\lambda_{\min}(\sum_{i \in S} x_i x_i^\top) \leq 1-{\color{black}\gamma}\epsilon$ and $A= (c I + \alpha \sum_{i \in S} x_i x_i^\top )^{-2}$, where $c \in \R$ is the unique number such that $A \succeq 0$ and $\tr[A] = 1$. For any $\alpha = \sqrt{d} {\color{black}\beta}/ \epsilon$ and $n \geq \frac{ {\color{black}6} }{  {\color{black}\gamma}-1-\beta }  d / \epsilon^2$, we have 
\begin{itemize}
    \item Part 1. There exists $i\in S$ such that $2\alpha x_i^\top A x_i<\beta$ and
\item Part 2. There exists $j\in \ov S$ such that
\end{itemize}
\end{lemma}
\begin{proof}
In this proof, we will extensively use Claim~\ref{cla:claim_2.11}, therefore, we pre-compute the value $d+\alpha\sqrt d$ and $\sqrt d/\alpha$ here for references. By the choice of our $\alpha$, we have
\begin{align}\label{eq:az_bound}
    d+\alpha\sqrt d = (1+\frac{\beta}{\epsilon})d & ~, \sqrt d/\alpha =  \frac{\epsilon}{\beta}.
\end{align}
We also define the quantity  $\nu:=\min_{i\in S, 2\alpha x_i^\top Ax_i<\beta}~B^-(x_i)$ which will be used throughout our proof.

The proof directly follows from combining Claim~\ref{cla:approximate_swapping_lemma_part1} and Claim~\ref{cla:approximate_swapping_lemma_part2}.
\end{proof}

\subsection{Approximate Swapping Lemma, Part 1}
\label{sec:zlsw_swap_1}
In this section, we will prove that as long as we enter the main while loop of the algorithm, we can always find an index $i\in S$ such that $B^-(x_i)$ is small.
\begin{claim}[Part 1 of Lemma~\ref{lem:lemma_2.8}]\label{cla:approximate_swapping_lemma_part1}
There exists $i\in S$ such that $2\alpha x_i^\top A x_i<\beta$ and
$B^-(x_i) \leq  \frac{1-\epsilon}{\beta n}$.
\end{claim}

\begin{proof}
To demonstrate the existence of such an $i$, it suffices to show that $\min_{i\in S,2\alpha x_i^\top Ax_i<\beta} B^-(x_i)\leq \frac{1-\epsilon}{\beta n}$, we use $\nu$ to denote this minimum value. Note that $\nu>0$, due to the fact $2\alpha x_i^\top Ax_i<\beta$ and $A$ is positive definite. To start off, we first show that there always exists an $i$ such that $2\alpha \langle A^{1/2}, x_i x_i^{\top} \rangle < 1$. Define $Z = \sum_{i \in S} x_i x_i^\top$, and by definition $A = (c I + \alpha \sum_{ i \in S } x_i x_i^\top )^{-2} = ( \alpha Z + c I )^{-2}$. Assume for the sake of contradiction that such $i$ does not exists. We have
\begin{align}\label{eq:2.16}
\sum_{i \in S} 2 \alpha \langle A^{1/2} , x_i x_i^\top \rangle = 2 \alpha \langle A^{1/2}, Z \rangle \geq |S| = n.
\end{align}
On the other hand, because $Z \succeq 0 $ and $\lambda_{\min}(Z) < 1$, invoking Claim~\ref{cla:claim_2.11} we get
\begin{align*}
2 \alpha \langle A^{1/2} , Z \rangle \leq 2 d + 2 \alpha \sqrt{d},
\end{align*}
which contradicts Eq.~\eqref{eq:2.16} given the choice of $\alpha$ and $n > 4d/\epsilon$. Thus, there must exist $i \in S$ such that $2\alpha \langle A^{1/2}, x_i x_i^\top \rangle < 1$. Since we set $\beta\geq 1$, this means we can always find an index $i$ such that $2\alpha \langle A^{1/2},x_ix_i^\top\rangle<\beta$ holds. By the same token, we also have $\sum_{i\in S}(\beta-2\alpha \langle A^{1/2},x_ix_i^\top\rangle)\geq 0$. We claim that
\begin{align*}
( {\color{black}\beta} -2\alpha \langle A^{1/2}, x_i x_i^\top \rangle ) \nu \leq \langle A, x_i x_i^\top \rangle, \text{~for~all~} i \in S, 
\end{align*}
because if $2\alpha \langle A^{1/2}, x_i x_i^\top \rangle \geq {\color{black}\beta}$ the LHS is non-positive while the RHS is always non-negative due to the positive semi-definiteness of $A$. Subsequently,
\begin{align*}
\nu \leq & ~ \frac{ \sum_{i \in S} \langle A , x_i x_i^\top \rangle }{ \sum_{i\in S} ({\color{black}\beta}-2\alpha \langle A^{1/2}, x_i x_i^\top \rangle ) } \\
\leq & ~ \frac{ \sqrt{d}/\alpha + \lambda_{\min} ( \sum_{i\in S} x_i x_i^\top ) }{ {\color{black}\beta}n - 2d - 2 \alpha \sqrt{d} } \\
\leq & ~ \frac{ \epsilon/{\color{black}\beta} + 1 - {\color{black}\gamma}\epsilon }{ {\color{black}\beta} n (1- \beta\epsilon/3) } \\
\leq & ~ \frac{ 1 - \epsilon }{ {\color{black}\beta}n }
\end{align*}
where the first step holds because the denominator is strictly positive as we have shown; the second step is due to Claim~\ref{cla:claim_2.11}; the third step has used our choices $\alpha$ and $n$ and our assumption $\lambda_{\min} ( \sum_{i \in S} x_i x_i^\top ) \leq 1-{\color{black}\gamma} \epsilon$; and the forth step has used $1-\beta\epsilon/3<1$. We have thus proved that $\nu \leq (1-\epsilon) / ( {\color{black}\beta}n )$. This proves the existence of the $i$ we want.
\end{proof}

\subsection{Approximate Swapping Lemma, Part 2}
\label{sec:zlsw_swap_2}
In this section, we prove the other key gradient for the swapping to proceed, i.e., there exists an $j\in \ov S$ such that $B^+(x_j)$ is large.
\begin{claim}[Part 2 of Lemma~\ref{lem:lemma_2.8}]\label{cla:approximate_swapping_lemma_part2}
 There exists $j\in \ov S$ such that
$B^+(x_j) \geq  \frac{1}{\beta n}$.
\end{claim}
\begin{proof}
 Define $t =  1/( {\color{black}\beta}n )$. To prove Part 2 it suffices to show that
\begin{align}\label{eq:2.17}
\sum_{j \in \ov{S}} \pi_j \langle A , x_j x_j^\top \rangle \geq t \cdot \sum_{j \in \ov{S}} \pi_j ( {\color{black}\beta} + 2\alpha  \langle A^{1/2} , x_j x_j^\top \rangle ),
\end{align} 
because $\pi_j \geq 0$ for all $j\in[m]$. Recall that $\sum_{j=1}^m \pi_j = n$, $\sum_{j=1}^m \pi_j x_j x_j^\top = I_d$. We then have
\begin{align*}
\sum_{ j \in \ov{S}} \pi_j ( {\color{black}\beta} + 2\alpha  \langle A^{1/2}, x_j x_j^\top \rangle ) 
\leq & ~ {\color{black}\beta}(n- \sum_{j \in S} \pi_j) + 2 \alpha \cdot \sum_{j \in \ov{S}} \pi_j \langle A^{1/2} , x_j x_j^\top \rangle \\
\leq & ~ {\color{black}\beta}( n - \sum_{j \in S} \pi_j ) + 2\alpha  \cdot \sum_{j=1}^m \pi_j \langle A^{1/2}, x_j x_j^\top \rangle \\
= & ~ {\color{black}\beta}n - {\color{black}\beta}\sum_{j \in S} \pi_j + 2 \alpha \langle I , A^{1/2} \rangle \\
= & ~ {\color{black}\beta}n - {\color{black}\beta}\sum_{j \in S} \pi_j + 2 \alpha  \cdot \tr[ A^{1/2} ].
\end{align*}
Similarly,
\begin{align*}
  \sum_{j \in \ov{S}} \pi_j \langle A, x_j x_j^\top \rangle 
= & ~ \langle I - \sum_{j \in S} \pi_j x_j x_j^\top , A\rangle \\
= & ~ \tr[ A ] - \sum_{j \in S} \pi_j \langle A , x_j x_j^\top \rangle
\end{align*}
Subsequently,
\begin{align}\label{eq:2.19}
	& ~ \sum_{ j \in \ov{S} } \pi_j \langle A , x_j x_j^\top \rangle - t \cdot \sum_{j \in \ov{S}} \pi_j( {\color{black}\beta} +2\alpha \langle A^{1/2} , x_j x_j^\top \rangle ) \notag \\
	\geq & ~ \tr[A] - \sum_{j \in S} \pi_j \langle A , x_j x_j^\top \rangle - t \cdot {\color{black}\beta}\cdot (n - \sum_{j \in S} \pi_j) - 2\alpha t \cdot \tr[A^{1/2}]  \notag \\
	\geq & ~ 1 - \sum_{j \in S} \pi_j \langle A , x_j x_j^\top \rangle - t \cdot {\color{black}\beta} \cdot(n - \sum_{j \in S } \pi_j ) - 2\alpha t \sqrt{d} \notag \\
	= & ~ 1- t{\color{black}\beta}n - 2t \alpha \sqrt{d} - \sum_{j \in S} \pi_j ( \langle A , x_j x_j^\top \rangle - t{\color{black}\beta} ) \notag \\ 
	\geq & ~ 1 - t{\color{black}\beta}n - 2t \alpha \sqrt{d} - \sum_{ j \in S} \max \{ \langle A , x_j x_j^\top \rangle - t{\color{black}\beta}, 0  \} \notag \\
	= & ~ 1 - t{\color{black}\beta}n -2t\alpha \sqrt{d} - \sum_{j \in S} ( \langle A, x_j x_j^\top \rangle - t{\color{black}\beta} ) - \sum_{j \in S} \max \{ ( t{\color{black}\beta} - \langle A , x_j x_j^\top \rangle ) , 0 \} \notag \\
	\geq & ~ 1 - 2t \alpha \sqrt{d}-  \sqrt{d}/\alpha - \lambda_{\min} ( \sum_{j\in S} x_j x_j^\top ) - \sum_{ j \in S } \max\{ (t{\color{black}\beta}-\langle A , x_j x_j^\top \rangle ) ,0 \} \notag \\
	\geq & ~ ({\color{black}\gamma}-{\color{black}\beta}) \epsilon - \frac{2d}{\epsilon n} - \sum_{j \in S}\max \{ t{\color{black}\beta} -\langle A, x_j x_j^\top \rangle ,0 \}
\end{align}
where the second step follows from  Fact~\ref{fac:trace_sqrt} and $\tr[A]=1$.
The forth step follows from $\pi_j \leq 1 $ for all $j$, the second-to-last step follows from we apply $\sum_{j \in S} \langle A  , x_j x_j^\top \rangle \leq \sqrt{d}/\alpha + \lambda_{\min} ( \sum_{j \in S} x_j x_j^\top )$ which comes from Claim~\ref{cla:claim_2.11}. The fifth step comes from the fact that $\max\{x,0\}-\max\{-x,0\}=x$. Finally, the last step comes from the choices of $\alpha,t$ and $\lambda_{\min} (\sum_{j \in S} x_j x_j^\top) \leq 1-{\color{black}\gamma}\epsilon$.

Furthermore, because $({\color{black}\beta}-2\alpha \langle A^{1/2} , x_i x_i^\top \rangle ) \nu \leq \langle A , x_i x_i^\top \rangle $ for all $i \in S$, using Claim~\ref{cla:claim_2.11} we have
\begin{align*}
	\sum_{ i \in S' } ( {\color{black}\beta}\nu - \langle A , x_i x_i^\top \rangle ) \leq \sum_{ i \in S' } 2 \nu \alpha \langle A^{1/2} , x_i x_i^\top \rangle \leq 2 \nu (d+ \alpha\sqrt{d}),
\end{align*}
for all $S' \subseteq S$.

Consider $S' = \{ i \in S : {\color{black}\beta}t - \langle A , x_i x_i^\top \rangle \geq 0 \}$. We then have
\begin{align}\label{eq:2.20}
\sum_{j \in S'} \max \{ {\color{black}\beta}t - \langle A, x_j x_j^\top \rangle , 0 \} 
= & ~ \sum_{j \in S'} ({\color{black}\beta}t-\langle A , x_j x_j^\top \rangle ) \notag \\
= & ~ {\color{black}\beta}(t-\nu) |S'|+  \sum_{j \in S'} ( {\color{black}\beta}\nu - \langle A,x_j x_j^\top \rangle ) \notag \\
\leq & ~ {\color{black}\beta}(t-\nu) n + 2 \nu (d+\alpha \sqrt{d}) \notag \\
\leq & ~ \epsilon + \frac{4d/\epsilon}{n}
\end{align} 
where the last two inequalities hold because $t-\nu = \epsilon/({\color{black}\beta}n) \geq 0$, $|S'| \leq |S| =n$, $\nu \leq 1/({\color{black}\beta}n)$ and the choice of $\alpha$. 

Combining Eqs.\eqref{eq:2.19} and \eqref{eq:2.20} we arrive at
\begin{align*}
\sum_{j \in \ov{S}} \pi_j \{ \langle A , x_j x_j^\top \rangle - t ({\color{black}\beta}+ 2\alpha \langle A^{1/2}, x_j x_j^\top \rangle ) \} \geq ({\color{black}\gamma}-1-{\color{black}\beta})\epsilon - \frac{6d}{\epsilon n}.
\end{align*}
By choice of $n$,
the RHS of the above inequality is non-negative, which finishes the proof of Eq.~\eqref{eq:2.17} and thus also the proof of Part 2.
\end{proof}

\subsection{Main Result}
In this section, we present the correctness and runtime analysis of Algorithm~\ref{alg:zlsw17}. The correctness follows from the approximate regret and swap lemma, while the runtime comes from the approximate $\minip$ data structure.
\label{sec:zlsw_main}
\begin{theorem}[Formal version of Theorem~\ref{thm:intro_zlsw_main}]\label{thm:zlsw_main}
Let $\pi\in [0,1]^m$ with $\|\pi\|_1\leq n$ and $\sum_{i=1}^m \pi_i x_ix_i^\top=I_d$. Let $\gamma\geq 3$ and $\epsilon\in (0,\frac{1}{\gamma}]$. Then, there exists a subset $S\subset [m]$ with $|S|\leq n$ such that
\begin{align*}
    \lambda_{\min} (\sum_{i\in S} x_ix_i^\top)\geq & ~ 1-{\color{black}\gamma}\cdot \epsilon.
\end{align*}
Let $\tau,\delta\in (0,1)$ and $c\in (\frac{1}{\gamma-1},1)$. If $n\geq \frac{6d/\epsilon^2}{\gamma-1-2/c}$ and $\alpha=\sqrt d/(c\epsilon)$, then there exists a randomized algorithm with success probability at least $1-\delta$ and running time ${\cal T}=\min\{{\cal T}_{{\rm SmallIter}}, {\cal T}_{{\rm LargeIter}} \}$ where
\begin{itemize}
    \item For ${\cal T}_{{\rm SmallIter}}$, we have $c\in (\tau,\frac{1.01\tau}{0.01+\tau})$ and
    \begin{align*}
         {\cal T}_{{\rm SmallIter}} = & ~ \wt O(\Tmat(m,d,d)+nd^2+\epsilon^{-1}n\cdot (d^\omega+n+(m-n)\cdot d^2)).
    \end{align*}
   \item For ${\cal T}_{{\rm LargeIter}}$, we have $c\in (\tau,\frac{400\tau}{399+\tau})$ and
    \begin{align*}
         {\cal T}_{{\rm LargeIter}} = & ~ \wt O(\Tmat(m,d,d)+(n^{1.01}+\nnz(X)) d^2+\epsilon^{-1}n\cdot (d^\omega+(n^{0.01}+z)\cdot d^2+(m-n)\cdot d^2)),
    \end{align*}
    where $z=\max_{i\in [m]}~\nnz(x_i)$.
\end{itemize}
\end{theorem}
\begin{proof}
We will show Alg.~\ref{alg:zlsw17} satisfies the properties in the theorem statement. Similar to the proof of Theorem~\ref{thm:alg4}, we need to scale down the query point by a factor of $\tau$. This means each query will return an index $i\in S_{t-1}$ such that
\begin{align*}
    \frac{x_i^\top A_tx_i}{(1-\epsilon)/n}+2\alpha x_i^\top A_t^{1/2}x_i \leq \frac{1}{c},
\end{align*}
Set $\beta=\frac{1}{c}$, note this is equivalent to find an index $i$ satisfying $B^-(x_i)\leq \frac{1-\epsilon}{\beta n}$.

On the other hand, we can search the index $j\in \ov S_{t-1}$ such that
\begin{align*}
    B^+(x_j) \geq & ~ \frac{1}{\beta n}
\end{align*}

This means that at each iteration, we either have
\begin{align*}
\lambda_{\min} (\sum_{i\in S}x_ix_i^\top)\geq 1-\gamma\epsilon,
\end{align*}
which we are done, or we can find $i_t$ and $j_t$ such that \begin{align*}
B^-(x_{i_t})-B^+(x_{j_t})\leq & ~ -\frac{\epsilon}{\beta n}.
\end{align*}

Combining this fact with Lemma~\ref{lem:regret} and Claim~\ref{clm:init_bound}, we have
\begin{align*}
    - \langle Z_0+\sum_{t=0}^{T-1}F_t,U\rangle \leq & ~ \sum_{t=0}^{T-1} \beta (B^-(x_{i_t})-B^+(x_{j_t}))+\frac{2\beta\sqrt d}{\alpha} \\
    \leq & ~ -T\cdot \frac{\epsilon}{\beta n}+2\epsilon,
\end{align*}
Since we can choose $U$ such that 
\begin{align*}
-\langle Z_0+\sum_{t=0}^{T-1}F_t,U\rangle=-\lambda_{\min}(Z_0+\sum_{t=0}^{T-1}F_t)=-\lambda_{\min}(\sum_{i\in S_T} x_ix_i^\top),
\end{align*}
this gives a lower bound on the desired eigenvalue we want:
\begin{align*}
    \lambda_{\min}(\sum_{i\in S_T} x_ix_i^\top) \geq & ~ T\cdot \frac{\epsilon}{\beta n}-2\epsilon.
\end{align*}
Since $T=\frac{\beta n}{\epsilon}$, it is lower bounded by $1-2\epsilon>1-\gamma\epsilon$, and we have completed the proof of correctness.

For the running time, we separately consider initialization and cost per iteration. In initialization phase,
\begin{itemize}
    \item Computing $X(X^\top \mathrm{diag}(\pi)X)^{-1/2}$ takes $O(\Tmat(m,d,d))$ time;
    \item The Initialization time for data structure with $n$ random points is either $\wt O(nd^2)$ (see Theorem~\ref{thm:aipe}) or $\wt((n^{1.01}+\nnz(X))d^2)$ (see Theorem~\ref{thm:robust_minip});
\end{itemize}
For each iteration, we perform the following:
\begin{itemize}
    \item Computing eigen-decomposition of $\sum_{i\in S_{t-1}}x_ix_i^\top$ takes $O(d^\omega)$ time;
    \item Using binary search to finding $c_t$ takes $O(d^\omega\log d/(c\epsilon))$ since the searching range is $O(\alpha+\sqrt d)$ and each search takes $d^\omega$ to form the matrix and compute its trace;
    \item The time of querying data structure is either $\wt O(n+d^2)$ (see Theorem~\ref{thm:aipe}) or $\wt O(n^{0.01})$ (see Theorem~\ref{thm:robust_minip});
    \item The brute force search for $j$ takes $O((m-n)\cdot d^2)$ if we pre-compute $A_t$ and $A_t^{1/2}$;
    \item The insertion and deletion of point $x_{j_t}$ takes either  $\wt O(d^2)$ (see Theorem~\ref{thm:aipe}) or $\wt O((n^{0.01}+\nnz(x_{j_t}))d^2)$ (see Theorem~\ref{thm:robust_minip})  time.
\end{itemize}
This concludes the proof of running time.
\end{proof}

\newpage
\clearpage

\newpage
\addcontentsline{toc}{section}{References}
\bibliographystyle{alpha}
\bibliography{ref}

\newpage
\appendix
\section*{Appendix}
\section{Approximate Furthest Neighbor Search Data Structure}
\label{app:afn}

In this section, we include the algorithm and correctness analysis of the $\afn$ data structure. 

\begin{itemize}
    \item In Section~\ref{sec:minip_alg}, we give a detailed description of the $\afn$ algorithm.
    \item In Section~\ref{sec:minip_success_failure_rp}, we analyze the success and failure probability of the random projections.
    \item In Section~\ref{sec:minip_dfn}, we prove the guarantee of a $\dfn$ data structure.
    \item In Section~\ref{sec:minip_afn}, we show how to solve $\afn$ via $\dfn$ data structure.
\end{itemize}

Throughout this section, we use $n$ to denote the number of data points, and $d$ denote the dimension of the data.

\subsection{Algorithm}\label{sec:minip_alg}

The $\afn$ data structure we are going to use has similar high-level idea as that of Indyk~\cite{i03}, but we give an improved analysis on the overall running time.

In this section, we present our algorithm that solves approximate $\minip$ efficiently. We start with presenting the \textsc{SortedList} data structure in Alg.~\ref{alg:sorted_list}.

\begin{algorithm}[H]
\caption{Helper data structure \textsc{SortedList}}
\label{alg:sorted_list}
\begin{algorithmic}[1]
\State {\bf data structure} \textsc{SortedList} \Comment{This data structure can be implemented via various self-balancing binary search trees}
\State \hspace{4mm} \textsc{Init}$(P\in (\R\times \R^d)^n)$ \Comment{$n$ points each has a real key and $d$ dimensional data points, $O(n\log n)$ time} 
\State \hspace{4mm} \textsc{Insert}$(p\in \R\times \R^d)$ \Comment{Insert a single key-value pair, $O(\log n)$ time}
\State \hspace{4mm} \textsc{Delete}$(p\in \R\times \R^d)$ \Comment{Remove a single key-value pair, $O(\log n)$ time}
\State \hspace{4mm} \textsc{SearchLeq}$(T\in \R)$ \Comment{Output a subtree with key less than or equal to $T$, $O(\log n)$ time}
\State \hspace{4mm} \textsc{SearchGeq}$(T\in \R)$ \Comment{Output a subtree with key greater than or equal to $T$, $O(\log n)$ time}
\State \hspace{4mm} \textsc{Max}() \Comment{Return max key-value pair, $O(\log n)$ time}
\State \hspace{4mm} \textsc{Min}() \Comment{Return min key-value pair, $O(\log n)$ time}
\State {\bf end data structure}

\end{algorithmic}

\end{algorithm}

Next, we introduce a task called $(\ov c,r)$-$\dfn$ defined in Task~\ref{task:c_dfn}. 

\begin{task}
\label{task:c_dfn}
Let $P\subset \R^d$ be an $n$-point dataset. Let $\ov c>1$ We define the $(\ov c,r)$-$\dfn$ problem as follows: given a point $q\in \R^d$ and $r>0$, if there exists a point $p\in P$ such that, if $\|p-q\|_2\geq r$, then the data structure reports a point $\wh p\in P$ such that $\|\wh p-q\|_2\geq r/\ov c$, otherwise, it reports ``Fail''.
\end{task}

The data structure for $(\ov c,r)$-$\dfn$ shown in Alg.~\ref{alg:ind03_1} and Alg.~\ref{alg:ind03_2} is the building block of our approximate $\minip$ algorithm.  

\begin{algorithm}[H]
\caption{Data structure $\dfn$: members, init, insert and delete}
\label{alg:ind03_1}
\begin{algorithmic}[1]
\State {\bf data structure} \textsc{DFN} \Comment{Theorem~\ref{thm:dfn}}
\State
\State {\bf members}
\State \hspace{4mm} $\ell\in \mathbb{N}_+$ \Comment{Number of random directions}
\State \hspace{4mm} $G\in \R^{\ell\times d}$ \Comment{Random Gaussian vectors}
\State \hspace{4mm} $\textsc{SortedList}   L_1,\ldots,L_\ell$ \Comment{$\ell$ sorted lists, Alg.~\ref{alg:sorted_list}}
\State \hspace{4mm} $t\in \R_+$ \Comment{Threshold parameter}
\State \hspace{4mm} $\ov c\in (1,\infty)$ \Comment{Approximation parameter}
\State {\bf end members}
\State
\Procedure{Init}{$A\in \R^{n\times d},\ov c\in (1,\infty)$}
\State $\ov c\gets \ov c$
\State $\ell\gets \Theta(n^{1/\ov c^2}\log^{(1-1/\ov c)/2}n)$
\State $t \leftarrow \Theta(\sqrt{\log n})$ \Comment{$t$ is the solution to $e^{t^2/2}/t=2n$}
\State $G_{i,j}\sim {\cal N}(0,1),\forall i\in [\ell],\forall j\in [d]$ \Comment{Each entry is a standard Gaussian}
\State \textsc{SortedList} $L_1,\ldots,L_\ell$ \Comment{Alg.~\ref{alg:sorted_list}}
\State Let $g_i$ denote the $i$-th row of $G$ and $A_i$ denote the $i$-th row of $A$
\For{$i=1 \to \ell$}
\State $P_i\gets \{(\langle g_i,a_j\rangle,a_j): j\in [n] \}$
\State $L_i$.\textsc{Init}$(P_i)$ \Comment{Alg.~\ref{alg:sorted_list}}
\EndFor
\EndProcedure
\State
\Procedure{Insert}{$p\in \R^d$}
\For{$i=1 \to \ell$}
\State $k\gets \langle g_i,p\rangle$
\State $L_i.$\textsc{Insert}$((k,p))$ \Comment{Alg.~\ref{alg:sorted_list}}
\EndFor
\EndProcedure
\State
\Procedure{Delete}{$p\in \R^d$}
\For{$i=1 \to \ell$}
\State $k\gets \langle g_i,p\rangle$
\State $L_i.$\textsc{Delete}$((k,p))$ \Comment{Alg.~\ref{alg:sorted_list}}
\EndFor
\EndProcedure
\State
\State {\bf end data structure}
\end{algorithmic}
\end{algorithm}

\begin{algorithm}[H]
\caption{Data structure $\dfn$: query}
\label{alg:ind03_2}
\begin{algorithmic}[1]
\State {\bf data structure} \textsc{DFN} \Comment{Theorem~\ref{thm:dfn}}
\State
\Procedure{Query}{$q\in \R^d, r\in \R_+$}
\State $T\gets rt/\ov c$
\State $i\gets 1$, $m\gets 0$
\State $S\gets \emptyset$
\While{$i\leq \ell$ and $m\leq 2\ell+1$}
\State $\mathrm{dist}\gets \langle g_i,q\rangle$
\State $T_1\gets L_i.$\textsc{SearchLeq}$(\mathrm{dist}-T)$
\State $T_2\gets L_i.$\textsc{SearchGeq}$(T+\mathrm{dist})$
\State \Comment{Search for the subtree such that $|\langle g_i,q-p\rangle|\geq T$}
\If{$m+|T_1|+|T_2|\leq 2\ell+1$}
\State $S\gets S\cup T_1\cup T_2$
\State $m\gets m+|T_1|+|T_2|$
\Else
\State Add points from $T_1$ and $T_2$ to $S$ until $|S|=2\ell+1$
\State $m\gets 2\ell+1$
\EndIf
\State $i\gets i+1$
\EndWhile
\For{$p\in S$}
\If{$\|p-q\|_2\geq r/\ov c$}
\State \Return $p$
\Else
\State \Return ``Fail''
\EndIf
\EndFor
\EndProcedure
\State
\State {\bf end data structure}
\end{algorithmic}
\end{algorithm}

Finally, in Alg.~\ref{alg:ind03_main} and Alg.~\ref{alg:ind03_main_2}, we present our algorithm that solves $\afn$ (see Definition~\ref{def:afn}). As $\afn$ is the dual problem of approximate $\minip$,  this algorithm could be used to solve approximate $\minip$. 

\begin{algorithm}[H]
\caption{$\afn$ Algorithm: members, init, insert and delete}
\label{alg:ind03_main}
\begin{algorithmic}[1]
\State {\bf data structure} \textsc{AFN} \Comment{Theorem~\ref{thm:afn}}
\State
\State {\bf members}
\State \hspace{4mm} $\epsilon\in (0,1)$
\State \hspace{4mm} $\delta\in (0,1)$
\State \hspace{4mm} $s\in \N_+$ \Comment{Number of data structures}
\State \hspace{4mm} \textsc{DFN} $\text{dfn}_1,\text{dfn}_2,\ldots,\text{dfn}_{s}$
\State \hspace{4mm} $\bw\in \R_+$ \Comment{Boxwidth of all points}
\State \hspace{4mm} \textsc{SortedList} $T_1,\ldots,T_d$ \Comment{Max/min value for each dimension}
\State {\bf end members}
\State 
\Procedure{Init}{$A\in \R^{n\times d},\ov c\in (1,\infty),\delta\in (0,1)$}
\State $\epsilon\gets \ov c -1$
\State $s\gets \Theta(\log\log (d/\delta))$ 
\State $\text{dfn}_i$.\textsc{Init}$(A,\ov c)$ for all $i\in [s]$ \Comment{Alg.~\ref{alg:ind03_1}}
\For{$j=1 \to d$}
\State $T_j$.\textsc{Init}$(A_{*,j})$ \Comment{Alg.~\ref{alg:sorted_list}}
\EndFor
\State $\bw\gets \max_{j\in [d]}~|T_j.\textsc{Max}()-T_j.\textsc{Min}()|$ \Comment{1d boxwidth}
\EndProcedure
\State
\Procedure{Insert}{$p\in \R^d$}
\State $\text{dfn}_i$.\textsc{Insert}$(p)$ for all $i\in [s]$ \Comment{Alg.~\ref{alg:ind03_1}}
\For{$j=1 \to d$}
\State $T_j.\textsc{Insert}(p_j)$ \Comment{Alg.~\ref{alg:sorted_list}}
\EndFor
\State $\bw\gets \max_{j\in [d]}~|T_j.\textsc{Max}()-T_j.\textsc{Min}()|$
\EndProcedure
\State
\Procedure{Delete}{$p\in \R^d$}
\State $\text{dfn}_i$.\textsc{Delete}$(p)$ for all $i\in [s]$ \Comment{Alg.~\ref{alg:ind03_1}}
\For{$j=1 \to d$}
\State $T_j.\textsc{Delete}(p_j)$ \Comment{Alg.~\ref{alg:sorted_list}}
\EndFor
\State $\bw\gets \max_{j\in [d]}~|T_j.\textsc{Max}()-T_j.\textsc{Min}()|$
\EndProcedure
\State
\State {\bf end data structure}
\end{algorithmic}
\end{algorithm}

\begin{algorithm}[H]
\caption{$\afn$ Algorithm: query}
\label{alg:ind03_main_2}
\begin{algorithmic}[1]
\State {\bf data structure} \textsc{AFN} \Comment{Theorem~\ref{thm:afn}}
\State
\Procedure{Query}{$q\in \R^d$}
\State $\mathrm{lo}\gets \bw/2$
\State $\mathrm{hi}\gets \sqrt d/\epsilon\cdot \bw$
\State Binary search over the range $[\mathrm{lo},\mathrm{hi}]$ to search for $r\in \R_+$, 
\State with the predicate $\text{dfn}_i$.\textsc{Query}$(q,r)$ for $i\in [s]$
\State \Comment{ $\Theta(\log(d/\epsilon\delta))$ rounds}
\For{$i=1\to s$}
\State $p\gets \text{dfn}_i.\textsc{Query}(q,r)$
\If{$p\neq \text{``Fail''}$}
\State \Return $p$
\EndIf
\EndFor
\State \Return ``Fail''
\EndProcedure
\State
\State {\bf end data structure}
\end{algorithmic}
\end{algorithm}

\subsection{Success and Failure Probability of Random Projection}\label{sec:minip_success_failure_rp}
In this section, we analyze both the success and failure probability of random projection. To start with, we supply a technical lemma that upper bounds the failure probability that two points are far in the random direction but close in the original space.
\begin{lemma}
\label{lem:ind03_fail_prob}
Let $t$ be the solution to $e^{t^2/2}/t=2n$ and $T=rt/\ov c$. Let $\wh p$ be a point such that $\|\wh p-q\|_2<r/\ov c$. Then
\begin{align*}
    \underset{{g\sim {\cal N}(0,I)}}{\pr} [|\langle g,\wh p\rangle-\langle g,q\rangle|\geq T] \leq & ~ 1/n.
\end{align*}
\end{lemma}

\begin{proof}
Observe that
\begin{align*}
    \pr[|\langle g,\wh p\rangle-\langle g,q\rangle|\geq T] = & ~ \pr\left[\frac{|\langle g,\wh p-q\rangle|}{\|\wh p-q\|_2}\geq T/\|\wh p-q\|_2\right] \\
    \leq & ~ \pr\left[\frac{|\langle g,\wh p-q\rangle|}{\|\wh p-q\|_2}\geq \frac{T}{r/\ov c}\right] \\
    = & ~ \pr\left[\frac{|\langle g,\wh p-q\rangle|}{\|\wh p-q\|_2}\geq t\right] \\
    \leq & ~ 2\exp(-t^2/2)/t  \\
    \leq & ~ 1/n.
\end{align*}
The second step follows from $\|\wh p-q\|_2<r/\ov c$. For the fourth step, note that since standard Gaussian is 2-stable (Fact~\ref{fact:2stable}), we know that $\frac{\langle g,\wh p-q\rangle}{\|\wh p-q\|_2}$ follows a standard Gaussian distribution, so we can apply part 1 of the Gaussian concentration bound (Fact~\ref{fact:gauss_concentration_bd}).
\end{proof}

Next, we provide a lower bound on the success probability that when two points are far away from each other in the random direction, then they are far away in the original space.

\begin{lemma}
\label{lem:ind03_succ_prob}
Let $t$ be the solution to $e^{t^2/2}/t=2n$ and $T=rt/\ov c$. Let $\ell=O(n^{1/\ov c^2}\log^{(1-1/\ov c)/2} n)$. Let $p$ be a point such that $\|p-q\|_2\geq r$. Then
\begin{align*}
    \underset{g\sim N(0,I)}{\pr}[|\langle g,p\rangle-\langle g,q\rangle|\geq T] \geq & ~ 1/\ell.
\end{align*}
\end{lemma}
\begin{proof}
The proof is similar to Lemma~\ref{lem:ind03_fail_prob}, consider
\begin{align*}
    \pr[|\langle g,p\rangle-\langle g,q\rangle|\geq T] = & ~ \pr\left[\frac{|\langle g,p-q\rangle|}{\|p-q\|_2}\geq \frac{T}{\|p-q\|_2} \right] \\
    \geq & ~ \pr\left[\frac{|\langle g,p-q\rangle|}{\|p-q\|_2}\geq\frac{T}{r} \right] \\
    = & ~ \pr\left[\frac{|\langle g,p-q\rangle|}{\|p-q\|_2}\geq\frac{t}{\ov c} \right] \\
    \geq & ~ 2B\cdot \exp(-(t/\ov c)^2/2)/(t/\ov c) \\
    = & ~ \frac{2B\cdot \ov c(\exp(-t^2/2)/t)^{1/\ov c^2}}{t^{1-1/\ov c^2}} \\
    = & ~ \frac{2B\ov c}{n^{1/\ov c^2}t^{1-1/\ov c^2}}.
\end{align*}
The second step follows from $\|p-q\|_2\geq r$, the fourth step follows from Fact~\ref{fact:gauss_concentration_bd}. Note that by picking $\ell=O(n^{1/\ov c^2}\log^{(1-1/\ov c^2)/2}n)$, we get our desired result.
\end{proof}

\subsection{Guarantees of \texorpdfstring{$\dfn$}{~} Data Structure}\label{sec:minip_dfn}
In this section, we setup the theoretical guarantees of Alg.~\ref{alg:ind03_1}. We state and prove the following theorem regarding our $\dfn$ data structure. 

\begin{theorem}
\label{thm:dfn}
Let $P\subset \R^d$ be an $n$-point dataset and $\ov c>1$. There exists a randomized dynamic data structure (Alg.~\ref{alg:ind03_1}, \ref{alg:ind03_2}) that solves $\ov c$-$\dfn$ task (see Task~\ref{task:c_dfn}) using $O(n^{1+1/\ov c^2}\log n+dn^{1/\ov c^2}\log n)$ space with the following operations:
\begin{itemize}
    \item \textsc{Init}: Preprocess $P$ in $O(n^{1+1/\ov c^2}\log^2 n+dn^{1/\ov c^2}\log n)$ time;
    \item \textsc{Query}: Given a point $q\in \R^d$ and $r>0$, either outputs a point $\wh p\in P$ such that $\|\wh p-q\|_2\geq r/\ov c$ with constant probability or outputs ``Fail'' in $O(n^{1/\ov c^2}(d+\log n)\log n)$ time;
    \item \textsc{Insert}: Insert a point $p\in \R^d$ into the data structure in $O(n^{1/\ov c^2}\log^2 n)$ time;
    \item \textsc{Delete}: Delete a point $p\in P$ in $O(n^{1/\ov c^2}\log^2 n)$ time.
\end{itemize}
\end{theorem}
\begin{proof}
We prove four corresponding parts of Theorem~\ref{thm:dfn} accordingly.

\paragraph{Space:}
 Storing the $\ell\times d$ standard Gaussian matrix takes $O(\ell d)$ space. Maintaining $\ell$ sorted list takes  $O(\ell n)$ space. Thus, the total space is
 \begin{align*}
     O(\ell (d+n))= O(n^{1+1/\ov c^2}\log^{(1-1/\ov c)/2} n+dn^{1/\ov c^2}\log^{(1-1/\ov c)/2} n).
 \end{align*}

\paragraph{Procedure \textsc{Init}:} By Alg.~\ref{alg:ind03_1}, the initiation needs to initialize an $\ell\times d$ standard Gaussian matrix, which takes $O(\ell d)$ time, processing all points into sorted lists takes $O(\ell (d+n\log n))$ time. Thus, the total time for \textsc{Init} is 
\begin{align*}
O(\ell (d+n\log n))=O(n^{1+1/\ov c^2}\log n\log^{(1-1/\ov c)/2} n+dn^{1/\ov c^2}\log^{(1-1/\ov c)/2} n).
\end{align*}

\paragraph{Procedure \textsc{Query}:} We first show the correctness. Our goal is to prove that with constant probability, our data structure retrieves a pair $(p,i)$ among the first $2\ell+1$ pairs where each point $p$ satisfies $|\langle g_i,p\rangle-\langle g_i,q\rangle|\geq T$ and at least one of the point $p$ has the guarantee that $\|p-q\|_2\geq r/\ov c$.

We first justify that picking $2\ell+1$ pairs suffices for at least one point has desired distance guarantee, with constant probability. Let $Y_{\wh p,i}$ denote the event that a pair $(\wh p,i)\in P\times [\ell]$ has the property that $\|\wh p-q\|_2<r/\ov c$ and $|\langle g_i,q\rangle-\langle g_i,\wh p\rangle |\geq T$. Then
\begin{align*}
    \E[\sum_{(\wh p,i)} Y_{\wh p,i}] = & ~ n\ell \cdot \pr[Y_{\wh p,i}] \\
    \leq & ~ n\ell \cdot \frac{1}{n} \\
    = & ~ \ell.
\end{align*}
The second step follows from Lemma~\ref{lem:ind03_fail_prob}. Note that $\E[\sum_{(\wh p,i)}Y_{\wh p,i}]$ is the expected total number of such pairs, this means via a Markov bound, with the probability at least 1/2, there are no more than $2\ell$ such pairs. Thus, if we retrieve exactly $2\ell+1$ such pairs, there must be at least one pair $(p,i)$ with $\|p-q\|_2\geq r/\ov c$. 
Next, we analyze the failure probability when picking $2\ell+1$ pairs. Note that for a point $p\in P$ with $|\langle g_i,p\rangle-\langle g_i,q\rangle|\geq T$, the probability that $\|p-q\|_2<r/\ov c$ is at most $1-1/\ell$, due to Lemma~\ref{lem:ind03_succ_prob}. This means the probability that $\emph{some i}$ among the first $2\ell+1$ pairs has the property that $\|p-q\|_2\geq r/\ov c$ is at least
\begin{align*}
    1-(1-1/\ell)^{2\ell+1} \geq & ~ 1-1/e,
\end{align*}
this means we have a constant probability of success. Thus, our $\dfn$ data structure has a constant probability to output a point which is \emph{not} within the distance of $r/\ov c$ from $q$.

For the running time, note that we do at most $\ell$ rounds of search, at each round, we search the sorted lists, so we pay a total of $O(\ell \log n)$ for searching the lists. Finally, we need to examine these $2\ell+1$ pairs for their distances, this takes $O(\ell d)$ time. Therefore, the total running time is 
\begin{align*}
O(n^{1/\ov c^2}\log n \log^{(1-1/\ov c^2)/2}n+dn^{1/\ov c^2}\log^{(1-1/\ov c^2)/2}n).
\end{align*}
\paragraph{Procedure \textsc{Insert} and \textsc{Delete}:} It is obvious that the running time of both procedures is $O(\ell(d+\log n))$, which is the same as the time of procedure \textsc{Query}.
\end{proof}

\subsection{Guarantees  of \texorpdfstring{$\afn$}{~} Data Structure}\label{sec:minip_afn}
In this section we provide an analysis for an $\afn$ data structure implemented via $\dfn$ data structure. The idea is to use binary search to find the correct distance $r$. The search range is determined via the notion of \emph{box width}.
\begin{definition}
Given a dataset $P\subset \R^d$, we define the box width of $P$, denoted as $\bw(P)$ or $\bw$ if $P$ is clear from context as 
\begin{align*}
    \bw(P):= & ~ \max_{i\in [d]}~|\max_{p\in P}(p_i)-\min_{p\in P}(p_i)|.
\end{align*}
Note that $p_i$ denotes the $i$-th coordinate of point $p$.
\end{definition}
We now proceed with the formal statement and proof. 
\begin{theorem}
\label{thm:afn}
Let $P\subset \R^d$ be an $n$-point dataset, $\ov c>1$, $r>0$ and $\delta>0$. Let $\epsilon=\ov c-1$. There exists a randomized dynamic data structure (Alg.~\ref{alg:ind03_main}, \ref{alg:ind03_main_2}) that solves $(\ov c+\delta,r)$-$\afn$ task using space $O((n^{1+1/\ov c^2}\log n+dn^{1/\ov c^2}\log n)\log \log(d/\epsilon\delta)+dn)$ with the following operations:
\begin{itemize}
    \item \textsc{Init}: Preprocess $P$ in $O((n^{1+1/\ov c^2}\log^2 n+dn^{1/\ov c^2}\log n)\log\log (d/\epsilon\delta))$ time;
    \item \textsc{Query}: Given a point $q\in \R^d$, returns a $(\ov c+\delta)$-approximate furthest neighbor $p\in P$ with constant probability in $O(n^{1/\ov c^2}(d+\log n)\log n\log (d/\epsilon\delta)\log\log (d/\epsilon\delta))$ time;
    \item \textsc{Insert}: Insert a point $p\in \R^d$ into the data structure in $O(n^{1/\ov c^2}\log^2 n\log\log (d/\epsilon\delta)+d\log n)$ time;
    \item \textsc{Delete}: Delete a point $p\in \R^d$ from the data structure in $O(n^{1/\ov c^2}\log^2 n\log\log (d/\epsilon\delta)+d\log n)$ time.
\end{itemize}
\end{theorem}

\begin{proof}
We start with the space complexity
\paragraph{Space:}
We note that there are $s=O(\log \log(d/\epsilon\delta))$ $\dfn$ data structures to initialize, each data structure takes $O(n^{1+1/\ov c^2}\log n+dn^{1/\ov c^2}\log n)$ space. Moreover. the $d$ different sorted lists for each dimension takes $O(dn)$ space. Therefore, the final space is
\begin{align*}
    O((n^{1+1/\ov c^2}\log n+dn^{1/\ov c^2}\log n)\log \log(d/\epsilon\delta)+dn)
\end{align*}

Next, we prove four parts separately.
\paragraph{Procedure \textsc{Init}:} We note that there are $s=O(\log \log(d/\epsilon\delta))$ $\dfn$ data structures to initialize, each data structure takes $O(n^{1+1/\ov c^2}\log^2 n+dn^{1/\ov c^2}\log n)$ time. To initialize $d$ different sorted lists for each dimension, it takes $O(dn\log n)$ time. Finally, computing boxwidth $\bw$ takes $O(\log n)$ time. Thus, the total time in initialization phase is
\begin{align*}
O((n^{1+1/\ov c^2}\log^2 n+dn^{1/\ov c^2}\log n)\log \log(d/\epsilon\delta)).
\end{align*}

\paragraph{Procedure \textsc{Query}:} We need to prove the runtime and correctness of the procedure. For the runtime, we note that \textsc{Query} makes $O(\log (d/\epsilon\delta))$ calls to binary search with $O(\log \log (d/\epsilon\delta))$ different data structures. Each call takes $O(n^{1/\ov c^2}(d+\log n)\log n)$ time by Theorem~\ref{thm:dfn}. This completes the proof of runtime.

For correctness, note that for any query $q\in \R^d$, if $p$ is its furthest neighbor then $\|p-q\|_2\geq \bw/2$ since $q$ must be further from one point defining boxwidth. On the other hand, if the distance from $q$ to the center of box is at least $2\sqrt d/\epsilon\cdot \bw$, then any point in $P$ is a $(1+\epsilon)$-approximate furthest neighbor. To see this, note that any point $p\in P$ is at most $\sqrt d/2\cdot \bw$ away from the center, so the nearest point from the box to the center is at least $(\frac{2}{\epsilon}-\frac{1}{2})\sqrt{d}\cdot \bw$. On the other hand, the furthest point on the box to $q$ has a distance $(\frac{2}{\epsilon}+\frac{1}{2})\sqrt d\cdot \bw$, it suffices to show that $(\frac{2}{\epsilon}+\frac{1}{2})\sqrt{d}\cdot \bw/(1+\epsilon)\leq (\frac{2}{\epsilon}-\frac{1}{2})\sqrt d\cdot \bw$, since the furthest neighbor to $q$ from dataset $P$ must have distance smaller than $(\frac{2}{\epsilon}+\frac{1}{2})\sqrt d\cdot \bw$. Note that
\begin{align*}
    (\frac{2}{\epsilon}+\frac{1}{2})/(1+\epsilon) = & ~ \frac{4+\epsilon}{2\epsilon(1+\epsilon)},
\end{align*}
On the other hand,
\begin{align*}
    \frac{2}{\epsilon}-\frac{1}{2} = & ~ \frac{4-\epsilon}{2\epsilon} \\
    = & ~ \frac{4+3\epsilon-\epsilon^2}{2\epsilon(1+\epsilon)}.
\end{align*}
Since $\epsilon\in (0,1)$, we always have $3\epsilon-\epsilon^2>\epsilon$, as desired.

This gives a lower and upper bound on binary search, namely we search the range $[\bw/2,2\sqrt d/\epsilon\cdot \bw]$, hence, we need $O(\log \frac{d}{\epsilon\delta})$ rounds to achieve a $\delta$-precision solution. This leads to a $\ov c(1+\epsilon)(1+\delta)=(1+\epsilon)^2(1+\delta)$-approximation furthest neighbor. By picking $\epsilon$ as $\epsilon/2$ and $\delta$ as $\delta/3$, this leads to a $(1+\epsilon+\delta)=(\ov c+\delta)$-approximate furthest neighbor. Finally, to amplify the success probability of each query, we need to use $O(\log \log \frac{d}{\epsilon\delta})$ different data structures. This completes the correctness analysis.

\paragraph{Procedure \textsc{Insert} and \textsc{Delete}:} Both of these procedures require to insert or delete a point to $s$ different data structures and update the sorted list for each dimension, then compute the new boxwidth. The insert/delete point step takes $O(n^{1/\ov c^2}\log^2 n\log\log \frac{d}{\epsilon\delta})$ time and update the sorted list takes $O(d\log n)$ time. This completes the proof.
\end{proof}

\end{document}